\documentclass[fleqn,usenatbib]{mnras}

\usepackage{newtxtext,newtxmath}

\usepackage[T1]{fontenc}
\usepackage{lineno} 
\DeclareRobustCommand{\VAN}[3]{#2}
\let\VANthebibliography\thebibliography
\def\thebibliography{\DeclareRobustCommand{\VAN}[3]{##3}\VANthebibliography}

\usepackage{graphicx}	
\usepackage{amsmath}	

\title[Galactic Open Clusters: Structure and Variability]{Unveiling Dynamics and Variability in Open Clusters: Insights from a Comprehensive Analysis of Six Galactic Clusters}

\author[K. Belwal et al.]{
Kuldeep Belwal,$^{1}$\thanks{E-mail: kuldeepbelwal1997@gmail.com}
D Bisht,$^{1}$\thanks{E-mail: devendrabisht297@gmail.com}
Ing-Guey Jiang,$^{2}$\thanks{E-mail: jiang@phys.nthu.edu.tw}
R. K. S. Yadav,$^{3}$
Ashish Raj,$^{1}$$^{,5}$
Geeta Rangwal,$^{3}$
\newauthor
Arvind K. Dattatrey,$^{3}$$^{,6}$
Mohit Singh Bisht,$^{1}$
and Alok Durgapal$^{4}$
\\
\\
$^{1}$Indian Centre for Space Physics,466 Barakhola, Singabari road, Netai Nagar,
Kolkata, 700099, India\\
$^{2}$Department of Physics and Institute of Astronomy, National Tsing-Hua University, Hsin-Chu, Taiwan\\
$^{3}$Aryabhatta Research Institute of Observational Sciences, Manora Peak, Nainital 263129, India\\
$^{4}$ Center of Advanced Study, Department of Physics, D. S. B. Campus, Kumaun University, Nainital 263002, India\\
$^{5}$Uttar Pradesh State Institute of Forensic Science (UPSIFS)
Aurawan, P.O. Banthra, Lucknow 226401, U.P, India\\
$^{6}$ Indian Institute of Astrophysics, 560034 Bangalore, India
}

\date{Accepted XXX. Received YYY; in original form ZZZ}

\pubyear{\the\year{}}

\begin{document}
\label{firstpage}
\pagerange{\pageref{firstpage}--\pageref{lastpage}}
\maketitle

\begin{abstract}

We present a kinematic and dynamical analysis of six Galactic open clusters—NGC~2204, NGC~2660, NGC~2262, Czernik~32, Pismis~18, and NGC~2437, using \textit{Gaia}~DR3. We used Bayesian and Gaussian Mixture Model (GMM) methods to identify cluster members, but chose GMM because it's more appropriate for low-mass stars. Estimated distances range from 1.76 to 4.20~kpc and ages from 0.199 to 1.95~Gyr, confirming their intermediate-age nature. King model fits indicate compact morphologies, with core radii of 1--10~arcmin and cluster radii of 5--24~arcmin. We identify 13 BSS and 3 YSS members, whose central concentrations suggest origins via mass transfer or stellar collisions. The mass function slopes (0.96--1.19) are flatter than the Salpeter value, which indicates that these clusters have undergone dynamical mass segregation.  Orbit integration within a Galactic potential indicates nearly circular orbits (eccentricities 0.02--0.10), vertical excursions within $\pm$132~pc, and guiding radii near the solar circle, suggesting disk confinement. These clusters likely formed in the thin disk and are shaped by Galactic tidal perturbations, facilitating the rapid loss of low-mass members. Additionally, twelve variable stars were found across four clusters using \textit{TESS} light curves, including $\gamma$~Doradus and SPB pulsators, eclipsing binaries, and a yellow straggler candidate. Periods were derived via Lomb-Scargle analysis. Two eclipsing binaries (TIC~94229743 and TIC~318170024) were modeled using PHOEBE, yielding mass ratios of 1.37 and 2.16, respectively. Our findings demonstrate that integrating orbital dynamics and variable star studies presents valuable insights into the evolutionary pathways of open clusters.

\end{abstract}




\section{Introduction} \label{sec: intro}

Open clusters (OCs) are fundamental for understanding stellar evolution and formation, as they are primarily located in the Galactic disc and span a wide range of ages, covering the full lifetime of the Galactic disc \citep{freeman2002new, twarog1997some}. Older clusters trace the thick disc, while younger ones are typically found in the spiral arms of the thin disc, where perturbations are more frequent. The motion and spatial distribution of these clusters within the Galaxy help in understanding its gravitational potential and the perturbations that shape its structure and dynamics \citep{soubiran2018open}. Most Galactic OCs evaporate within approximately 100 million years \citep{wielen1971age}. However, clusters older than 1 Gyr are thought to have survived due to their orbital properties, which keep them away from the disruptive forces of the Galactic plane \citep{friel1995old}. The dynamical evolution of stars in OCs is driven by both internal stellar processes and external influences, such as tidal forces from the Galactic disc and encounters with molecular clouds. Key dynamical processes contributing to the disruption of OCs include internal interactions among members, stellar evolution, encounters with molecular clouds, and gravitational disturbances from the Galactic potential \citep{gieles2006star}. OCs that survive are typically more massive, have higher central concentration, and follow orbits around the Galactic center that minimize exposure to disruptive forces \citep{friel1995old, gustafsson2016gravitational}. Intermediate- and old-age OCs are especially valuable for testing theoretical stellar evolution models. Most stars in the Galaxy are believed to form clusters \citep{portegies2010young}, making OCs natural laboratories for studying stellar development. However, since cluster stars are often contaminated by foreground and background stars, it is essential to accurately distinguish cluster members from field stars to derive reliable cluster parameters. Recent studies have focused on membership analysis in OCs, investigating various properties \citep{cantat2020painting, castro2019hunting, bisht2019mass, rangwal2023investigating}. 

OCs host various populations of variable stars, including rotating variables, pulsating stars, eclipsing binaries, and non-periodic variables spanning a wide range of stellar masses. The detection of variable stars delivers important constraints for stellar pulsation and evolutionary models. Pulsating variables exhibit magnitude variations ranging from a few millimagnitudes (mag) to hundreds of mag, with periods from hours to days, offering valuable insights into stellar interiors and evolution as their variability arises from radial and non-radial pulsation \citep{gautschy1993non, kang2007variable}. Intermediate-age clusters are excellent tools for analyzing short-period variables. The period-luminosity relations of pulsating stars, such as Cepheids and RR Lyrae, are highly effective tools for distance measurement and studying the structure of the Galaxy \citep{iorio2019shape, skowron2019three}. Eclipsing binaries are valuable for testing theoretical stellar models \citep {torres2010accurate}. The evolutionary link between Algol-type and W UMa-type binaries offers insights into angular momentum loss, tidal locking, and mass transfer processes in binary systems \citep{jiang2014detached, chen2020zwicky}.


This study used the six OC samples NGC 2204, NGC 2660,  NGC 2262, Czernik 32, Pismis 18, and NGC 2437. These clusters are located very close to the Galactic plane. Their diverse ages, distances, and metallicities help us to trace the Galactic disk's chemical enrichment, radial abundance gradients, and star formation history, delivering fundamental insights into stellar and Galactic dynamics. The cluster identiﬁcation chart is taken from the Digitized Sky Survey (DSS) {\footnote[1]{https://archive.stsci.edu/cgi-bin/dss}} and shown in figure \ref{fig:ID_chart}. The available information about these clusters in the literature is discussed here:

\begin{figure*}
    \centering
    \vspace{-0.1cm}
    \includegraphics[width=5.4cm,height=5.2cm]{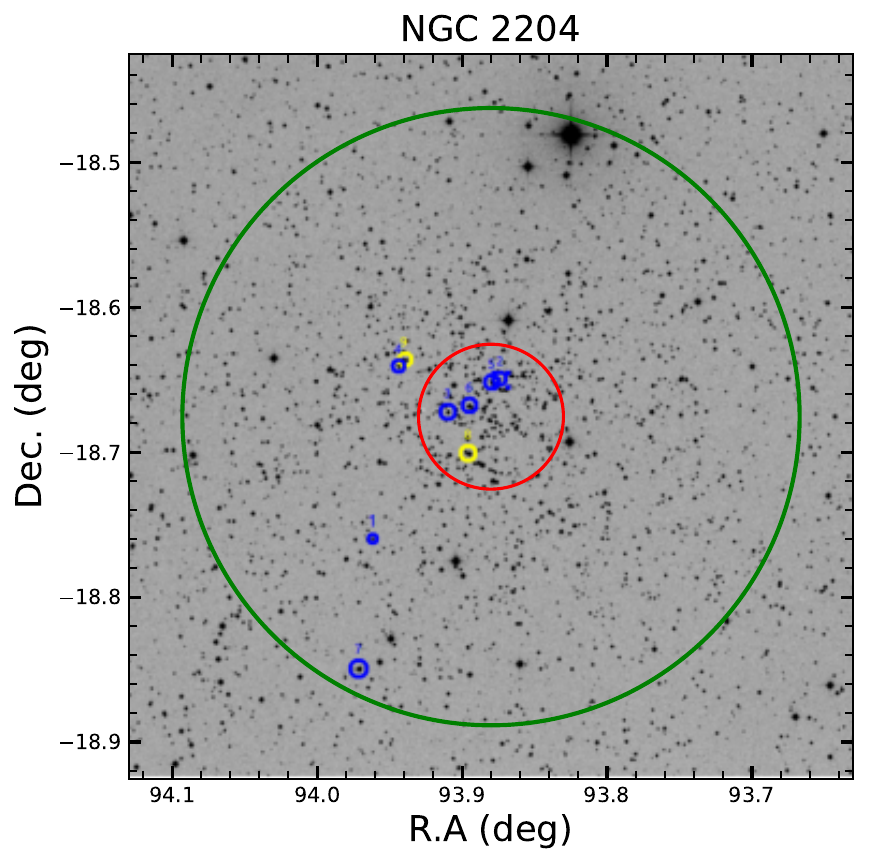}
    \includegraphics[width=5.4cm,height=5.2cm]{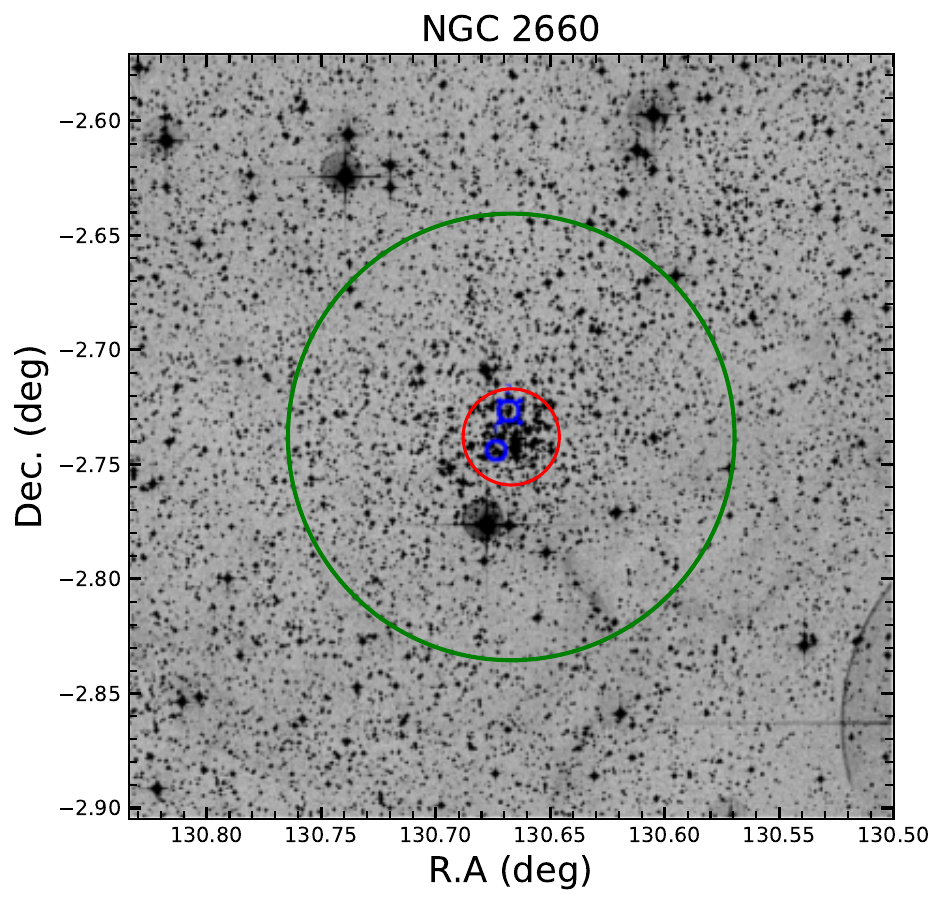}
    \includegraphics[width=5.4cm,height=5.2cm]{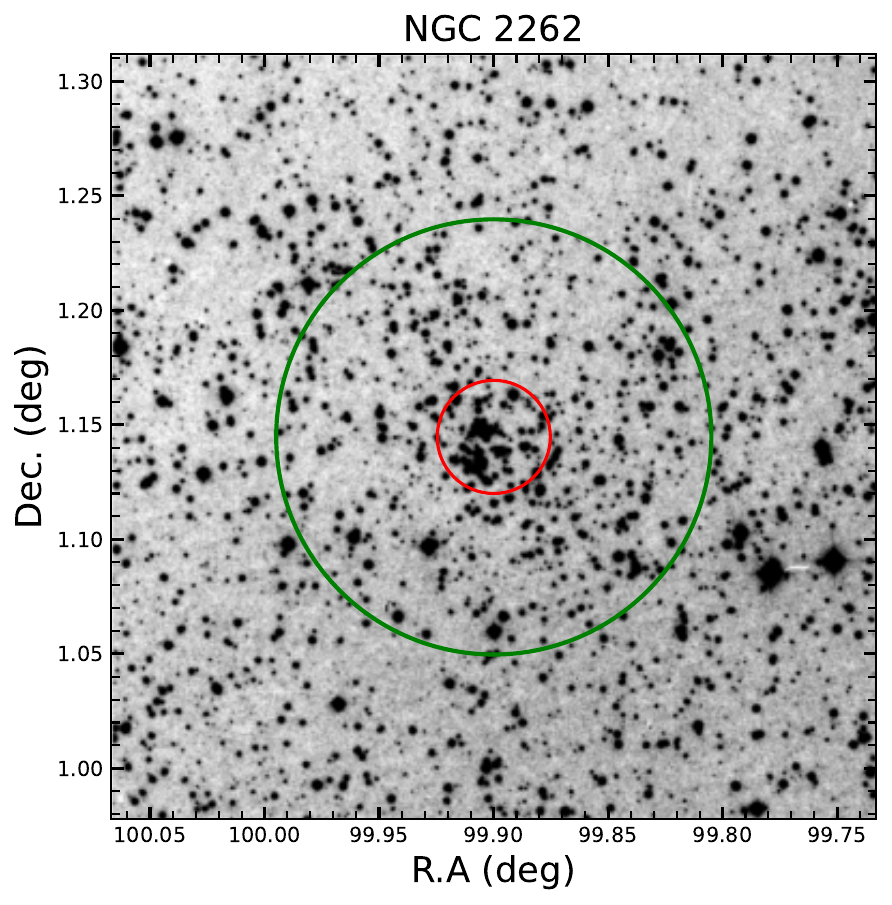}
    \includegraphics[width=5.4cm,height=5.2cm]{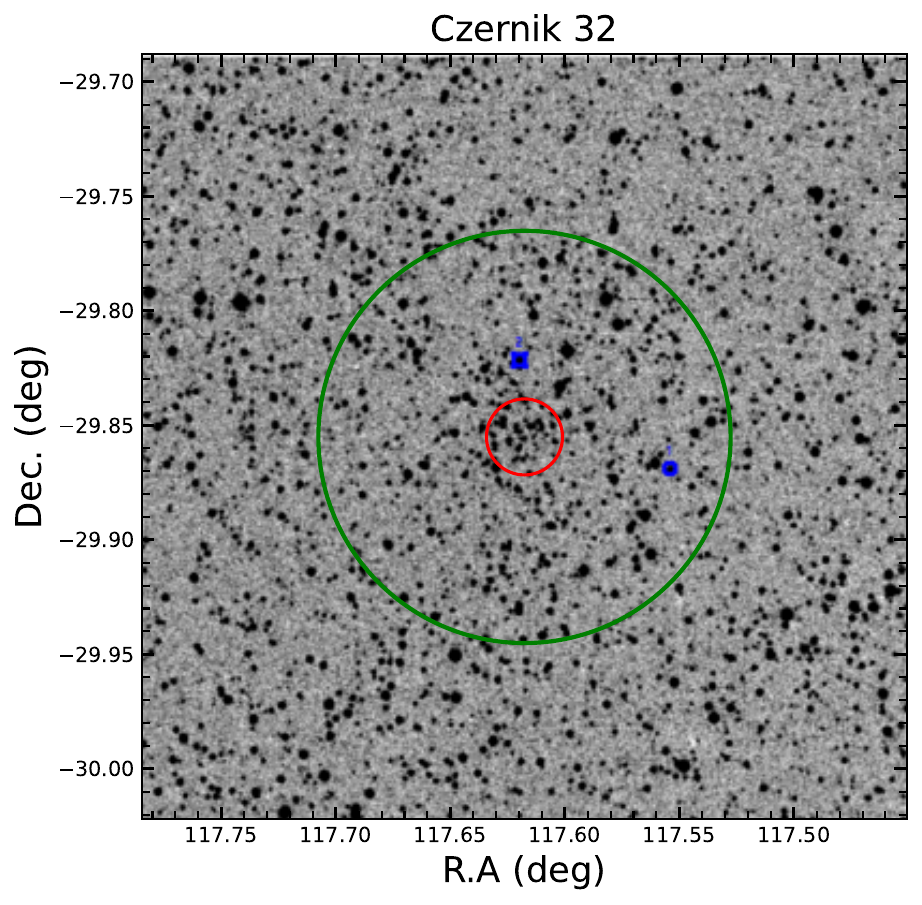}
    \includegraphics[width=5.4cm,height=5.2cm]{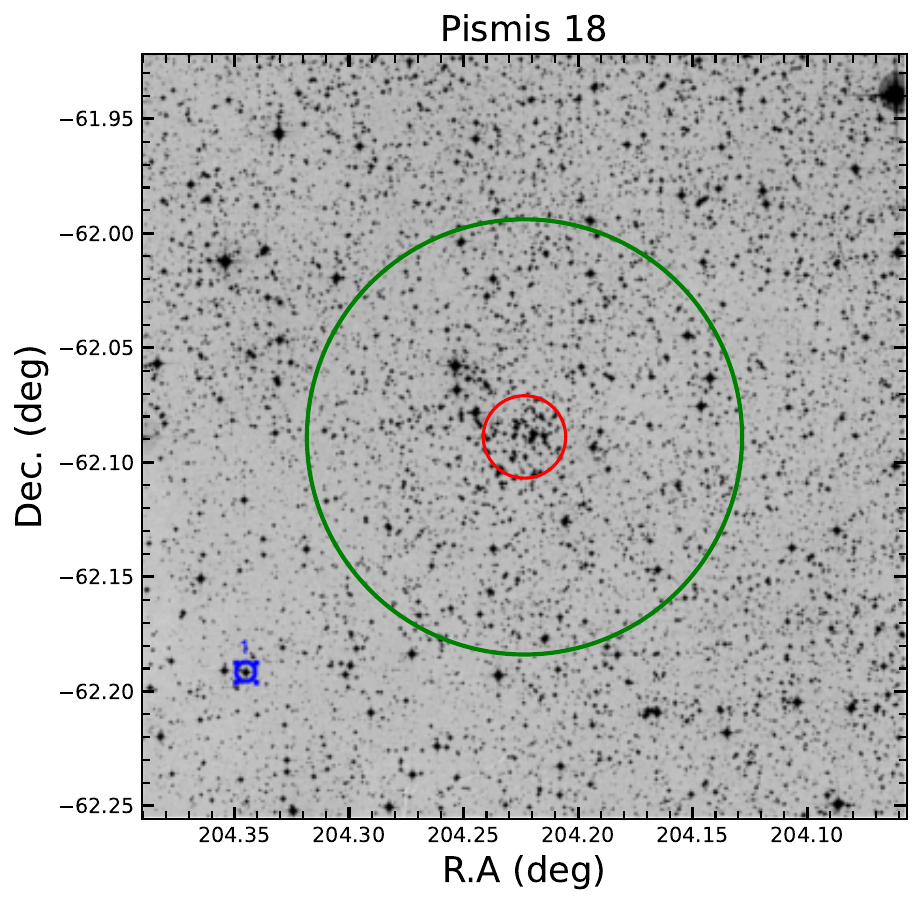}
    \includegraphics[width=5.4cm,height=5.2cm]{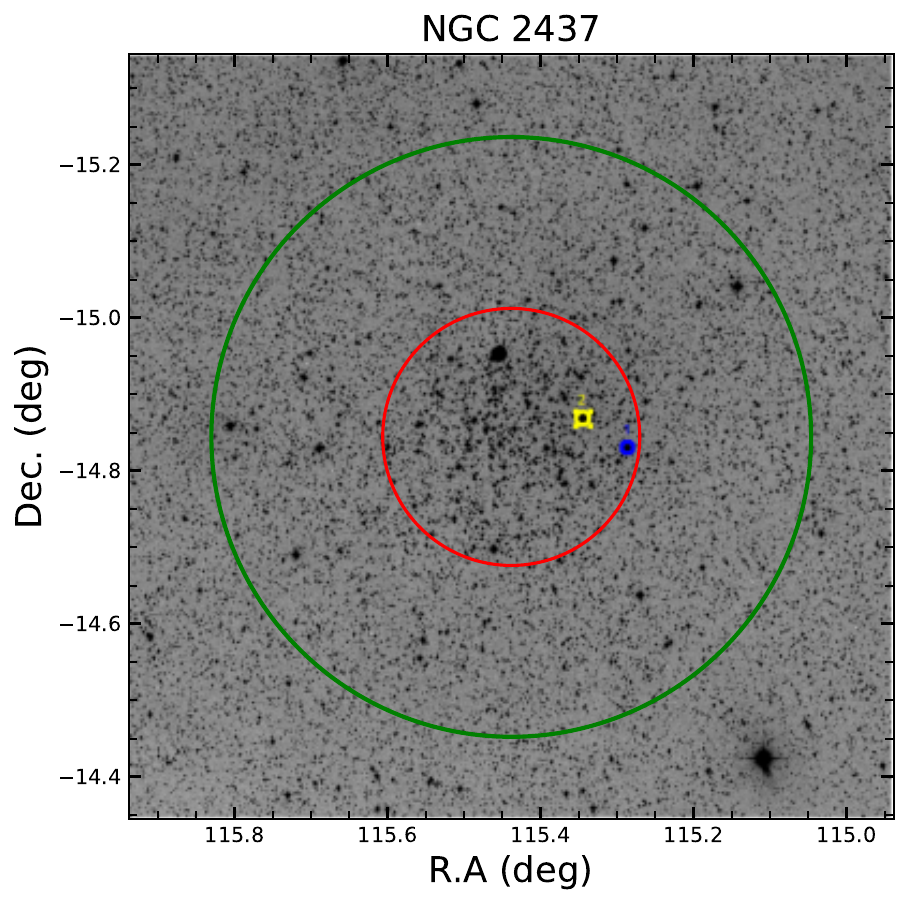}
    \caption{Identification chart of six OCs taken from the  STScI Digitized Sky Survey, oriented in the northeast direction (north is up and east is left). The inner red circle represents the core radius, while the outer green ring denotes the cluster radius. The blue and yellow open circles represent the BSS and YSS, respectively. }
    \label{fig:ID_chart}
\end{figure*}

{\bf {NGC 2204}}

NGC 2204 (C0613-186) is located in the constellation of Canis Major ($\alpha_{2000}=06^{h}15^{m}33^{s}, \delta_{2000}= -18^{\circ}39{'}54{''})$ and corresponding Galactic coordinates ($l = 226.^{\circ}0208, b = -16^{\circ}.1143$). The calculated parameters in the literature for the age, reddening, and distance range vary from 1.85 to 2.08 Gyr, 0.1 to 0.2 mag, and 3.90 to 4.20 kpc, respectively. \citet{angelo2023enlightening} estimated the cluster's core radius and distance modulus to be $4.81 \pm 0.82$ arcminutes and 13.03 $\pm$ 0.15 mag respectively, and \citet{anthony2024li} reported the radial velocity  90.9 $\pm$ 1.94 km $s^{-1}$. In this cluster, \citet{jadhav2021high} found 6, and \citet{rain2021new} found 7  Blue straggler stars (BSS) and 2 Yellow straggler stars (YSS).


{\bf NGC 2660}

NGC 2660 (C0840 - 469) is situated in the constellation of Vela ($\alpha_{2000}=08^{h}42^{m}38^{s}, \delta_{2000}= -47^{\circ}12{'}00{''}$) and corresponding Galactic coordinates ($l = 265^{\circ}.931, b = -03^{\circ}.015$). The estimated parameters in literature age, reddening, and distance range vary from 0.93 - 1.64 Gyr, 0.4 - 0.5 mag, and 2.6 - 2.8 kpc, respectively \citep{sandrelli1999intermediate,jeffery2016bayesian, cantat2020painting, dias2021updated}. \citet{angelo2023enlightening} estimated the cluster's core radius, half-mass radius, and distance modulus to be $0.86 \pm 0.21$ arcminutes and $2.04 \pm 0.35$ parsec and 11.96 $\pm$ 0.25 mag, respectively. In this cluster \citet{jadhav2021high} found 1 and \citet{rain2021new} found 2 BSS. In the study by \citet{bragaglia2024gaia}, the radial velocity of this cluster is reported to be 22.47 $\pm$ 0.27 km $s^{-1}$.

{\bf {NGC 2262}}

NGC 2262 (C0635+012) is located in the constellation of Puppis ($\alpha_{2000}=06^{h}39^{m}38^{s}, \delta_{2000}= +01^{\circ}08{'}36{''})$ and corresponding Galactic coordinates ($l = 245.^{\circ}.8931, b = -1^{\circ}.7398$). The estimated parameters age, reddening, distance modulus, and distance range vary from 0.645 - 1.17 Gyr, 0.55 - 0.63 mag, 12.54 - 14.50 mag, and 2.51 - 3.6 kpc, respectively, given by  \citet{carraro2005intermediate,hoq2015open,dias2021updated, cantat2020painting}. The radial velocity of this cluster has been reported as 53.64 $\pm$ 2.05 km $s^{-1}$ in the study conducted by \citet{zhong2020exploring}. 

{\bf {Czernik 32}}

Czernik 32 (C0748-297) is located in the constellation of Puppis $(\alpha_{2000}=07^{h}50^{m}30^{s}, \delta_{2000}= -29^{\circ}51{'}00{''})$ and corresponding Galactic coordinates ($l = 245.^{\circ}893, b = -1^{\circ}.7398$). The estimated parameters age, reddening, distance modulus, distance, and Galactic distance range vary from 1.08 - 1.35 Gyr, 0.66 - 0.85 mag, 12.5 - 15.7 mag, 3.16 - 4.14 kpc, and 9.75 - 10.80 kpc, respectively, given by \citet{bica2005properties,carraro2005intermediate,angelo2020characterizing, cantat2020painting}. In this cluster \citet{jadhav2021high} found 2 BSS.

{\bf {Pismis 18}}

Pismis 18 (C1333-619) is located in the constellation of Centaurus ($\alpha_{2000}=13^{h}36^{m}55^{s}, \delta_{2000}= -62^{\circ}05{'}36{''}$), with corresponding Galactic coordinates ($l = 308^{\circ}.226, b = 0^{\circ}.3134$). As an intermediate-age open cluster, its estimated distance, reddening $(E(B-V))$, and age fall within the ranges of 1.79 - 2.47 kpc, 0.50 - 0.61 mag, and 0.70 - 1.2 Gyr, respectively, given by \citet{piatti1998photometric, cantat2020painting, dias2021updated, hatzidimitriou2019gaia}. \citet{tadross2008main} estimated the cluster's radius to be 5.6 arcminutes, while \citet{hatzidimitriou2019gaia} reported a radius of approximately 5.0 arcminutes and a metallicity [Fe/H] of $0.23 \pm 0.05$ dex. Using 26 high-confidence members, \citet{hatzidimitriou2019gaia} estimated the radial velocity to be $-27.5 \pm 2.5$ km s$^{-1}$. 


{\bf {NGC 2437}}

NGC 2437 (C0739$-$147) is located in the constellation of Puppis ($\alpha_{2000}=07^{h}41^{m}46^{s}, \delta_{2000}= -14^{\circ}48^{\prime}36^{\prime \prime}$), and corresponding Galactic coordinates ($l = 231^{\circ}.889, b = 4^{\circ}.0493$). A young rich cluster with an estimated age, reddening, and distance ranging from 0.199 - 0.302 Gyr, 0.12 - 0.20 mag, and 1.5 - 1.7 kpc, respectively, given by \citet{sharma2006wide,davidge2013open, cantat2020painting, jadhav2021high}. \citet{sanchez2020catalogue} estimated the cluster’s radius to be around 25 arcminutes. In this cluster \citet{rain2021new} found 1 BSS and 1 YSS. The cluster may also be physically associated with the nearby planetary nebula NGC 2438 \citep{bonatto2008discovery}. 

The selection of this sample was guided by several considerations. First, these are relatively rich, intermediate-age to old OCs that had been studied in the pre-Gaia era. At that time, measurements of proper motion and parallax were significantly less precise. Revisiting these clusters with Gaia DR3 allows us to derive far more accurate astrometric and photometric parameters. This leads to an improved characterization of their kinematical and dynamical properties. Second, the clusters lie close to the Galactic plane, where their fields are strongly contaminated by foreground and background stars. Gaia’s high-precision data is essential for robust membership determination. Third, TESS observations serve as an independent dataset that strengthens our characterization of stellar variability. This is particularly useful for detecting eclipsing binaries and variable stars that serve as additional probes of the stellar populations. Finally, the chosen systems span a range of ages, total masses, and Galactocentric positions. This offers a representative set to explore how structural and dynamical properties evolve under different conditions within the Galactic disk. Together, these criteria ensure that the present study provides accurate and comprehensive insights into the internal dynamics and evolutionary states of Galactic open clusters.\\

This study presents a photometric analysis of six OCs, focusing on their spatial structure, fundamental parameters, and age to gain insight into their dynamical evolution. Membership probabilities were estimated for stars within the clusters down to G $\sim$ 20 mag. Highly probable members were employed to analyse the mass function and investigate mass segregation.

The paper is organised as follows: Section \ref{sec:data} describes the data used in this study. Section \ref{sec:Membership probability} discusses the determination of stellar membership within the clusters. Section \ref{sec:Structural Parameters of the Cluster} derives the structural parameters of the clusters. Section \ref{sec:Distance Estimated Based on Parallax} presents distance estimates based on parallax measurements, while physical parameters such as age and photometric distance are derived using theoretical isochrone fitting and the location of the BSS within the cluster, as discussed in Section \ref{sec: Age and Distance from Isochrone Fitting}. Section \ref{sec: dynamical study} examines the dynamical evolution of the clusters. Section~\ref{sec: Variable_stars_classification} discusses the identification and classification of variable stars, as well as light curve modelling. Finally, we present a comparative discussion of the six clusters in Section~\ref{sec: discussion}, and the main findings are summarized in Section~\ref{sec: Conclusion}.


\section{Data Sources and Selection Criteria} \label{sec:data}
\subsection{Gaia~DR3}
\begin{figure}
    \centering
    \hspace{-0.9cm}
    \includegraphics[width=9cm,height=7.0cm]{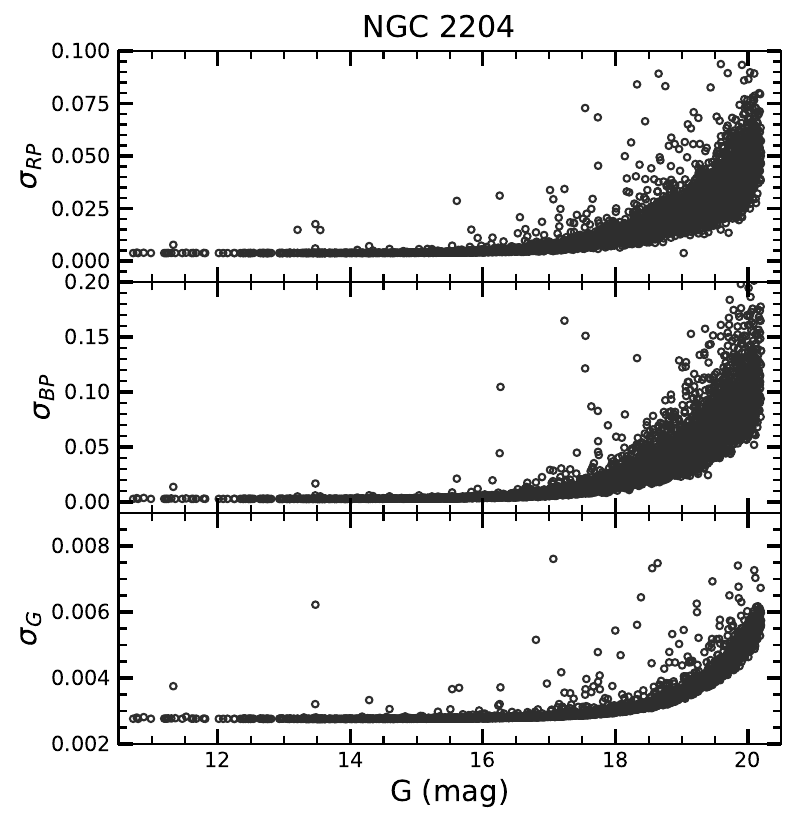}
    \caption{Photometric uncertainties in the \textit{G}, $G_{\mathrm{BP}}$, and $G_{\mathrm{RP}}$ bands plotted as a function of \textit{G} magnitude for stars in the open cluster NGC~2204.}
    \label{fig:mag_error}
\end{figure}

\emph{Gaia} DR3 \citep{collaboration2023gaia} is used for the photometric and astrometric study of all clusters and for determining their structural parameters. \emph{Gaia} DR3 provided celestial positions and {\textit{G}}-band magnitudes for a vast dataset of around 1.8 billion sources, with a magnitude measurement extending up to 21 mag. Additionally, \emph{Gaia} DR3 provides valuable parallax, proper motion, and color information ($G_{BP}$ – $G_{RP}$) for a subset of this dataset, specifically for 1.5 billion sources. The uncertainties in parallax values are $\sim$ 0.02 - 0.03 milli arcsecond (mas) for sources at  {\textit{G}} $\leq$ 15 mag and $\sim$ 0.07 mas for sources with  {\textit{G}} $\sim$ 17 mag.
The adopted search radii were chosen with reference to cluster sizes reported in the literature. Specifically, the reported radii are 9.3 arcmin for NGC 2204 \citep{kharchenko2013global}, 4.3 arcmin for NGC 2660 \citep{sanchez2020catalogue}, 1.0 arcmin for NGC 2262 \citep{carraro2005intermediate}, 4.75 arcmin for Czernik 32 \citep{sanchez2020catalogue}, and for Pismis 18, values of 5.6 arcmin and 5.0 arcmin are given by \citet{tadross2008main} and \citet{hatzidimitriou2019gaia}, respectively. For NGC 2437, a radius of 25.5 arcmin is reported by \citet{sanchez2020catalogue}. These literature values guided our choice of extraction radii, which ranged from 10 to 30 arcmin, adopted in this study.
We applied several selection criteria to collect the data, including a Renormalised Unit Weight Error (RUWE) factor of $\leq$ 1.4 \citep{lindegren2021gaia}, parallax $\geq$ 0, proper motion error $\leq$ 0.6, and G magnitude $\leq$ 20.2. The uncertainties in the corresponding proper motion components are $\sim$ 0.01 - 0.02 mas $yr^{-1}$ (for  {\textit{G}} $\leq$ 15 mag), $\sim$0.05 mas $yr^{-1}$ (for  {\textit{G}} $\sim$ 17 mag), $\sim$0.4 mas $yr^{-1}$ (for  {\textit{G}} $\sim$ 20 mag) and $\sim$1.4 mas $yr^{-1}$ (for $ G$ $\sim$ 21 mag). For cluster Pismis 18, we observed that at the fainter end of the G magnitude range (17 to 20), the associated photometric error exceeds 0.01 mag. In contrast, the error remains below 0.01 mag within the same magnitude range in the G-band for the remaining clusters. To illustrate this trend, a reference plot for NGC 2204 is presented in Figure \ref{fig:mag_error}.\\
We cross-matched our cluster members with the Gaia DR3 variability catalogue. Among all stars analyzed in this work, only a single source (TIC 318170024) is flagged as variable in Gaia, while no variability information is available for the remaining stars. We also examined the astrometric quality indicator RUWE for all selected members. In every case, the RUWE values are $\leq$ 1.4, confirming that the stars included in our study have reliable Gaia astrometry. This ensures that spurious or poor-quality solutions do not compromise our membership selection.

\subsection{TESS Data}

Transiting Exoplanet Survey Satellite (TESS) is an almost all-sky photometric survey primarily focused on detecting transiting exoplanets \citep{ricker2015transiting}. TESS observes each 24$^{\circ} \times$ 96$^{\circ}$ sector for approximately 27 days of near-continuous monitoring. It is equipped with four identical cameras, each covering a 24$^{\circ} \times$ 24$^{\circ}$ field of view. One camera contains four CCDs that generate 2048 $\times$ 2048 pixel images with a pixel scale of 21$^{\prime\prime}$. During the primary mission (Sectors 1--26), full-frame images (FFIs) were acquired with a 30-minute cadence. In the first extended mission (Sectors 27--55), the cadence was shortened to 10 minutes. Beginning with Cycle 5 (Sectors 56 and beyond), the duration was further reduced to 200 seconds. Photometric data are available for sources brighter than approximately 17th magnitude in the TESS bandpass, which spans 600-1000 nm and is centered on the Cousins I-band. In the present study, we utilise TESS photometric data to investigate the stellar variability and properties of the open clusters NGC 2204, NGC 2262, Pismis 18, and NGC 2437. Specifically, we use the 30-minute cadence FFIs from Sectors 1--26 and the 10-minute cadence FFIs from Sectors 27--39, processed by the TESS Science Processing Operations Centre (SPOC) FFI pipeline\footnote{\url{https://heasarc.gsfc.nasa.gov/docs/tess/documentation.html}}.

\section{Membership probability}
\label{sec:Membership probability}
\subsection{Bayesian Method}
OCs, mostly in the Galactic disk, are usually contaminated by foreground and background stars. To derive the actual parameters of the cluster, we need to separate the cluster stars from the field stars. The stars of clusters have the same kinematical properties as well as the same heliocentric distance. Hence, proper motion and parallax of stars can be used to differentiate cluster members from field stars. We utilized parallax and proper motion data from the Gaia DR3 catalog to select member stars of the clusters in this study. We used the maximum likelihood method described by \cite{balaguer1998determination} for these clusters. Previous studies \cite{bisht2021detailed,bisht2021deep,bisht2022comprehensive,panwar2024low,sariya2021comprehensive} used this method in their analysis. This method relies entirely on the proper motion of stars, with \emph{Gaia} DR3 providing precise measurements of astrometric parameters.

\begin{figure}
    \centering
    \includegraphics[width=9cm,height=8.5cm]{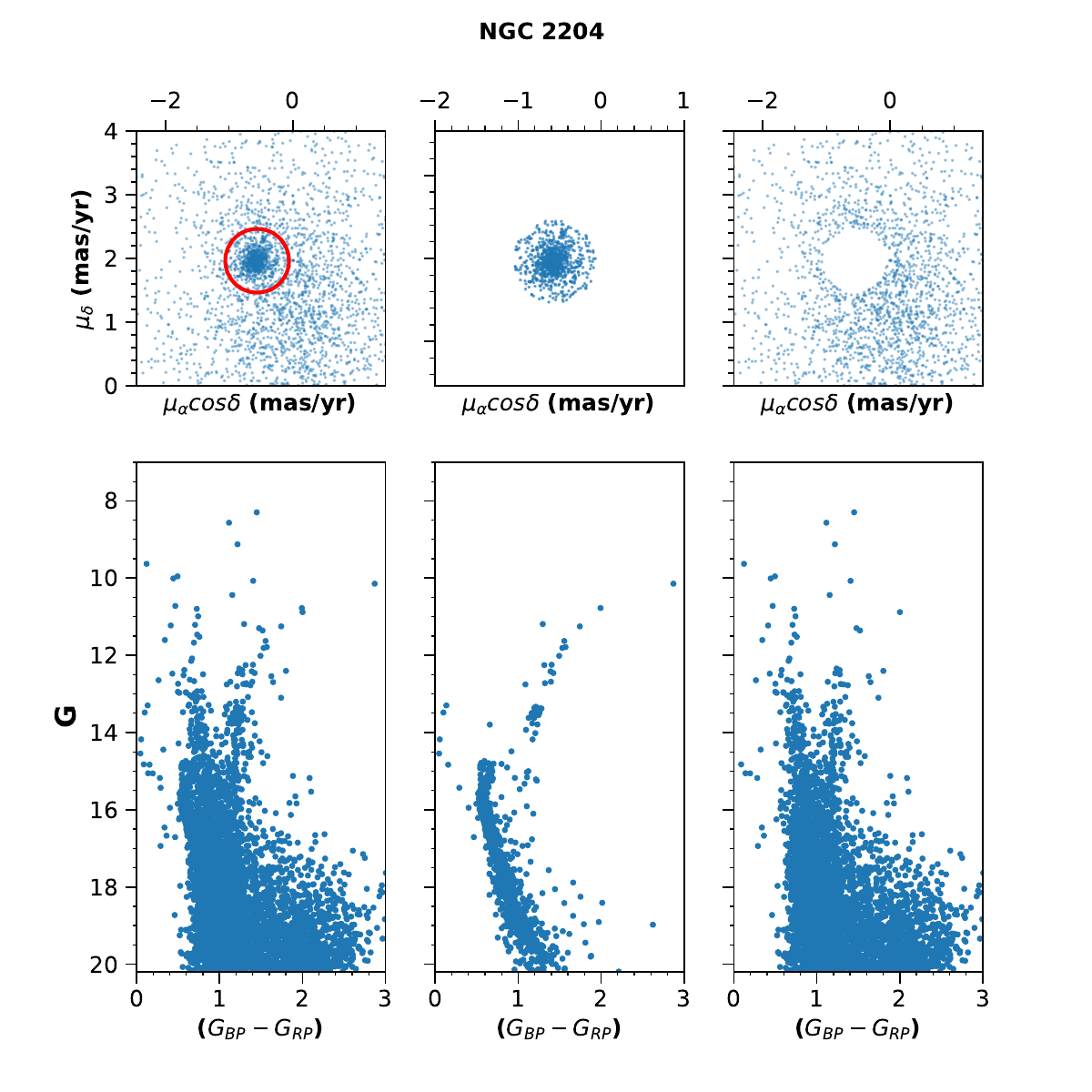}
    \caption{This figure illustrates the initial cluster member separation method based on Gaia's proper motions for the cluster. The top panel displays the VPDs, while the bottom panels show the Gaia CMDs for the total stars, cluster members, and field stars. In the top panel, a circle with a radius of 0.5 mas yr$^{-1}$ represents the member stars within the cluster field.}
    \label{fig:vpd}
\end{figure}

We discussed this method in \citet{belwal2024exploring} to calculate the membership probability of stars in the cluster region. We define a circular region around the concentrated clustering region in proper motion space; the radius of the circle is based on $r = \sqrt{r_{0}^{2} + \sigma^{2}}$, where $r_{0}$ is based on visual inspection and $\sigma$ is an error in proper motion. We have chosen the radii of each circle as 0.6, 0.6, 0.55, 0.45, 0.5, and 0.5 mas yr$^{-1}$ for NGC 2204, NGC 2660, NGC 2262, Czernik 32, Pismis 18, and NGC 2437, respectively. In Figure \ref{fig:vpd}, we present the vector point diagram (VPDs) for NGC 2204.

After selecting stars within the circle of the VPDs, we applied an additional selection criterion based on the mean parallax value. We calculated the arithmetic mean of the parallaxes for the stars within the VPD circle with G magnitudes brighter than 20 mag. A star is considered a most probable cluster member if its parallax angle is within $3\sigma$ of the mean parallax value. In this way, we get the actual number of probable cluster members. We plotted histograms for each cluster to estimate the mean proper motion values. A Gaussian function was fitted to the histograms of proper motions ($\mu_{\alpha} \cos \delta$ and $\mu_{\delta}$), yielding the mean values and standard deviations for each cluster. The estimated mean proper motions (in mas~yr$^{-1}$) are as follows: $-0.56 \pm 0.31$, $1.96 \pm 0.32$ for NGC~2204; $-2.74 \pm 0.20$, $5.20 \pm 0.27$ for NGC~2660; $0.28 \pm 0.11$, $0.11 \pm 0.17$ for NGC~2262; $-2.95 \pm 0.21$, $2.49 \pm 0.15$ for Czernik~32; $-5.69 \pm 0.25$, $-2.25 \pm 0.30$ for Pismis~18; and $-3.87 \pm 0.18$, $0.41 \pm 0.19$ for NGC~2437. Our estimated values of PMs are in good agreement with those reported by \citet{cantat2020painting} and \citet{dias2021updated}.  The corresponding dispersions in proper motions ($\sigma_c$) are 0.09, 0.06, 0.08, 0.09, 0.11, and 0.10~mas~yr$^{-1}$, assuming distances of 4.20, 2.95, 3.29, 4.02, 2.91, and 1.76~kpc, respectively. 
We adopted a radial velocity dispersion of 1~km~s$^{-1}$, following \citet{girard1989relative}. This choice is motivated by observational evidence that intermediate-age and old OCs typically exhibit internal velocity dispersions in the range 0.5–2~km~s$^{-1}$ (e.g., \citet{geller2010wiyn} ). Using the relation $\sigma^{2}=GM/R$ , where G is the gravitational constant, M is the cluster mass, and R is the half mass radius, we independently estimated the radial velocity dispersion for our clusters. We obtained values close to $\sim$1~km~s$^{-1}$, consistent with \citet{girard1989relative}. Tests with alternative dispersions within this range (0.5–2~km~s$^{-1}$) indicate that the membership probabilities remain robust, although adopting significantly larger values increases contamination from field stars.

For field stars, the proper motion centers ($\mu_{xf}$, $\mu_{yf}$) are (0.11, 1.04), ($-3.25$, 3.78), ($-0.22$, $-0.03$), ($-1.50$, 2.02), ($-6.50$, $-1.62$), and ($-1.42$, 1.18), while the corresponding dispersions ($\sigma_{xf}$, $\sigma_{yf}$) are (3.24, 4.05), (3.30, 3.91), (0.64, 0.98), (0.95, 1.06), (3.34, 1.50), and (3.46, 4.95), respectively. We have used the equation provided by \citep{yadav2013proper} to calculate the membership probability. Based on a membership probability threshold of $\geq 70\%$, we identified 1326, 1078, 560, 680, 1137, and 2470 member stars in the respective clusters. In Figure~\ref{fig: gmm_plot_with_member}, the Bayesian 70\% member stars are represented by blue open circles, showing their spatial distribution compared to the GMM-based 70$\%$ probable members.

Regarding the membership probability threshold, we adopted a cutoff of 70$\%$ to select high-confidence cluster members. This threshold is widely used in the literature (e.g., \citet{cantat2020painting}; \citet{castro2019hunting}; \citet{bisht2020comprehensive}) as it provides a balance between maximizing completeness and minimizing field-star contamination. We also tested thresholds between 50$\%$ and 90$\%$, finding that lower thresholds include more faint members but increase contamination, whereas higher thresholds reduce contamination but exclude some genuine faint members. Thus, 70$\%$ provides an optimal compromise for our analysis.

In the Bayesian membership estimation, we found residual contamination from field stars, particularly at faint magnitudes (G $\sim$ 18–20). For example, applying a probability threshold of P $\geq$ 0.7 left $\sim$ 15–16$\%$ of stars lying outside the expected cluster sequence in the CMD, indicative of field contamination. By contrast, the Gaussian Mixture Model (GMM) method reduced this fraction to $\sim$4–5$\%$, producing a much cleaner main sequence, especially at the faint end as shown in Fig. \ref{fig:comparsion_plot}. To mitigate this issue, we therefore adopted the GMM approach to refine our membership determination. A detailed description of the GMM methodology is provided in the following section.

\subsection{Gaussian Mixture Model}

This approach does not require any prior information about cluster parameters. Instead, it relies on modeling the cluster and field stars as two distinct Gaussian distributions in proper motion and parallax space. We have adopted the method described in \citet{agarwal2021ml}, which has also been widely used in previous studies, as referenced by \citet{gao2018machine}, \citet{qiu2024deeper}, and \citet{sheikh2025statistical}. First, we extracted sample sources from the entire dataset by applying the k-nearest neighbours (knn) algorithm and removing field star noise. Then, we normalised the proper motion and parallax data before using the GMM.

\subsubsection{Source Extraction}
First, we downloaded data from Gaia DR3 for a cone-shaped region centered on each cluster, using different radii based on literature values: 30 arcminutes for NGC 2437, 20 arcminutes for NGC 2204, and 10 arcminutes for NGC 2260, NGC 2262, Pismis 18, and Czernik 32. We then removed field star noise from the downloaded dataset by applying the following criteria:
\begin{enumerate}
    \item Each source must have all five astrometric parameters: right ascension, declination, proper motion in right ascension and declination, and parallax, as well as photometric magnitudes in three bands:  $G$, $G_{BP}$, $G_{RP}$.
    \item Parallax must be positive, the proper motion error must be $\leq 1.0$, and the \emph{G}-band magnitude error must be $\leq$ 0.008.
\end{enumerate}

To estimate the mean values of proper motion and parallax for the cluster members, we employed the k-Nearest Neighbors (kNN) algorithm \citep{cover1967nearest}. This algorithm is based on Euclidean distance and calculates the distance of each point to its kNN for classification and regression purposes. The data point with the smallest total distance, defined as the sum of distances to its k nearest neighbours, was used to make a preliminary estimation of the cluster parameters. We applied the kNN algorithm in the 3D space of proper motion and parallax to determine the initial mean parameters of this space. For this, we considered stars within a 10 arcminutes radius, where the cluster population dominates over field stars. The estimated mean parameters show good agreement with those reported by \citep{cantat2020painting, hunt2023improving}.\\

The sample selection range of 1.5 to 3.5 mas yr$^{-1}$ was iteratively adjusted by sliding the window in proper motion to converge on the most likely mean value, as determined by the kNN algorithm. Similarly, the parallax range was selected using an initial window whose width depended on the estimated cluster parallax mean value from kNN—wider for nearby clusters and narrower for distant ones. The final parallax range spanned a width between 0.4 and 1.5 mas. As a result, we obtained a refined sample of sources for the subsequent steps.

\subsubsection{Normalisation and GMM Implementation}

We applied the GMM \citep{mclachlan2000finite}, an unsupervised clustering algorithm, to separate the cluster stars from the sample sources. First, the selected data for parallax and proper motion parameters were normalised. Given $N$ sample sources, each with $m$ parameters $x={x^{1},x^{2},x^{3},…,x^{m}}$, the normalization for the  $j^{th}$ dimension was defined as  $X_{i}^{j}$ \citep{agarwal2021ml}.

\begin{equation}
    X_{i}^{j} = \frac{x_{i}^{j} - \mu^{j}}{\sigma^{j}} \hspace{0.3cm} (i = 1,2,...,N; j = 1,2,...m)
\end{equation}

where $x_{i}^{j}$ initial parameter and $\mu^{j}$ is the median of $x^{j}$ and $\sigma^{j}$ is its standard deviation.\\
A GMM is a statistical technique that models data drawn from a combination of several Gaussian distributions, each with its own set of unknown parameters. Unlike K-Means clustering \citep{macqueen1967some, lloyd1982least}, which focuses only on the mean $(\mu)$ to define clusters, GMM also accounts for the covariance $(\Sigma)$, capturing the shape and orientation of each cluster in the data. The model is fitted by maximising the likelihood estimates of the distribution parameters using the Expectation-Maximisation (EM) algorithm \citep{dempster1977maximum}. Given $N$ data points $x={x_{1},x_{2},x_{3},…,x_{N}}$  in a parameter space of $M$ dimensions, the $K$-component GMM is defined as:

\begin{equation}
P(x) = \sum_{i=1}^{K} w_i G(x \mid \mu_i, \Sigma_i), \quad \text{such that} \quad \sum_{i=1}^{K} w_i = 1,
\end{equation}

Here, $P(x)$ represents the probability distribution of the data points $x$, $w_i$ denotes the weight of the $i$th Gaussian component in the mixture, and $G(x \mid \mu_i, \Sigma_i)$ is the Gaussian distribution associated with the $i$th component.
\begin{equation}
G(x \mid \mu_i, \Sigma_i) = 
\frac{\exp\left[-\frac{1}{2}(x - \mu_i)^{T} \Sigma_i^{-1} (x - \mu_i)\right]}{(2\pi)^{M/2} {\sqrt{|\Sigma_i|}}}
\end{equation}

Here, $\mu_i$ and $\Sigma_i$ designate the mean vector and the full covariance matrix of the $i^{\text{th}}$ Gaussian component, respectively. The GMM delivers a soft membership probability for each data point, reflecting the likelihood of its connection with each cluster.

To reliably estimate the membership probability of cluster members within the sample sources, we adhered to the following initial conditions: (1) using high-precision datasets, (2) ensuring a high proportion of cluster members relative to field sources, i.e., a high cluster-to-field source ratio, and (3) ensuring distinct peak locations in the distributions of cluster and field stars \citep{de1979astrometric}. If these conditions are not met, the estimated membership probabilities may lose reliability and significance. We utilised the exceptional quality of Gaia DR3 data to fulfil the first condition in our analysis. We ensured a high proportion of cluster stars for the sample selection and then applied the GMM to the selected sources. As shown in the histogram plot in Figure \ref{fig: gmm_plot_with_member} and the other figures in the appendix \ref{appendix: appendixA1}, the parameters exhibit distinct distributions for cluster and field stars, confirming that all three conditions were fully satisfied. We employed a two-component GMM to represent the cluster and field populations in the normalised three-dimensional parameter space ($\mu^{*}_{\alpha}, \mu_{\delta},\Bar{\omega}  $). This method assumes that the proper motions and parallaxes of the sample sources follow a two-component Gaussian distribution. Positional data were excluded from the parameter space to avoid confining cluster members within a fixed ellipsoidal boundary, allowing us to study the cluster's morphology and identify potential escapers \citep{agarwal2021ml}. We considered only stars with a membership probability $\geq $70$\%$ for further analysis in this study. We used 70$\%$ of the member stars to estimate the mean values of right ascension, declination, parallax, and proper motion for the cluster stars listed in Table \ref{tab:mean_values}, showing good agreement with the values reported by \citep{cantat2020painting, hunt2023improving}.

\begin{table*}
    \centering
    \begin{tabular}{|c c c c c c c|} 
    \hline
    Cluster &  R.A. & DEC. & pmRA & pmDEC & Parallax & RVs\\
            &  deg. & deg. & mas/yr & mas/yr & mas & km s$^{-1}$ \\
    \hline
     NGC 2204 & 93.880 $\pm$ 0.050 & -18.675 $\pm$ 0.067 & -0.58 $\pm$ 0.07  & 1.96 $\pm$ 0.12  &    0.202 $\pm$ 0.07 & 92.32 $\pm$  2.29\\
     NGC 2660    &130.667 $\pm$ 0.06 & -47.196 $\pm$ 0.04  & -2.74 $\pm$ 0.20 & 5.20 $\pm$ 0.27 &  0.33 $\pm$ 0.10 & 22.66 $\pm$ 5.43 \\
     NGC 2262    & 99.899 $\pm$ 0.039 & 1.145 $\pm$ 0.035  & 0.30 $\pm$ 0.11 & 0.11 $\pm$ 0.10  &    0.26 $\pm$ 0.11 & 63.94 $\pm$ 3.76 \\
     Czernik 32  & 117.617 $\pm$ 0.059 & -29.855 $\pm$ 0.070 & -2.96 $\pm$ 0.18 & 2.48 $\pm$ 0.15  & 0.22 $\pm$ 0.12  &  72.29 $\pm$ 4.69 \\
     Pismis 18  & 204.223 $\pm$ 0.134 & -62.089 $\pm$ 0.064 & -5.69 $\pm$ 0.22 & -2.35 $\pm$ 0.21 &   0.35 $\pm$ 0.10 &  -24.70 $\pm$ 8.50 \\
     NGC 2437   & 115.438 $\pm$ 0.174 & -14.844 $\pm$ 0.175  & -3.87 $\pm$ 0.18 & 0.41 $\pm$ 0.19 &  0.60 $\pm$ 0.07 & 45.15 $\pm$ 15.75 \\
     \hline
    \end{tabular}
    \caption{Mean astrometric and kinematic parameters of selected open clusters derived using stars with a membership probability greater than 70\% as identified by the GMM method. The parameters include the cluster center coordinates (R.A. and Dec), proper motions in right ascension (pmRA) and declination (pmDEC), parallax, and radial velocities (RVs). The uncertainties represent the standard error of the mean.}
    \label{tab:mean_values}
\end{table*}

\begin{figure*}
    \centering
    \vspace{-0.3cm}
    \includegraphics[width=18cm, height= 6.2cm]{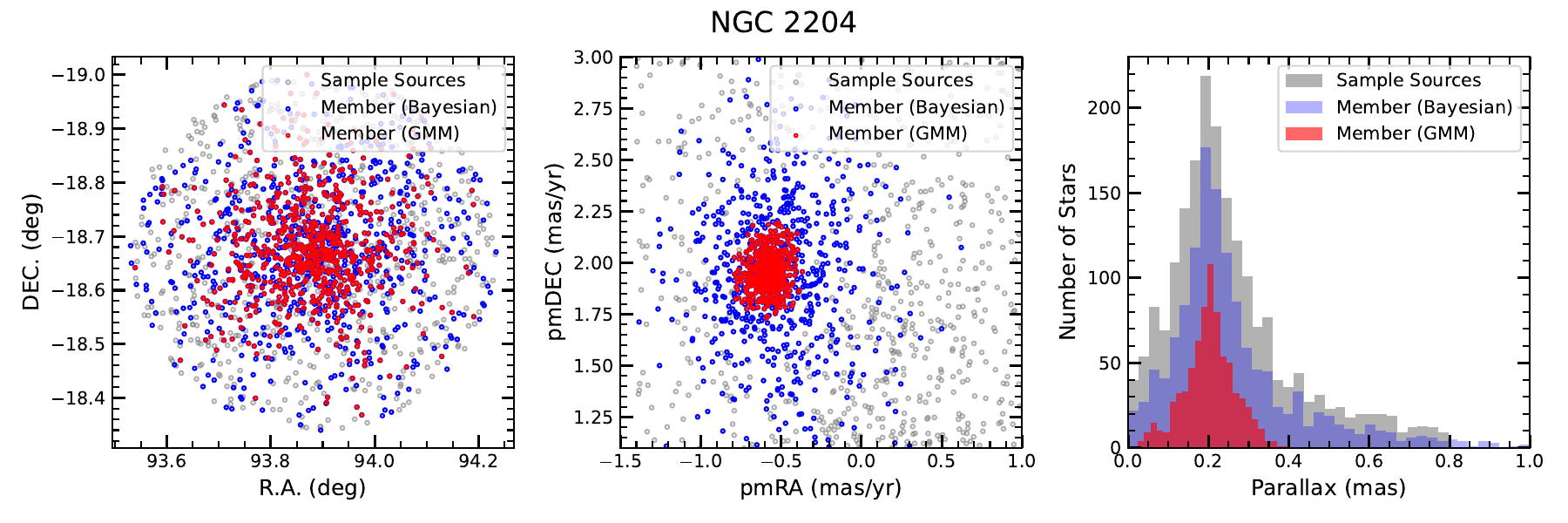}
    \caption{Membership analysis of the open cluster NGC~2204. Panel (a) shows the spatial distribution of stars in R.A. and Dec., panel (b) displays the proper motion distribution in the $\mu_{\alpha}\cos\delta$ vs. $\mu_{\delta}$ plane, and panel (c) presents the parallax distribution. The results from the GMM-based method and the Bayesian approach (ML-MOC) are compared, with red and blue dots indicating high probability members identified by each method, respectively, and grey points representing the full sample.}
    
    \label{fig: gmm_plot_with_member}
\end{figure*}

\begin{figure}
    \hspace{-0.7cm}
    \includegraphics[width=4.5cm, height = 5.2 cm]{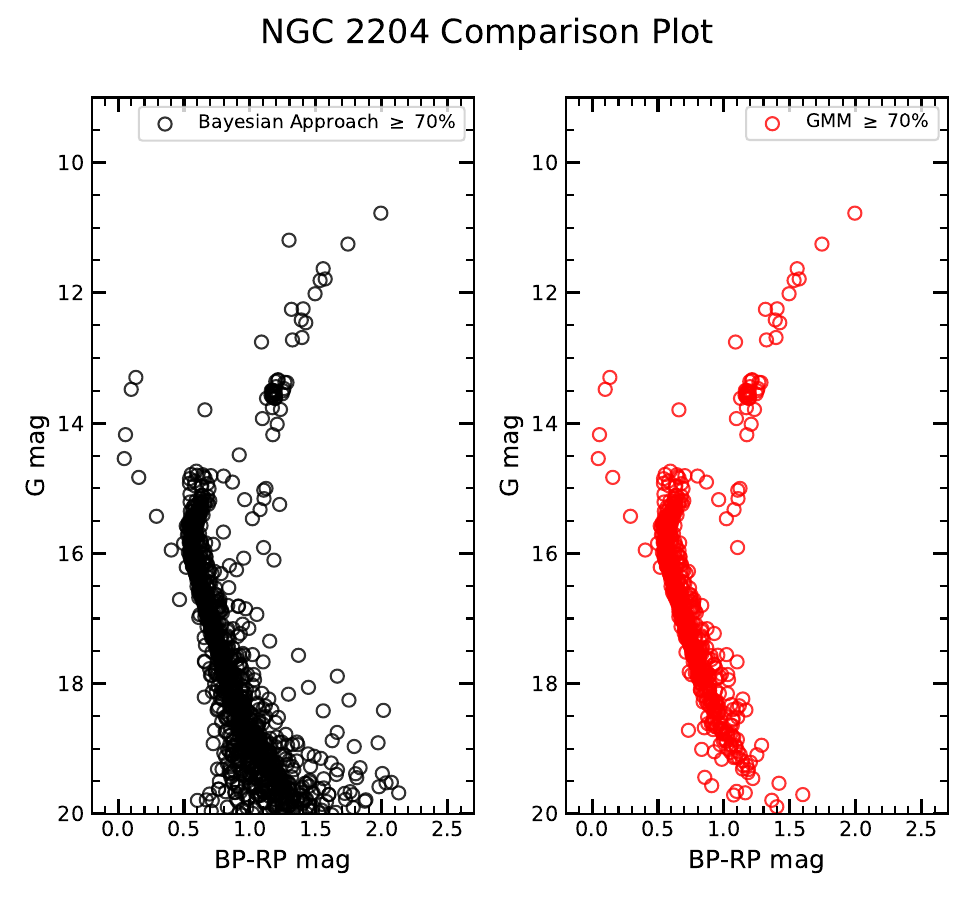}
    \vspace{-0.3cm}
    \includegraphics[width=4.8cm,height=4.0cm]{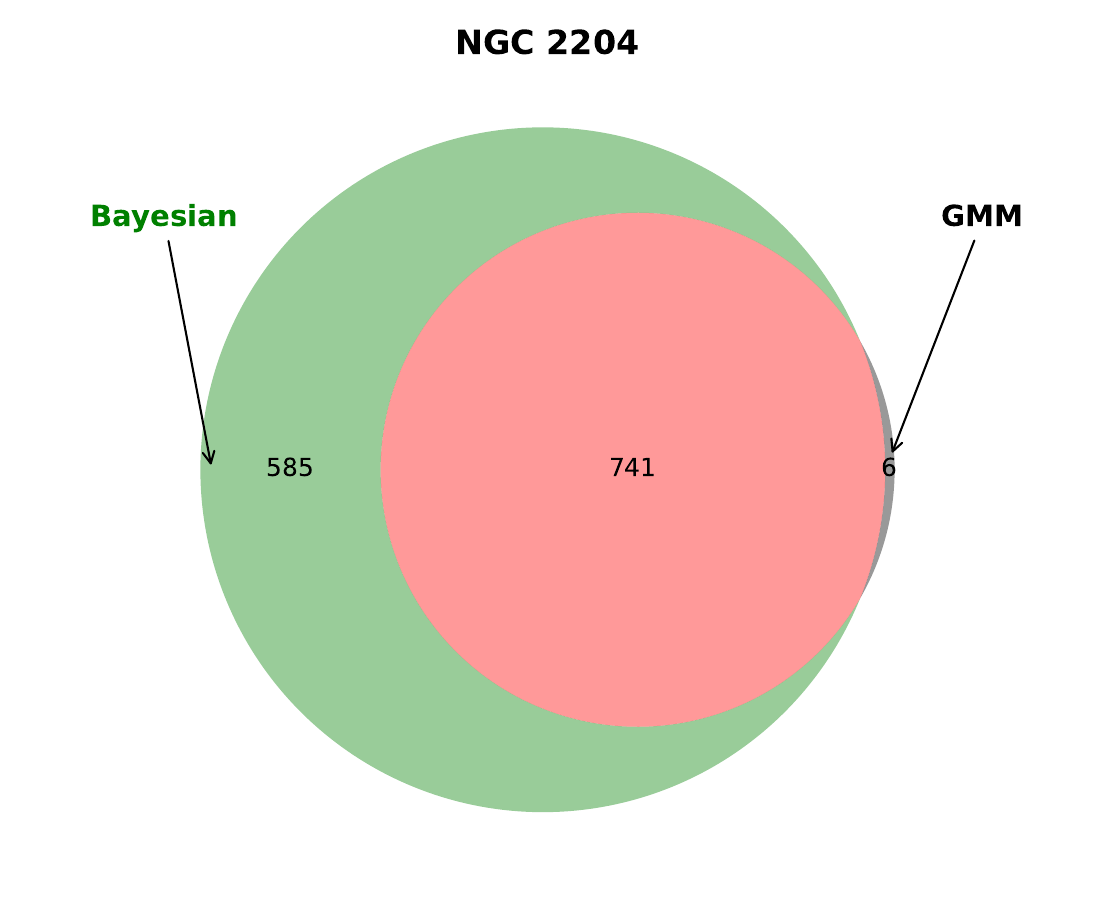}
    \caption{A comparison of the CMDs from both methods is shown on the left, displaying stars with a membership probability greater than 70$\%$. The Venn diagram for the cluster NGC 2204 is presented on the right.}
    \label{fig:comparsion_plot}
\end{figure}

To estimate cluster memberships, we applied both the Bayesian and the GMM method. A comparison of the two methods is shown in Figure \ref{fig:comparsion_plot}, for NGC 2204, with results for the other clusters provided in Appendix \ref{appendix: appendixA2}. The left panel of Figure \ref{fig:comparsion_plot} presents the CMD distributions of members identified by each method. In contrast, the right panel shows a Venn diagram illustrating their overlap, following the approach of \citet{qiu2024deeper}. For NGC 2204, 741 of 747 stars (99.1$\%$) selected by the GMM are also identified by the Bayesian method, whereas only 585 of 1326 stars (44.1$\%$) from the Bayesian sample are recovered by the GMM. This pattern is consistent across all clusters, with GMM overlaps ranging from 85–99$\%$ compared to 36–79$\%$ for the Bayesian method. The Bayesian approach includes a larger number of stars, particularly at faint magnitudes (G $\ge$ 18 mag) and in the cluster periphery, owing to its weaker selection criteria, which often admit field contamination. In contrast, the GMM employs stricter thresholds that effectively reduce field star contamination. Therefore, we adopt the GMM-derived memberships for the subsequent mass function and structural analyses, as they provide a more reliable separation between cluster members and field stars. Our analysis yields similar outcomes, confirming the conclusion of \citet{qiu2024deeper} that the Bayesian approach admits more field stars, especially at fainter magnitudes.

\begin{figure*}
    \centering
    \includegraphics[width=15cm,height=4.4cm]{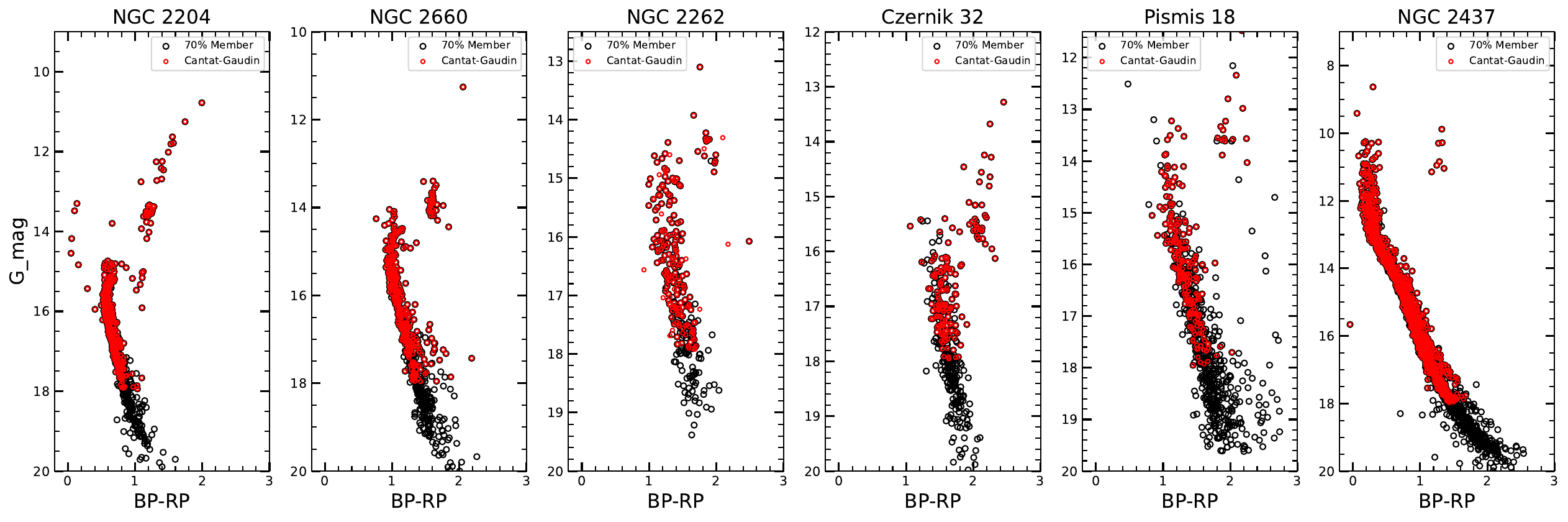}
    \caption{Color–magnitude diagrams (CMDs) for the clusters NGC 2204, NGC 2060, NGC 2262, Czernik 32, Pismis 18, and NGC 2437. Black points represent the stars from our analysis, while red points indicate the matched members from the \citet{cantat2020painting} catalog. The good agreement between datasets confirms our membership results.}
    \label{fig: comparsion}
\end{figure*}

 We compared GMM-selected member stars with those identified by \citet{cantat2020painting}, as shown in Figure~\ref{fig: comparsion}, where the matched stars are highlighted using red open circles. The comparison shows a strong agreement between the estimated parameters, thereby further validating our member selection method. Through cross-matching, we identified 500, 370, 203, 155, 230, and 1,365 common member stars for the clusters NGC 2204, NGC 2660, NGC 2262, Czernik 32, Pismis 18, and NGC 2437, respectively.

\section{Structural Analysis of Clusters}\label{sec:Structural Parameters of the Cluster}

\begin{figure*}
    \centering
    \vspace{-0.5cm}
    \includegraphics[width=5.2cm,height=5.0cm]{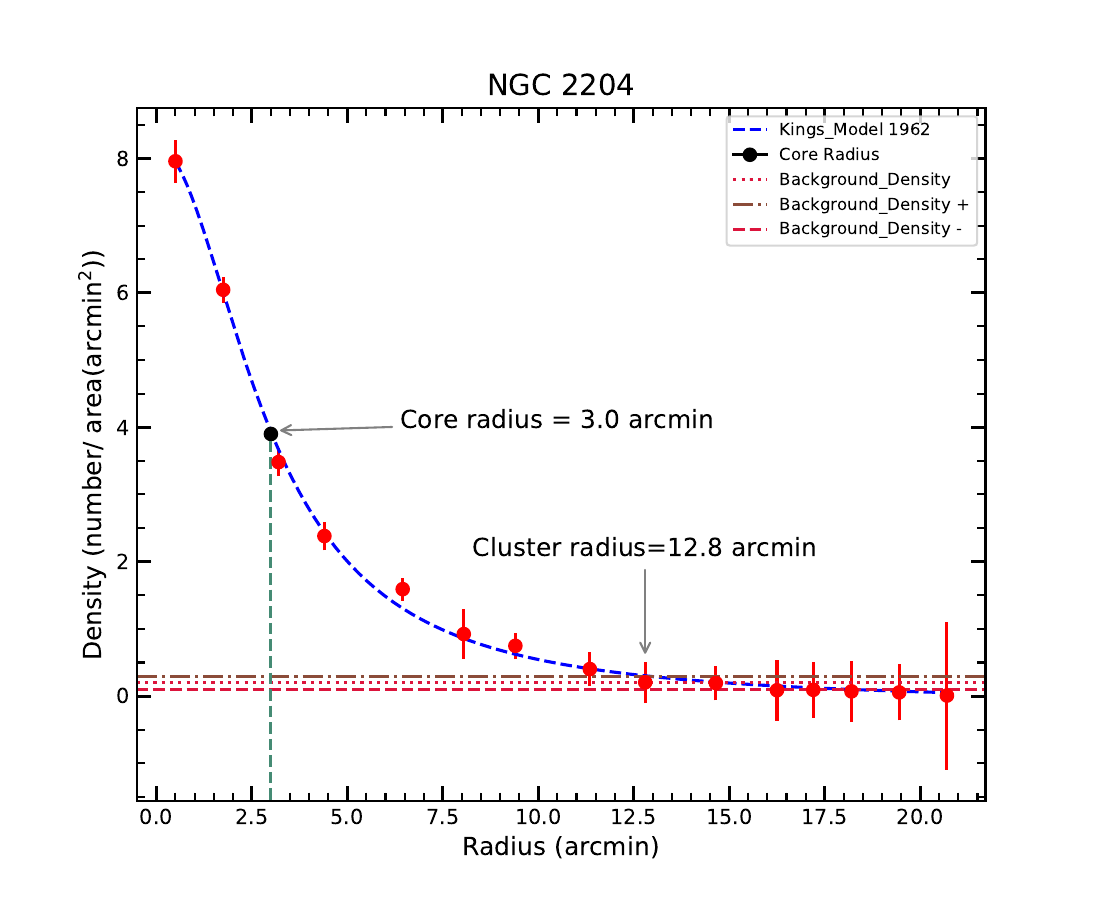}
    \includegraphics[width=5.2cm,height=5.0cm]{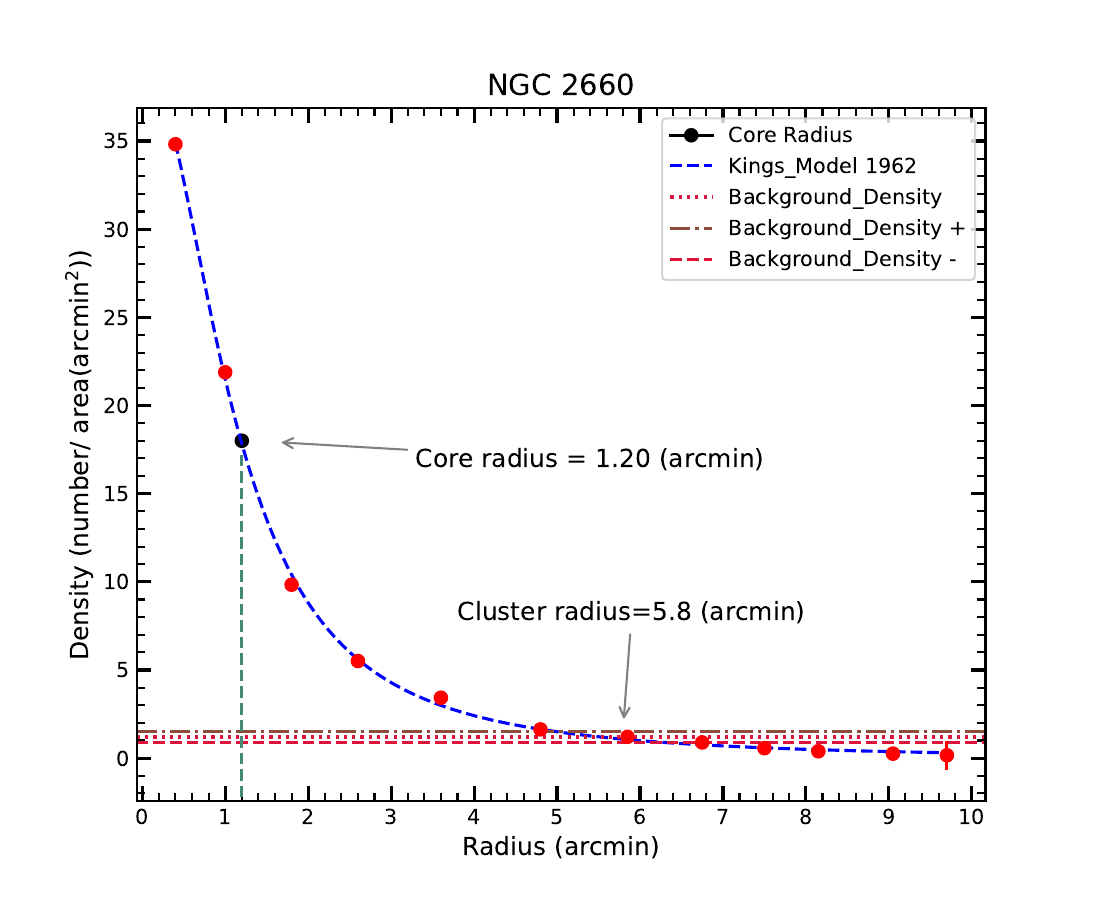}
    \includegraphics[width=5.2cm,height=5.0cm]{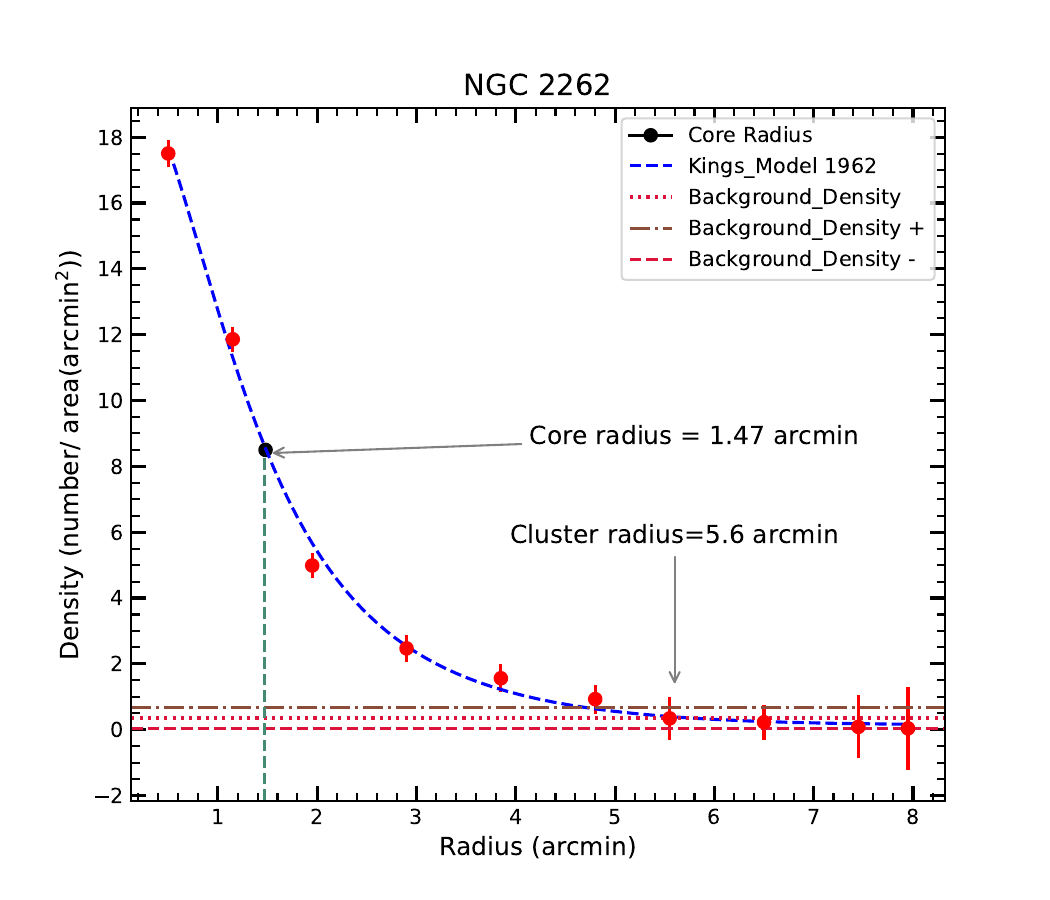}
    \includegraphics[width=5.2cm,height=5.0cm]{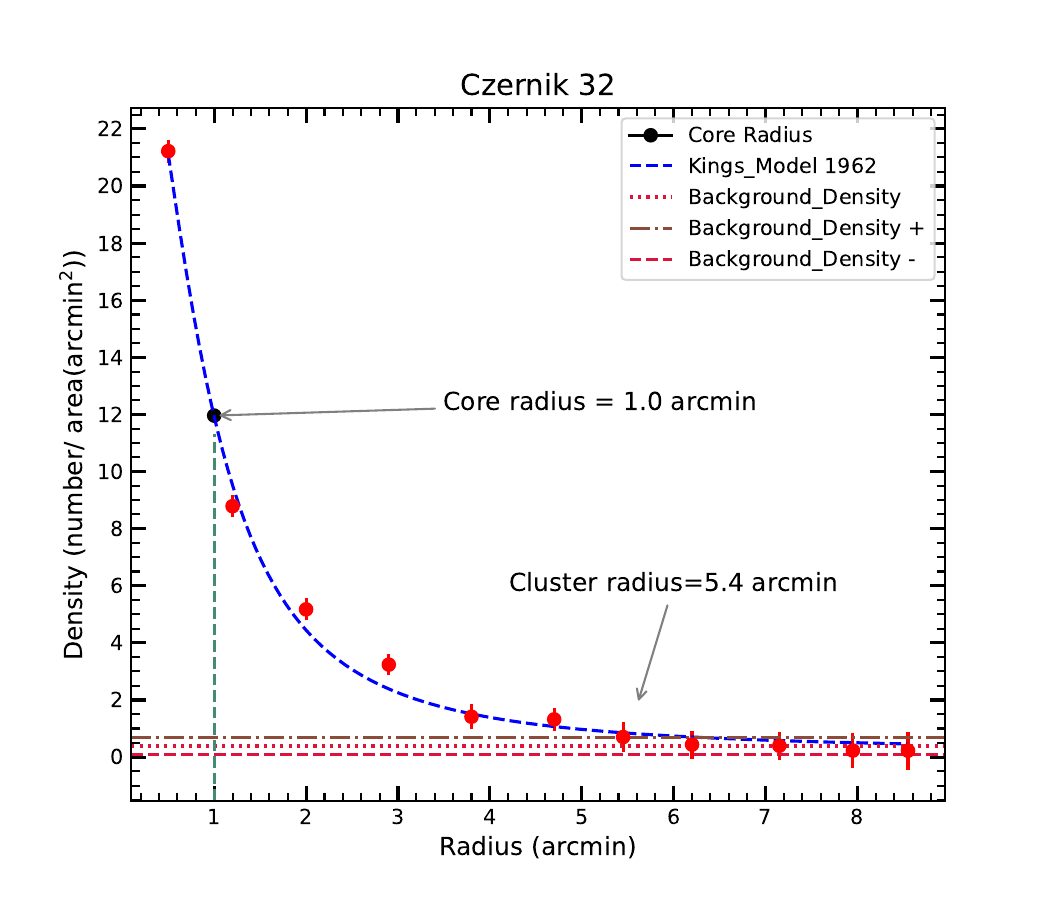}
    \includegraphics[width=5.2cm,height=5.0cm]{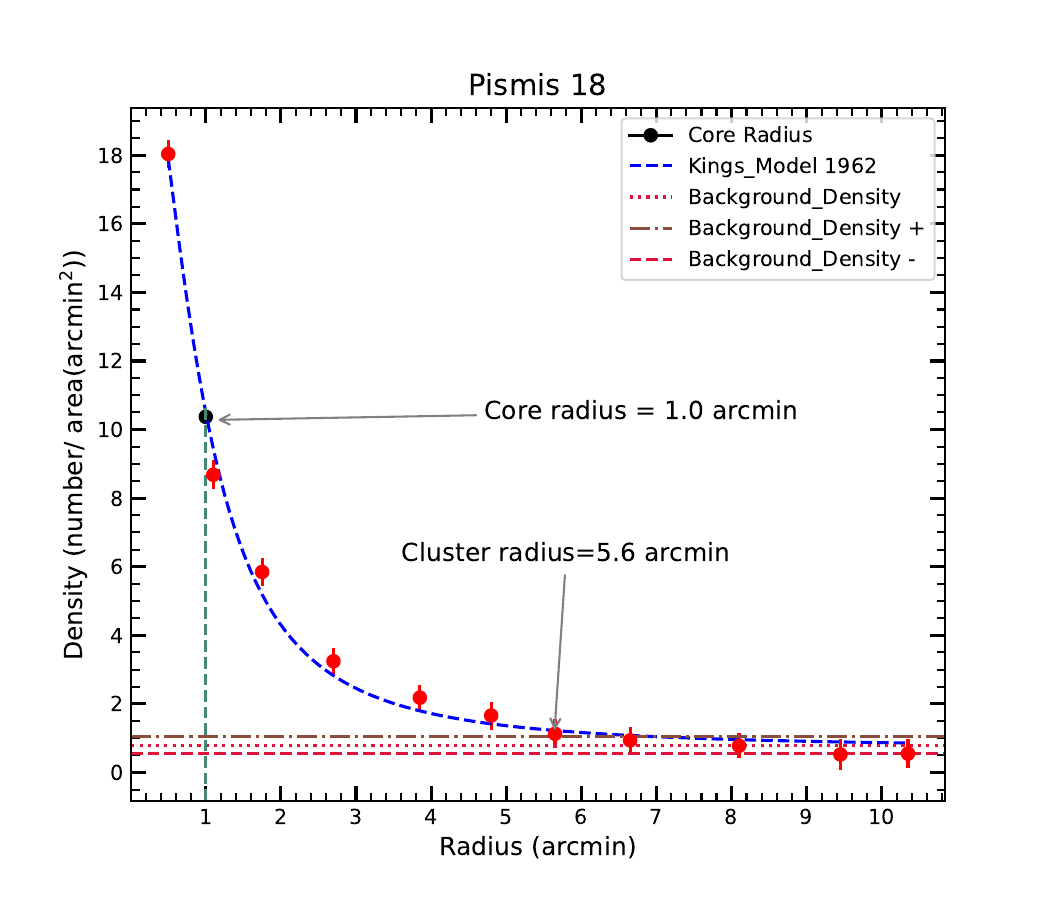}
    \includegraphics[width=5.2cm,height=5.0cm]{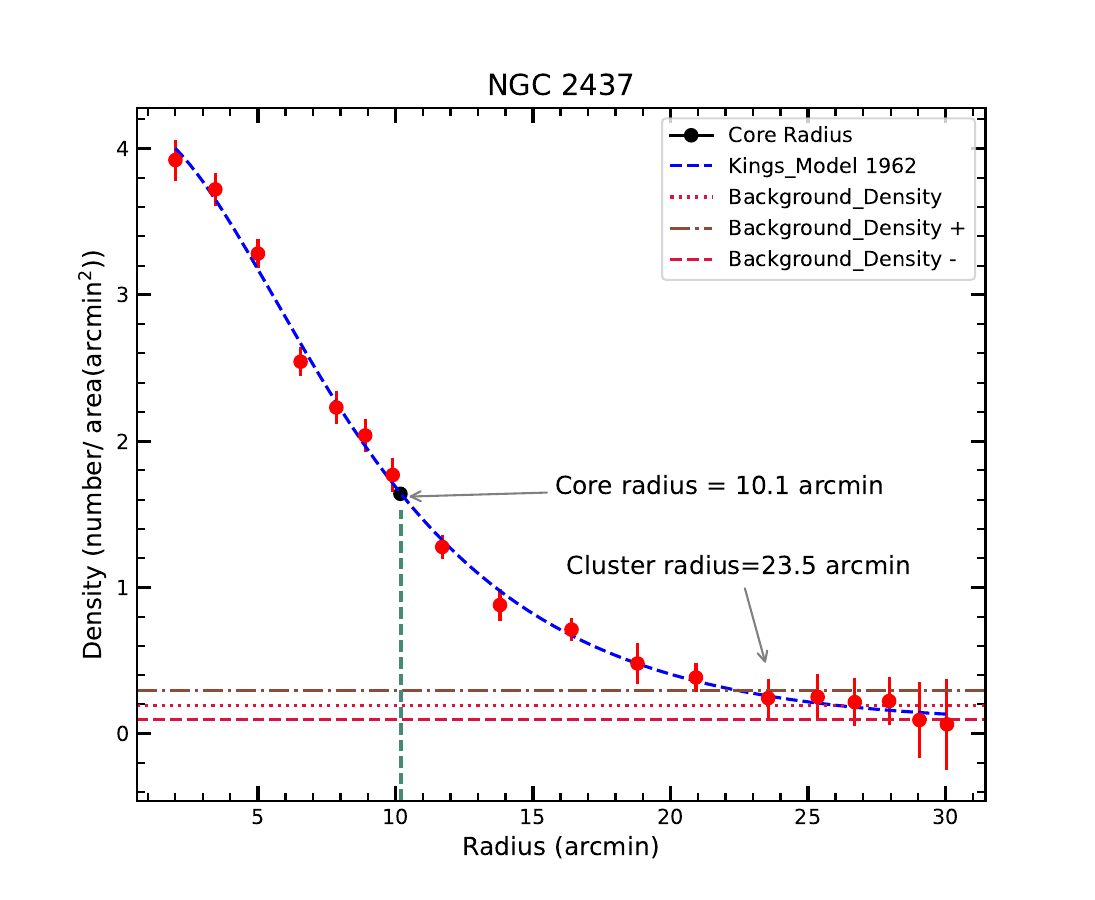}
    \caption{Radial density profiles of the studied clusters fitted with the empirical King (1962) model, shown as the solid blue curve. The core radius ($r_c$) and cluster radius ($r_{cl}$) are estimated from the fit. Horizontal dashed lines represent the background stellar density level, with shaded regions indicating the associated uncertainties. Each panel corresponds to one cluster as labeled.}
    
    \label{fig: Radial_density}
\end{figure*}

OCs are loosely bound systems with thinly dispersed stars, but the stellar density is observed to be highest at the center. Determine the cluster center using the right ascension and declination of the 70$\%$ most probable members and estimate the mean values through Gaussian curve fitting.
 It provides the mean value and standard deviation used as a cluster center, as shown in Table \ref{tab: kings model parameters}. We collected data for each cluster at various scales to determine the true extent of the cluster radius and derive the radial density profile (RDP). We divided the entire cluster area into concentric rings of 1 arcminute width, using the cluster center as the reference point. The stellar density in each ring is estimated by dividing the number of stars in the ring by the area of that specific ring $\rho_{i} = N_{i}/A_{i}$. Where N$_{i}$ is number of stars and A$_{i}$ area of the i$^{th}$ zone. The structural parameters of the cluster are derived by fitting the \cite{king1962structure} profile to the RDP, as shown in Figure \ref{fig: Radial_density}. The King's profile is given by 
\begin{equation}
    f(r) = f_{0} \Biggl[\frac{1}{\sqrt{1 + (r/r_{c})^{2}}} - \frac{1}{\sqrt{1+(r_{t}/r_c)^{2}}}\Biggr]^{2} + f_{b}
\end{equation}

Where $f_{b}$ represents the background density, $r_{c}$ denotes the core radius, $f_{0}$ is the central density, and $r_{t}$ corresponds to the tidal radius. The cluster radius ($r_{cl}$) is defined as the distance from the cluster center at which the stellar density equals the background density, ranging from 5.4 to 23.5 arcminutes. The core radius is the distance from the cluster center where the density drops to half of the central density, varying from 1.0 to 10.1 arcminutes. Table \ref{tab: kings model parameters} provides all the calculated structural parameter values. Our estimated values of the core radius and cluster radius are consistent with the reported uncertainties in the literature \citep{angelo2023enlightening, sanchez2020catalogue, tadross2008main, hatzidimitriou2019gaia}.

We have calculated the density contrast parameter, $\delta_{c}$, for each cluster using the formula described in \citet{bonatto2007open}. This parameter provides insight into the compactness and sparseness of the clusters. The calculated values range from $ 21 \leq \delta_{c} \leq 70$. All the clusters analysed in this study are compact based on these values.\\

\begin{table*}
\centering
\vspace{-0.3cm}
\caption{Structural parameters of the open clusters under study, derived from stellar radial density profiles. Listed parameters include central stellar density ($f_0$), background density ($f_b$), core radius ($r_c$), cluster radius ($r_{cl}$), and tidal radius ($r_t$), with associated uncertainties. These values provide insights into the stellar concentration and spatial extent of each cluster.}
\begin{tabular}{|c|c|c|c|c|c|c|}
\hline
Parameters & NGC 2204 & NGC 2660 & NGC 2262  & Czernik 32  & Pismis 18 & NGC 2437 \\
\hline
 Central Density$(f_{0})$ number/arcmin$^{2}$ & 8.46 $\pm$ 0.48 & 41.90 $\pm$ 3.04 & 28.65 $\pm$ 5.16 &  28.32 $\pm$ 1.46 & 22.98 $\pm$ 1.59 & 7.30 $\pm$ 1.44  \\
 Background Density$(f_{b})$ number/arcmin$^{2}$ & 0.20 $\pm$ 0.12 & 1.20 $\pm$ 0.30 &  0.41 $\pm$ 0.30    & 0.45 $\pm$ 0.28  & 0.70 $\pm$ 0.24 & 0.21 $\pm$ 0.10  \\
 Core Radius$(r_{c})$ arcmin & 3.00 $\pm$ 0.10 & 1.20 $\pm$ 0.20  & 1.47 $\pm$ 0.20    &  1.00 $\pm$ 0.20 & 1.00  $\pm$ 0.10 & 10.1 $\pm$ 0.20     \\
Core Radius$(r_{c})$ parsec &  3.70 $\pm$ 0.15 & 1.02 $\pm$ 0.17 & 1.41 $\pm$ 0.20 &  1.17 $\pm$ 0.24 & 0.85 $\pm$ 0.1 & 5.17 $\pm$ 0.20    \\
Cluster Radius$(r_{cl})$ arcmin & 12.80 $\pm$ 1.20  & 5.80 $\pm$ 0.50  & 5.60 $\pm$ 0.50  &  5.40 $\pm$ 0.70 & 5.60  $\pm$  0.60  & 23.5 $\pm$ 2.10    \\
Cluster Radius$(r_{cl})$ parsec & 15.64 $\pm$ 1.45 &  4.97 $\pm$ 0.43 & 5.35 $\pm$ 0.48 & 6.31 $\pm$ 0.82 & 4.74 $\pm$ 0.51 & 12.03 $\pm$ 1.50 \\ 
Tidal Radius$(r_{t})$ arcmin  & 17.28 $\pm$ 2.20 & 12.21 $\pm$ 2.05  &  8.80 $\pm$ 1.80   & 8.39 $\pm$ 1.11  &   11.78  $\pm$  1.71 & 34.73 $\pm$3.50     \\
Tidal Radius$(r_{t})$ parsec & 20.63 $\pm$ 2.50 & 10.48 $\pm$ 1.76 & 8.28 $\pm$ 1.95 & 9.81 $\pm$ 1.30 & 9.98 $\pm$ 1.45 & 17.78 $\pm$ 1.80 \\
\hline
\end{tabular}
\label{tab: kings model parameters}
\end{table*}

\section{Parallax-Based Distance Estimation} \label{sec:Distance Estimated Based on Parallax}
Parallax is a key parameter of an open cluster, and the relationship between distance and parallax is straightforward: the distance can be calculated by inverting the parallax under ideal conditions without accounting for measurement uncertainties. Measurement errors are mandatory, and determining the distance from parallax values becomes a complex task that requires careful consideration of uncertainties \citep{bailer2018estimating}. In this analysis, we consider the exponentially decreasing space density (EDSD) before the distance \citep{bailer2018estimating},
\begin{align}
P(r|L) = \begin{cases}
\frac{1}{2L^{3}}r^{2}\exp\left(-\frac{r}{L}\right) & \text{if } r > 0 \\
0 & \text{otherwise}
\end{cases}
\label{eq:bj}
\end{align}
where L $>$ 0 is a length scale. The prior shows a single peak at 2L, and L changes based on Galactic longitude (l) and latitude (b). According to the model, which accounts for expected variations in the distribution of stellar distances in the Gaia-observed Galaxy \citep{bailer2018estimating}.
\begin{align}
& P^{*}(r|\overline{\omega},\sigma_{\overline{\omega}}, L_{\text{sph}}(l,b)) \nonumber\\
&= \begin{cases}
    r^{2}\exp\left[-\frac{r}{L_{\text{sph}}(l,b)}-\frac{1}{2\sigma_{\overline{\omega}}^2}(\overline{\omega}-\overline{\omega}_{zp}-\frac{1}{r})^2\right] & \text{if } r > 0 \\
    0 & \text{otherwise}
  \end{cases}
\label{eq:bj2}
\end{align}
For a star's actual distance $r$, the parallax is $1/r$, with the measured parallax ($\bar{\omega}$) subject to observational noise of $1/r$. By considering the Gaussian likelihood of the parallax ($\bar{\omega}$) with a standard deviation ($\sigma_{\bar{\omega}}$) and using the EDSD prior from Eq. \ref{eq:bj}, we can obtain the unnormalized posterior over the distance to a source \citep{bailer2018estimating}. The three physical parameters - finite parallax ($\bar{\omega}$), positive standard deviation ($\sigma_{\bar{\omega}}$), and positive scale length L${sph}$—ensure a proper (i.e., normalizable) density function \citep{bailer2018estimating}. The accuracy of parallax measurements is limited, especially for distant stars, due to the precision of fractional values. \cite{bailer2018estimating} opted to estimate distances without assuming specific properties of individual stars or the level of extinction affecting them. Following this method, we estimated the distance to each cluster by first applying a zero-point offset of $-$0.021 mas to the parallax values, as suggested by \citet{groenewegen2021parallax}. We then used Gaussian curve fitting to determine the mean and standard deviation of the parallax distribution for each cluster. The calculated distance values are as follows: for NGC 2204, $4.20^{+0.75}_{-0.56}$; for NGC 2660, $2.95^{+0.50}_{-0.67}$; for NGC 2262, $3.29^{+0.51}_{-0.39}$; for Czernik 32, $4.02^{+0.34}_{-0.30}$; for Pismis 18, $2.91^{+0.56}_{-0.79}$; and for NGC 2437, $1.76^{+0.74}_{-0.40}$ kpc. Table \ref{tab:mean_values} shows the values of parallaxes. Our estimated values show fair agreement with \citet{cantat2020painting, dias2021updated}.

\section{Color-Magnitude Diagram Analysis and Cluster Characterization} \label{sec: Age and Distance from Isochrone Fitting}
The  CMDs demonstrate the relationship between the surface temperature and absolute magnitude of stars, highlighting the corresponding colors and evolutionary tracks of stars with different initial masses in the cluster. Analysing the morphology of CMDs reveals key features, such as the main sequence, the turnoff point, and giant stars, which are crucial for deriving model-based estimates of the mass, age, and distance of each star \citep{bisht2019mass}.

\begin{figure*}
    \centering
    \vspace{-0.2cm}
    \includegraphics[width=4.5cm,height=5.9cm]{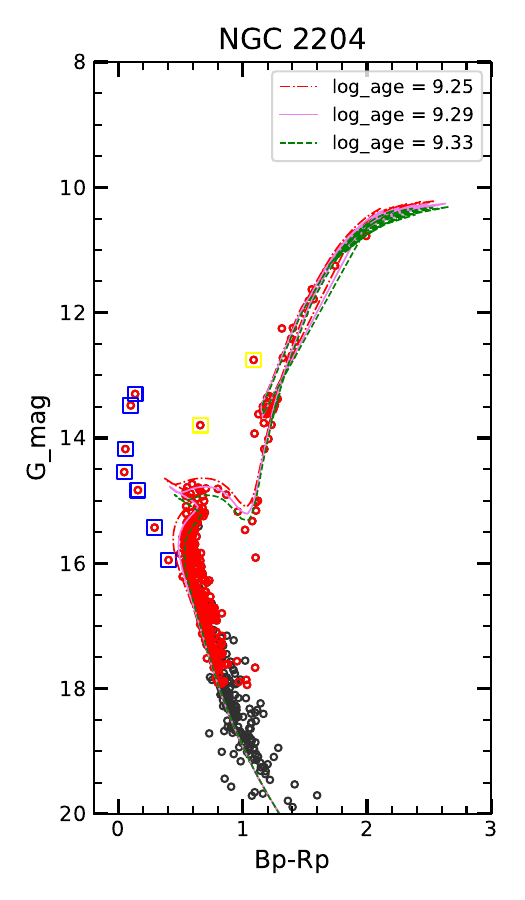}
    \includegraphics[width=4.5cm,height=5.9cm]{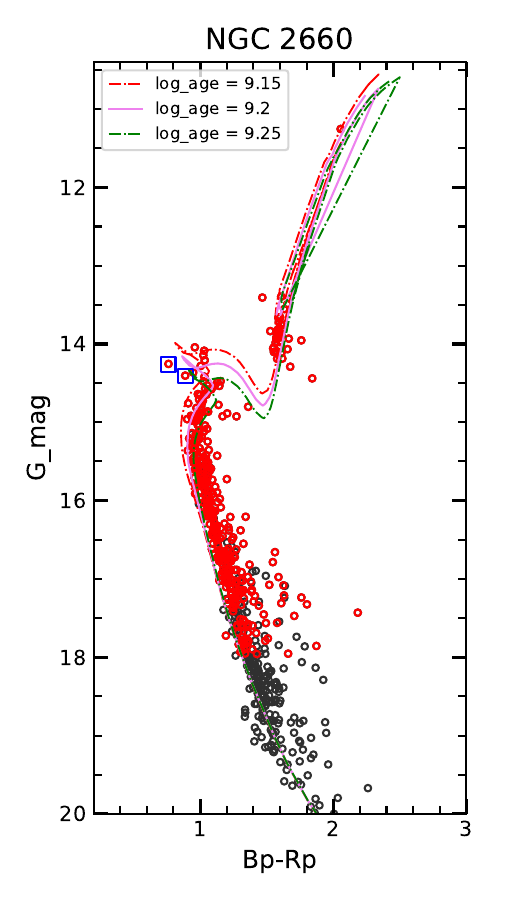}
    \includegraphics[width=4.5cm,height=5.9cm]{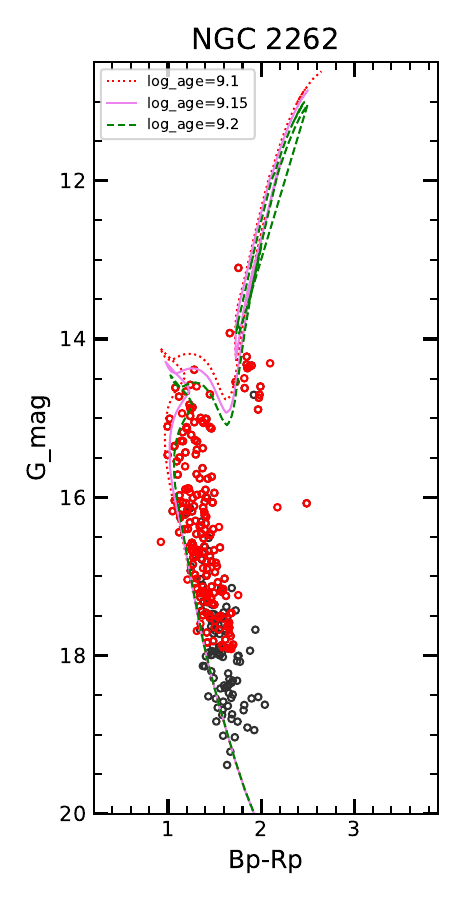}
    \includegraphics[width=4.5cm,height=5.9cm]{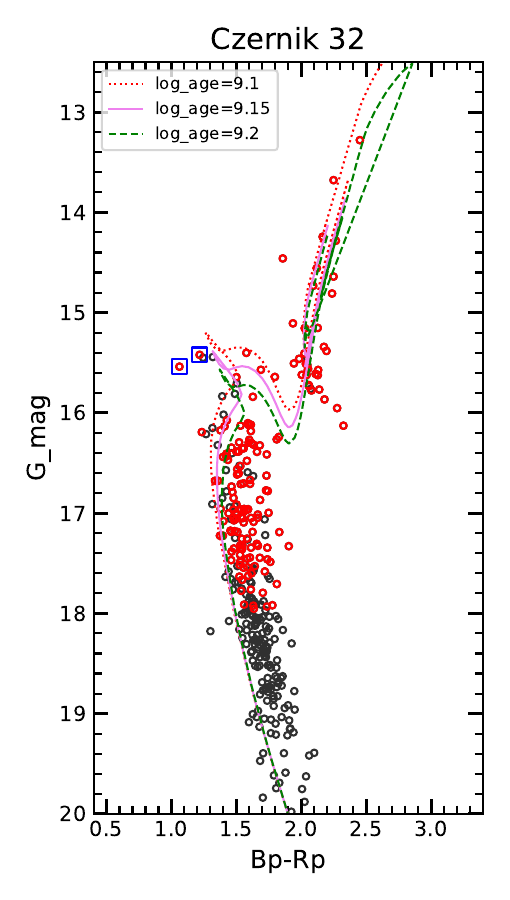}
    \includegraphics[width=4.5cm,height=5.9cm]{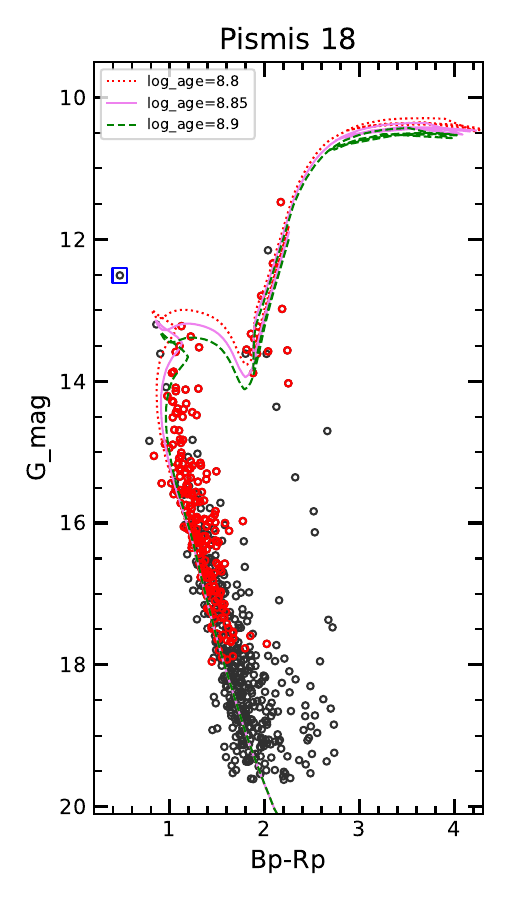}
    \includegraphics[width=4.5cm,height=5.9cm]{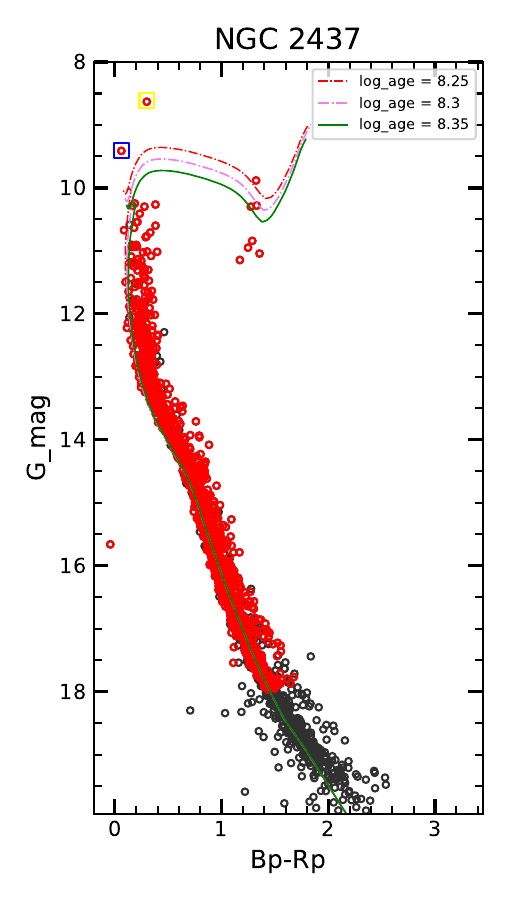}
    \caption{The Gaia DR3 color-magnitude diagram of the studied cluster displays isochrones for the log age values indicated in the plot legend based on models from \citet{marigo2017new}. Black circles represent the probable cluster members identified in this study, while red circles denote stars matched with the \citet{cantat2020painting} catalog. Blue and yellow squares mark the blue straggler stars (BSS) and yellow straggler stars (YSS), respectively, which were matched with the \citet{rain2021new} catalog.}
    \label{fig:Matched_CMD}
\end{figure*}

The interstellar reddening for each cluster was quantified by determining the color excess \(E(G_{\mathrm{BP}} - G_{\mathrm{RP}})\) through isochrone fitting to the observed color-magnitude diagram (CMD). This method enabled us to estimate the color shift caused by interstellar dust extinction. The derived values of \( E(G_{\mathrm{BP}} - G_{\mathrm{RP}}) \) range from 0.14 to 0.80. To estimate the extinction in the Gaia G-band, we applied the empirical relation \( A_G = 1.86 \times E(G_{\mathrm{BP}} - G_{\mathrm{RP}}) \), as given by \citet{casagrande2018use}, resulting in \( A_G \) values between 0.80 and 1.35 mag. The corresponding color excess in the Johnson system, \( E(B - V) \), was derived using the transformation equation \( E(B - V) = 0.72 \times E(G_{\mathrm{BP}} - G_{\mathrm{RP}}) \), yielding values in the range 0.15 to 0.58. Using the computed values of \( A_G \) and \( E(B - V) \), the extinction ratio \( R_G = A_G / E(B - V) \) was calculated for each cluster. The derived values of \( R_G \) are 2.47 for NGC~2204, 2.45 for NGC~2660, 2.56 for NGC~2262, 2.55 for Cz~32, 2.53 for Pismis~18, 2.57 for NGC 2437. These results were compared with the standard extinction ratio \(R_G \approx 2.74 \) reported by \citet{casagrande2018use}, suggesting that the extinction toward these clusters follows the typical interstellar reddening law, with no significant anomalies in the dust properties along their lines of sight.
In this study, the distances obtained from Gaia parallax measurements were used as fixed inputs during the isochrone fitting process, as shown in Figure \ref{fig:Matched_CMD}. We did not treat distance as a free parameter in the fitting. Instead, the primary objective was to estimate the interstellar reddening and the ages of the clusters by fitting theoretical isochrones from \citet{marigo2017new} to the Gaia $G$ versus $(G_{\mathrm{BP}} - G_{\mathrm{RP}})$ color-magnitude diagrams (CMDs). We determined the metallicities of the clusters through theoretical isochrone fitting. The obtained values are $0.006 \pm 0.004$ (NGC~2204), $0.010 \pm 0.003$ (NGC~2660), $0.010 \pm 0.003$ (NGC~2262 and Czernik~32), $0.020 \pm 0.004$ (Pismis~18), and $0.017 \pm 0.003$ (NGC~2437), respectively. They show that some clusters are close to solar metallicity ($Z_\odot \approx 0.0142$; \citealt{asplund2009chemical}), while others exhibit slightly subsolar values. The modest differences correlate with the clusters’ Galactocentric radii. This pattern aligns with the well-established radial metallicity gradient of the Galactic disk (e.g., \citealt{friel1995old, netopil2016metallicity}). In particular, the subsolar metallicities of NGC 2204, NGC 2262, and Czernik 32 are consistent with their older ages and larger Galactocentric distances. The near-solar values of Pismis 18 and NGC 2437 reflect their positions closer to the Galactic center. 
For NGC 2204, we cross-matched our member stars with the Apache Point Observatory Galactic Evolution Experiment (APOGEE) dataset. We identified 27 stars with available metallicities, obtaining a mean value of [Fe/H] $= -0.26 \pm 0.04$ ($Z = 0.007 \pm 0.0008$). This is broadly consistent with the high-resolution spectroscopic study of \citet{anthony2024li}, who reported [Fe/H] $= -0.40 \pm 0.12$ ($Z = 0.005 \pm 0.002$) based on 45 stars. For NGC 2660 and Pismis 18, \citet{magrini2023gaia} derived [Fe/H] $= -0.05$ ($Z = 0.012$) and $+0.14$ ($Z = 0.0193$), respectively, from Gaia-ESO survey data. For NGC 2437, \citet{davidge2013open} adopted a metallicity of $Z = 0.019$, while for NGC 2262 and Czernik 32, \citet{carraro2005intermediate} assumed $Z = 0.008$ in their isochrone-based analysis. Overall, these spectroscopic and literature values are in good agreement with our isochrone-based estimates, further supporting the robustness of our metallicity determinations.

The derived ages for the clusters span from 0.199 to 1.95 Gyr. The isochrone fitting results, including the estimated ages and distances for each cluster, are summarized in Table~\ref{tab: Isochrone table}. To assess the reliability of our fits, we applied the reduced chi-square test for isochrone fitting, following the methodology outlined in \citet{valle2021goodness} and \citet{bisht2024berkeley}, to determine the optimal values of the fitting parameters. The reduced chi-square ($\chi_r^2$) is a widely used method for evaluating the goodness of fit, which quantifies how well the model (isochrone) describes the observed data. The optimal values of the distance modulus and colour were determined by minimising the $\chi_r^2$ statistic, which compares the difference between the observed and fitted colour-magnitude diagrams (CMDS), considering the observational errors. The resulting minimised reduced chi-square values are 0.75 for NGC 2204, 0.70 for NGC 2660, 0.65 for NGC 2262, 0.59 for Czernik 32, 0.62 for Pismis 18, and 0.77 for NGC 2437. These values indicate that our fitting procedure is robust and appropriate for the data.\\
\citet{magrini2023gaia}

Blue Straggler Stars (BSSs) are observed to be both brighter and bluer than the main sequence turn-off, following an extension of the main sequence in the CMDs of star clusters \citep{sandage1953color}. Their formation is primarily attributed to two mechanisms: mass transfer in binary systems \citep{mccrea1964extended} and direct stellar collisions \citep{hills1976stellar}. Additionally, recent studies suggest that BSSs may result from the merger of main-sequence stars within hierarchical triple systems, driven by the eccentric Kozai-Lidov mechanism \citep{naoz2014mergers}, which significantly contributes to BSS formation in OB associations. Extensive research has characterised BSS populations and explored their formation scenarios in open and globular clusters \citep{ferraro2009two, dattatrey2023globules, yadav2024uocs}.

In this analysis, we identified both blue straggler stars (BSS) and yellow straggler stars (YSS), which are consistent with the findings of \citet{rain2021new} and \citet{jadhav2021high}. Specifically, we detected seven BSS and two YSS in NGC 2204, two BSS in NGC 2660, two BSS in Czernik 32, and one BSS along with one YSS in NGC 2437. The matched stars are highlighted in Figure \ref{fig:Matched_CMD} using blue and yellow square boxes. Furthermore, our analysis uncovered one previously unreported BSS in Pismis 18, marked with blue square boxes in the figure \ref{fig:Matched_CMD}.

In NGC 2204, we identified five BSS within a radius of 5 arcminutes from the cluster center, while the remaining two are located at distances of 7 and 11 arcminutes. Additionally, two YSS are positioned at 1.83 and 4.27 arcminutes from the center. In NGC 2660, two BSS are situated within the core radius of the cluster. For NGC 2437, one BSS is found at a distance of 9 arcminutes from the center, while one YSS is located 5.9 arcminutes away. In Czernik 32, we detected two BSS at distances of 2.17 and 4.3 arcminutes from the cluster center. In Pismis 18, one newly identified BSS is located 7.2 arcminutes from the cluster center. All BSS identified in this study have a membership probability exceeding 90$\%$. The possible formation mechanisms suggest that the five blue straggler stars (BSS) in the core region of NGC 2204 likely formed through stellar collisions, while the two BSS in the halo region were probably formed via mass transfer in binary systems \citet{mapelli2006radial}. In general, the higher stellar density in the central region of a cluster makes the collisional channel more favorable there, while the lower-density halo environment is more conducive to binary evolution pathways. Nevertheless, we note that location alone is not definitive, as dynamical interactions and mass segregation can alter the positions of BSSs. Further spectroscopy or spectral energy distribution (SED) analysis will be necessary to distinguish formation mechanisms.

We have determined the Galactocentric coordinates of the clusters (\( X_{\text{GAL}}, Y_{\text{GAL}}, Z_{\text{GAL}} \)) and estimated their Galactocentric distance (\( R_{\text{GC}} \)), as listed in Table \ref{tab: Isochrone table}. The five clusters are located close to the Galactic plane, with their vertical distances ranging from $-$0.155 to 0.124 kpc. The remaining cluster is positioned farther from the plane, at $-$1.168 kpc below the Galactic plane. The calculated parameters show good agreement with the values reported by  \citet{cantat2020painting}. This information is crucial for understanding the cluster’s environment, its motion within the Galaxy, and its interactions with other Galactic components. Furthermore, Galactocentric coordinates serve as a foundation for broader astrophysical studies, contributing to a deeper understanding of the cluster’s role in the dynamic structure of the Milky Way.

\begin{table*}
\centering
\vspace{-0.4cm}
\caption{Comparison of derived cluster parameters from the present study with those reported in the literature for the six open clusters analyzed.}
\small
\resizebox{\textwidth}{!}{%
\begin{tabular}{|c c c c c c c c c c|}

\hline

Author & $\mu_{\alpha}cos(\delta)$ & $\mu_{\delta}$ & Parallax  & $d_{plx}$ &  Age & R$_{GC}$ & X$_{GAL}$ & Y$_{GAL}$ &  Z$_{GAL}$ \\
      & mas yr$^{-1}$ & mas yr$^{-1}$ & mas  &  (kpc)  & (Gyr) & (kpc) & (kpc) & (kpc) & (kpc) \\

\hline

&  &  &  &  & \textbf{NGC 2204}   &  &  &  \\
Present study &  $-0.58 \pm 0.07$ & $1.96 \pm 0.12$ & 0.202 $\pm$ 0.07 & 4.20$^{+0.75}_{0.56}$ & 1.950 $\pm$ 0.18 & 11.678 $\pm$ 0.302 & -2.801 $\pm$ 0.234 & -2.905 $\pm$ 0.324 & -1.168 $\pm$ 0.223  \\ 
\cite{cantat2020painting} & -0.560  & 1.96 &  0.21   &   3.991 & 2.089 &  11.344 & -2.663 & -2.759 & -1.107    \\
\citet{angelo2023enlightening} & -0.57 $\pm$ 0.06 & 1.95 $\pm$ 0.05   &  --  & --  & 1.995 $\pm$ 0.230 & 11.1 $\pm$ 0.50 & -- & -- & --  \\
\citet{dias2021updated} & -0.56 $\pm$ 0.14 & 1.98 $\pm$ 0.15 & 0.212 $\pm$ 0.07 &  4.11 $\pm$ 0.12  & 2.090 $\pm$ 0.144 & -- & -- & -- & -- \\

\hline

&  &  &  &  & \textbf{NGC 2660}    &  &  &  \\
Present study & $-2.74 \pm 0.20$ & $5.20 \pm 0.27 $ & 0.33 $\pm$ 0.10  & $2.95^{+0.50}_{-0.67}$  & 1.585 $\pm$ 0.183 & 9.192 $\pm$ 0.528 & -0.209 $\pm$ 0.120 & -2.932 $\pm$ 0.288 & -0.155 $\pm$ 0.219  \\
\cite{cantat2020painting} & -2.76 & 5.17 & 0.31 & 2.788 & 0.933 & 8.978 & -0.197 & -2.777 & -0.146  \\
\citet{dias2021updated}& -2.74 $\pm$ 0.14 & 5.17 $\pm$ 0.16 & 0.31 $\pm$ 0.06 &  2.640 $\pm$ 0.11 & 1.202$\pm$0.195 & -- & -- & -- & -- \\

\hline

&  &  &  &  & \textbf{NGC 2262}   &  &  &  \\
Present study & $0.30 \pm 0.11$ & $0.11 \pm 0.10$ & 0.26 $\pm$ 0.11  &  $3.29^{+0.51}_{-0.39}$ & 1.413 $\pm$ 0.163 & 11.454 $\pm$ 0.442 & -2.831 $\pm$ 0.332 & -1.672 $\pm$ 0.225 & -0.122 $\pm$ 0.229 \\
\cite{carraro2005intermediate}  & -- & -- & -- & 3.6   &  1.000 $\pm$ 0.2 & 11.7  & -1.8  &  3.1  &  -0.13  \\
\cite{cantat2020painting} & 0.26  & 0.11  & 0.22  & 3.552  & 0.646 & 11.539 & -3.056 & -1.805 & -0.130  \\
\citet{dias2021updated} & 0.29 $\pm$ 0.18 & 0.11 $\pm$ 0.16 & 0.23 $\pm$ 0.11 & 2.91 $\pm$ 0.18 & 1.023 $\pm$ 0.118 & -- & -- & -- & --  \\

\hline
&  &  &  &  & \textbf{Czernik 32}  &  &  &  &  \\
Present study &  $-2.96 \pm 0.18$ & $2.48 \pm 0.15$ & 0.22 $\pm$ 0.12  & $4.02^{+0.34}_{-0.30}$  & 1.413 $\pm$ 0.163 & 10.784 $\pm$ 0.524 & -1.641 $\pm$ 0.256 & -3.668 $\pm$ 0.645 & -0.122 $\pm$ 0.211  \\ 
\cite{carraro2005intermediate}  &  &  &  & 4.1 & 1.000 $\pm$ 0.3 &  10.8  & -1.7  &  3.7  & -0.12 \\
\cite{cantat2020painting} & -2.96 & 2.48  & 0.196  & 4.139  & 1.122 & 10.717 & -1.690 & -3.777 & -0.125 \\
\cite{bica2005properties} & -- & -- & --  & 4.0 $\pm$ 0.2  & 1.120 $\pm$ 200 & 10.3 $\pm$ 0.3 & -- & -- & -0.121 $\pm$ 0.08 \\
\citet{dias2021updated}   & -2.96$\pm$ 0.12  & 2.47 $\pm$ 0.12 & 0.194 $\pm$ 0.06 & 3.67 $\pm$ 0.21 & 1.349 $\pm$ 0.124 & -- & -- & --  & -- \\ 
\hline

&  &  &   &  &  \textbf{Pismis 18}   &  &  &   \\

Present study  &  $-5.69 \pm 0.22$ & $-2.35 \pm 0.21$ & 0.353 $\pm$ 0.10  &  $2.91^{+0.56}_{-0.79}$  & 0.708 $\pm$ 0.082 & 7.079 $\pm$ 0.542 & 1.799 $\pm$ 0.520 & -2.284 $\pm$ 0.421 & 0.016 $\pm$ 0.394 \\
\cite{cantat2020painting}  & -5.66 & -2.29 & 0.332 & 2.860  & 0.575 & 6.943 & 1.770 & -2.247 & 0.015 \\
\cite{hatzidimitriou2019gaia}  & -5.65 $\pm$ 0.08  & -2.29 $\pm$ 0.11 & 0.335 $\pm$ 0.054  & 2.47 & 0.700 & 6.800 & --  & -- & --\\
\citet{dias2021updated} & -5.67 $\pm$ 0.11 & -2.29 $\pm$ 0.12 & 0.328 $\pm$ 0.07  & 2.19 $\pm$ 0.08  & 0.676 $\pm$ 0.078 & -- & -- & -- & -- \\

\hline  

&  &  &  &  &  \textbf{NGC 2437}    &  &  &  \\

Present study & $-3.87 \pm 0.18$ & $0.41 \pm 0.19$ & 0.60 $\pm$ 0.07  &  $1.76^{+0.74}_{-0.40}$ & 0.199 $\pm$ 0.046 & 9.682 $\pm$ 0.340 & -1.083 $\pm$ 0.422 & -1.381 $\pm$ 0.198 & 0.124 $\pm$ 0.050   \\
\cite{cantat2020painting} & -3.83  & 0.36  & 0.603  & 1.511  & 0.302 & 9.345 & -0.930 & -1.185 & 0.106 \\
\hline

\end{tabular}%
}
\label{tab: Isochrone table}
\end{table*}

\section{Dynamical Study}
\label{sec: dynamical study}
To study the luminosity function (LFs) and mass function (MFs), the first crucial step is to eliminate field star contamination from the sample of stars in the cluster region. A statistical field star subtraction method is employed, assuming that field stars in the cluster and surrounding areas have a similar distribution \citep{sagar1998mass} and \citep{phelps1993young}.

\subsection{Luminosity Function}

\begin{figure*}
    \centering
    \vspace{-0.2cm}
    \includegraphics[width=4.5cm,height=4.2cm]{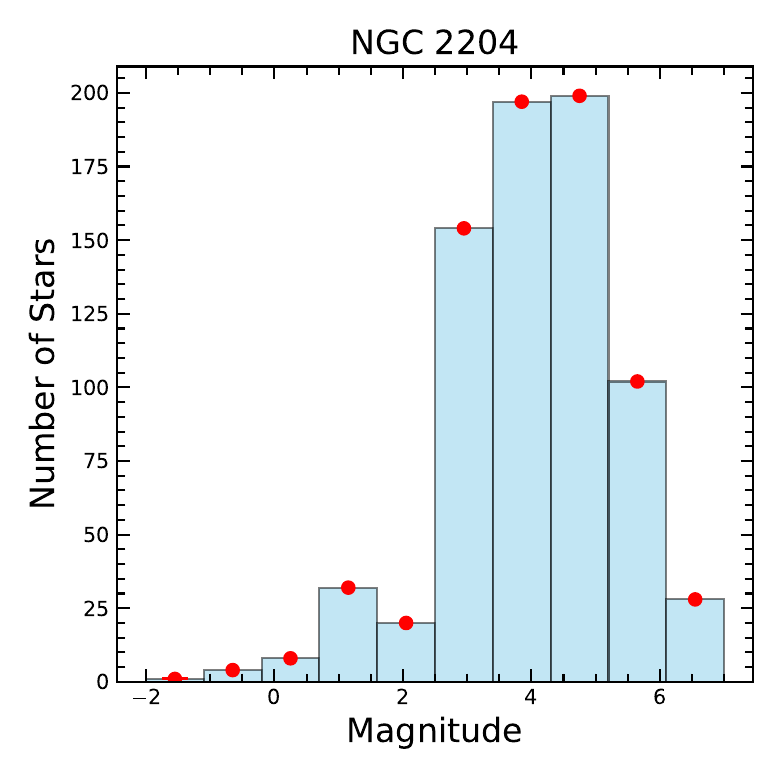}
    \includegraphics[width=4.5cm,height=4.2cm]{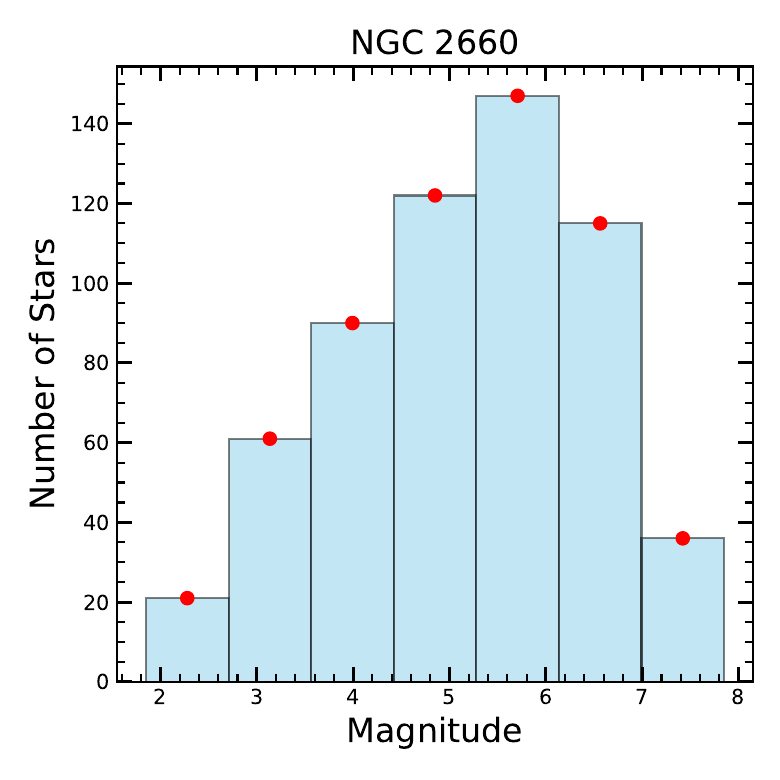}
    \includegraphics[width=4.5cm,height=4.2cm]{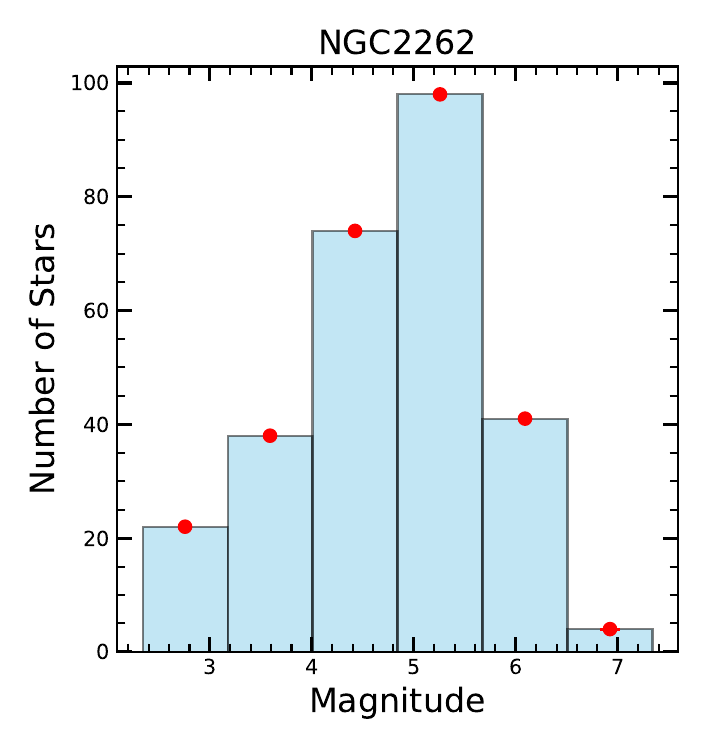}
    \includegraphics[width=4.5cm,height=4.2cm]{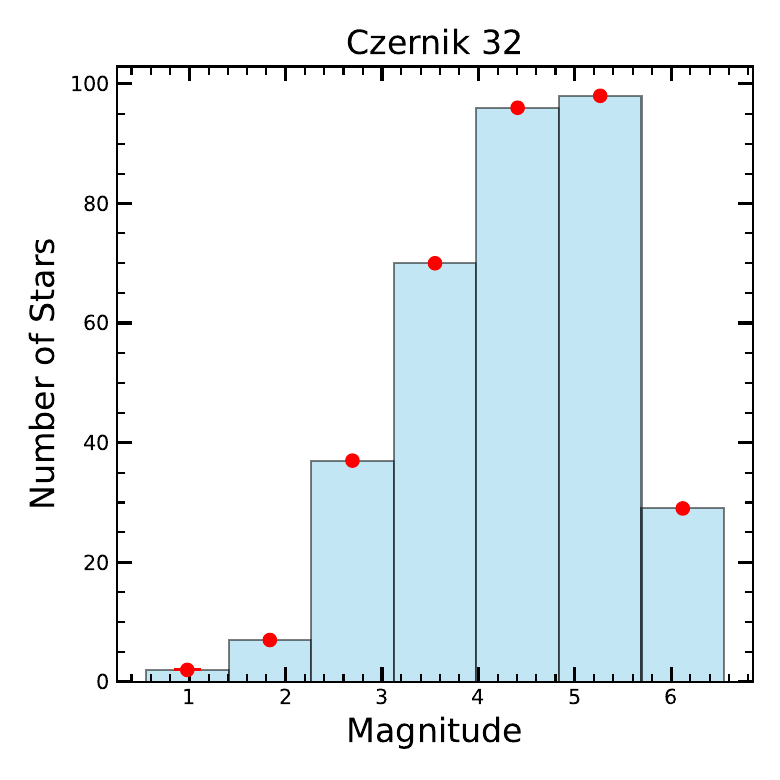}
    \includegraphics[width=4.5cm,height=4.2cm]{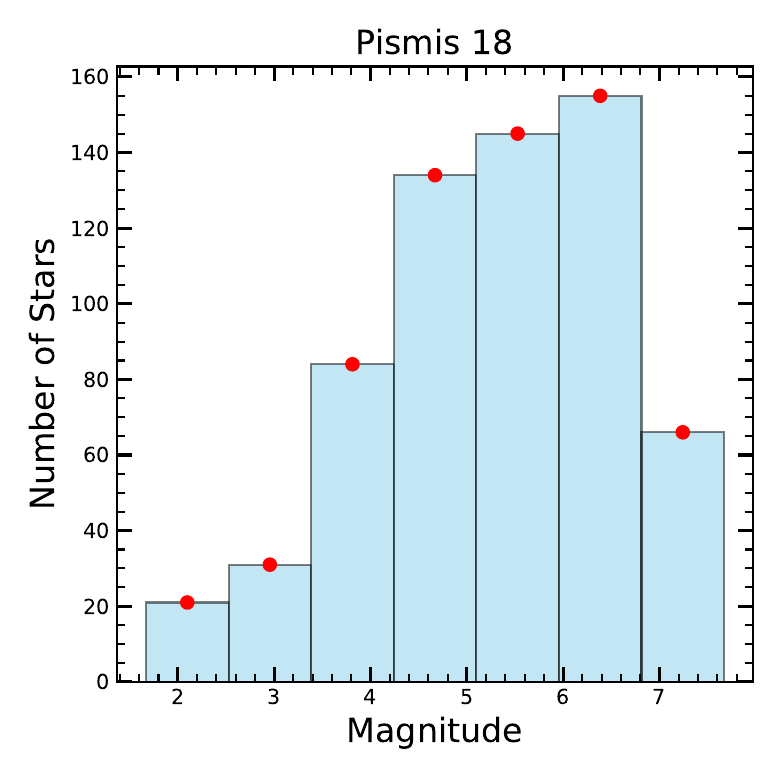}
    \includegraphics[width=4.6cm,height=4.2cm]{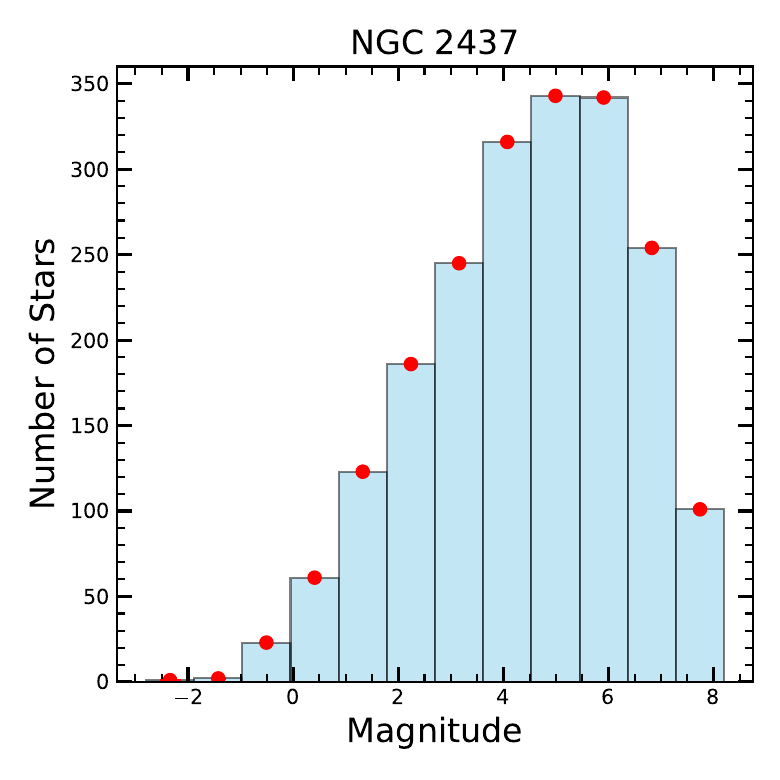}
    \caption{Histograms show the luminosity distribution of stars in each open cluster as a function of absolute magnitude. The blue bars represent the number of stars in each magnitude bin, and the red dots indicate the corresponding Poisson uncertainties.}
    
    \label{fig: Luminosity Function}
\end{figure*}

The LF represents the distribution of stars in a cluster based on their luminosity. For main-sequence stars in each cluster, the LFs is derived from the true cluster stars identified in section \ref{sec:Membership probability}. To construct the LFs histogram, we used the stars from the CMDs ($G$ vs. $G_{BP} - G_{RP}$) as targets. The apparent $G$ magnitudes of the cluster stars were converted into absolute magnitudes using the distance modulus and A$_G$ (extinction in the $G$ band). A binning interval of 1 magnitude was chosen to ensure a sufficient number of stars for good statistical analysis. The constructed histogram is illustrated in the figure \ref{fig: Luminosity Function}. The mean absolute magnitudes were found to be 4.3, 5.3, 4.8, 4.5, 5.5, and 5.8 for NGC 2204, NGC 2660, NGC 2262, Czernik 32, Pismis 18, and NGC 2437, respectively. These values represent the central tendency of the luminosity function for the main-sequence members in each cluster. Building upon this, we next examine the mass function, which provides a deeper understanding of the stellar mass distribution and the dynamic state of each cluster.\\

\subsection{Mass Function}
\begin{figure*}
    \centering
    \vspace{-0.2cm}
    \includegraphics[width=4.5cm,height=5.0cm]{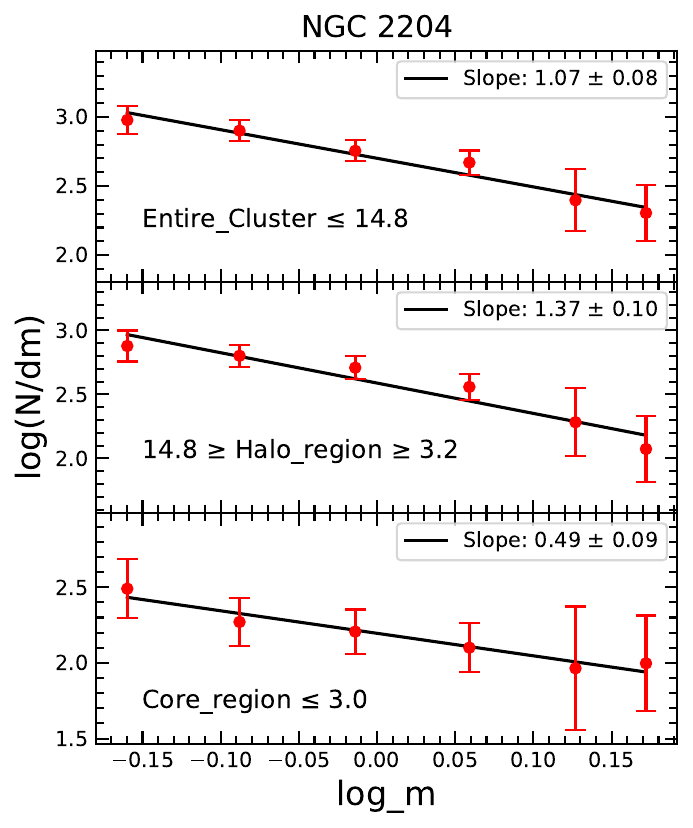}
    \includegraphics[width=4.5cm,height=5.0cm]{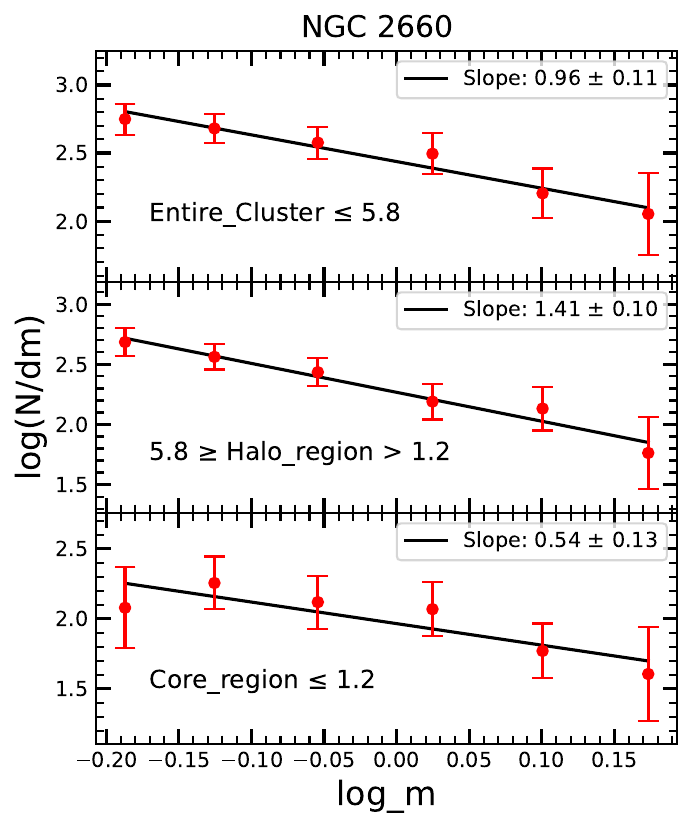}
    \includegraphics[width=4.5cm,height=5.0cm]{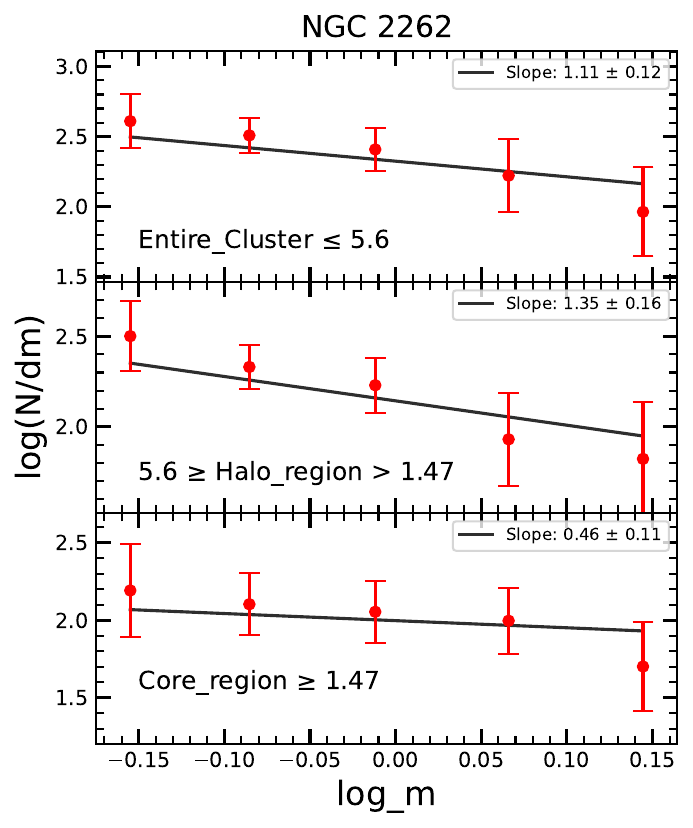}
    \includegraphics[width=4.5cm,height=5.0cm]{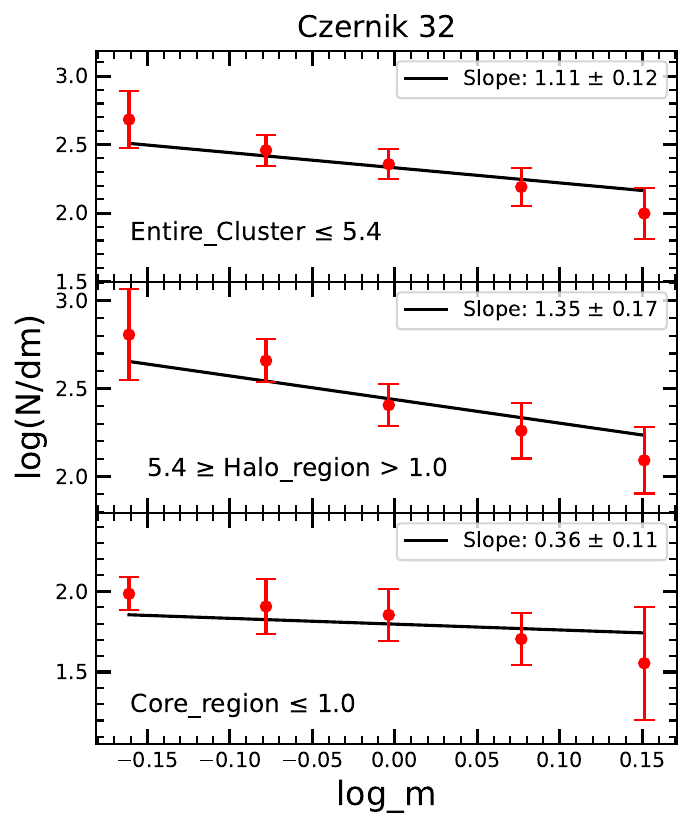}
    \includegraphics[width=4.5cm,height=5.0cm]{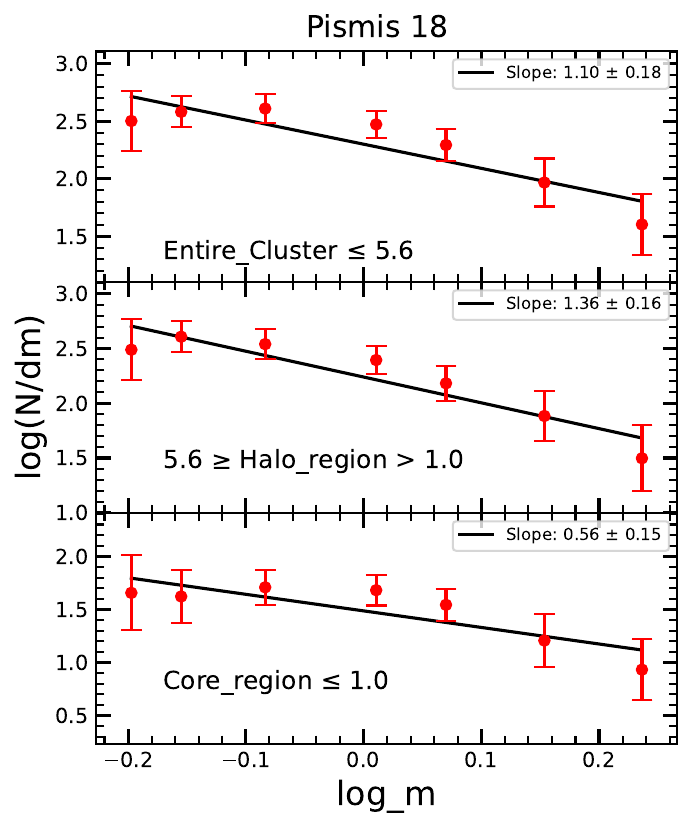}
    \includegraphics[width=4.5cm,height=5.0cm]{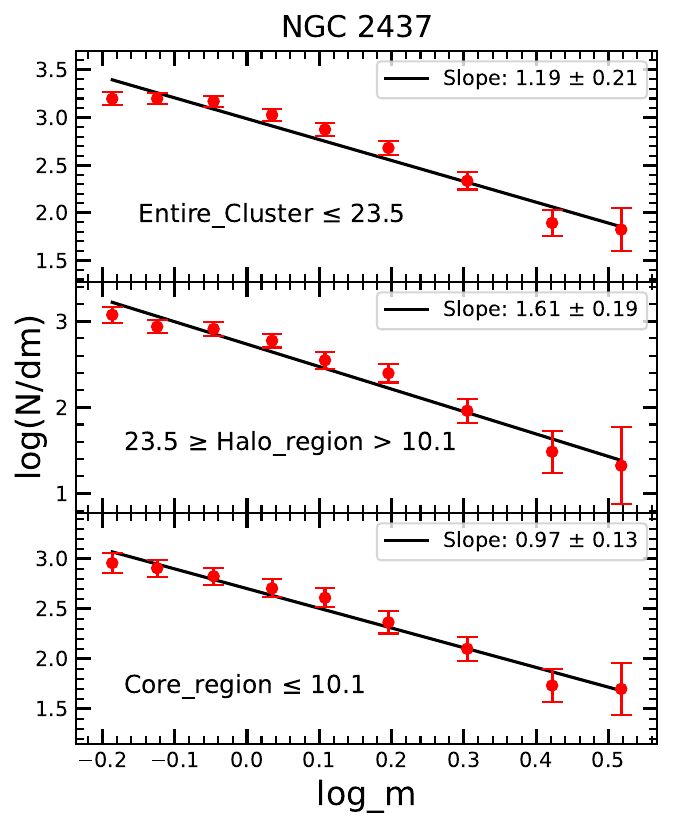}
    \caption{Mass function (MF) slopes for the six open clusters—NGC~2204, NGC~2660, NGC~2262, Czernik~32, Pismis~18, and NGC~2437—evaluated separately for the entire cluster region, halo region, and core region. The MF slope (x) is derived by applying a least-squares linear fit to the $\log(N/\Delta \log m)$ vs \ $\log m$ distribution, shown as solid black lines in each panel. Error bars represent Poisson uncertainties. Variations in the MF slopes across different regions provide insights into the internal dynamical evolution and degree of mass segregation within each cluster.}
    \label{fig: Mass Function Slope}
\end{figure*}

The mass function (MF) describes the distribution of stellar masses within a cluster per unit volume. To derive the MF, the observed luminosity function (LF) is converted using a mass-luminosity relation that depends on the cluster’s fundamental parameters, such as reddening, distance, and age. For this purpose, we employed theoretical isochrones from \citet{marigo2017new}. The observed stars were grouped into magnitude bins, and each bin was associated with a corresponding mass interval derived from the adopted isochrone. This enabled the transformation of the LF into the MF. The resulting mass function plots are shown in Figure  \ref{fig: Mass Function Slope}. The slope of the resulting MF was determined by fitting a power-law to the mass distribution of cluster members using a least-squares fitting approach, as described by Equation~\ref{eq: mass_equation}.
:
\begin{equation}{\label{eq: mass_equation}}
    log \frac{dN}{dM} = -(1+x) \times log(M) + constant
\end{equation}
$dN$ represents the number of stars within a specific mass bin $dM$, while $M$ denotes the central mass of that bin. The slope of the MFs, $x$, can be derived from the equation mentioned earlier. We calculate the MFs slope for three distinct regions of each cluster: the core, the halo, and the entire cluster region. The radius values for these cluster regions are referenced from Section \ref{sec:Structural Parameters of the Cluster}. \\

We calculated the MF slopes for the stellar populations in each cluster, with values ranging from 0.36 to 0.97 in the core regions, 1.35 to 1.61 in the halos, and 0.96 to 1.19 for the entire clusters. The total stellar masses of the clusters range between 268~$M_{\odot}$ and 2453~$M_{\odot}$, where NGC~2262 is the least massive, and NGC~2437 is the most massive cluster in our sample. On average, the typical stellar mass across all clusters is close to 1~$M_{\odot}$, reflecting the dominance of low- and intermediate-mass main-sequence stars in these OCs. 
The stellar mass ranges adopted for the mass function calculations are as follows: 1.58–0.56M$_{\odot}$ for NGC 2204, 1.61–0.60M$_{\odot}$ for NGC 2660, 1.94–0.56 M$_{\odot}$for NGC 2262, 1.93–0.61M$_{\odot}$ for Czernik 32, 2.63–0.60M$_{\odot}$for Pismis 18, and 3.97–0.60M$_{\odot}$ for NGC 2437. These ranges and mass-function related parameters listed in Table \ref{tab: mass function slope} were derived from the isochrone-based mass–luminosity relations of the member stars.

The MF slopes are shallower in the core region, whereas in the entire cluster region, it is slightly lower than the Salpeter value of x = 1.354 \citep{salpeter1955luminosity}.
In contrast, the halo region shows good agreement with the Salpeter value. Moreover, the MF slopes tend to increase toward the outer regions of the cluster compared to the core. This trend may be a result of the concentration of massive stars in the cluster core. In contrast, low stars migrate toward the halo and outer regions, indicating possible mass segregation, as discussed in the next section.
\begin{table*}
\centering
\caption{The mass-function slopes for each region, along with the total mass, mean mass, relaxation time, and half-mass radius values for each cluster in this study.}
\small
\begin{tabular}{|c c c c c c c c c|}
\hline
Cluster &  Mass  & & Mass function slope(x)&  & Total Mass  & $\bar{m}$ & $R_{h}$  & $T_{R}$  \\
\hline
        & range M$_{\odot}$  & Core  & Halo  & Entire region & M$_{\odot}$   &   M$_{\odot}$  & Parcsec  &  Myr   \\
\hline
NGC 2204 & 1.58-0.56 & 0.49 $\pm$ 0.09 & 1.37 $\pm$ 0.10 & 1.07 $\pm$ 0.08 & 682 $\pm$ 80  & 0.91 & 4.26 $\pm$ 0.25 & 90.1 $\pm$ 8.1 \\
NGC 2660 & 1.61-0.60 & 0.54 $\pm$ 0.13 & 1.41 $\pm$ 0.12 & 0.96 $\pm$ 0.11  & 566 $\pm$ 60   & 0.96 & 1.73 $\pm$ 0.28 &  21.1 $\pm$ 5.4 \\
NGC 2262 & 1.94-0.56 & 0.46 $\pm$ 0.11 & 1.35 $\pm$ 0.16 & 1.11 $\pm$ 0.12 &  268 $\pm$ 40   & 0.95 & 1.93 $\pm$ 0.34 & 19.9  $\pm$ 5.5 \\
Czernik 32 & 1.93-0.61 & 0.36 $\pm$ 0.09 & 1.35 $\pm$ 0.17 & 1.11 $\pm$ 0.12  & 351 $\pm$ 50 & 1.03 & 1.79 $\pm$ 0.30 & 18.1 $\pm$ 4.8\\
Pismis 18 & 2.63-0.60 & 0.56 $\pm$ 0.15 & 1.36 $\pm$ 0.16 & 1.10 $\pm$ 0.18 & 763 $\pm$ 110  & 1.19 & 1.54 $\pm$ 0.20 & 16.4 $\pm$ 3.3\\
NGC 2437 &  3.97-0.60 & 0.97 $\pm$ 0.13 & 1.61 $\pm$ 0.19 & 1.19 $\pm$ 0.21 & 2453 $\pm$ 220  & 1.23 & 6.17 $\pm$ 0.30  &  188.7 $\pm$  15.8 \\

\hline
\end{tabular}
\label{tab: mass function slope}
\end{table*}

\subsection{Mass Segregation}
\begin{figure*}
    \centering
    \vspace{-0.3cm}
    \includegraphics[width=4.0cm,height=4.0cm]{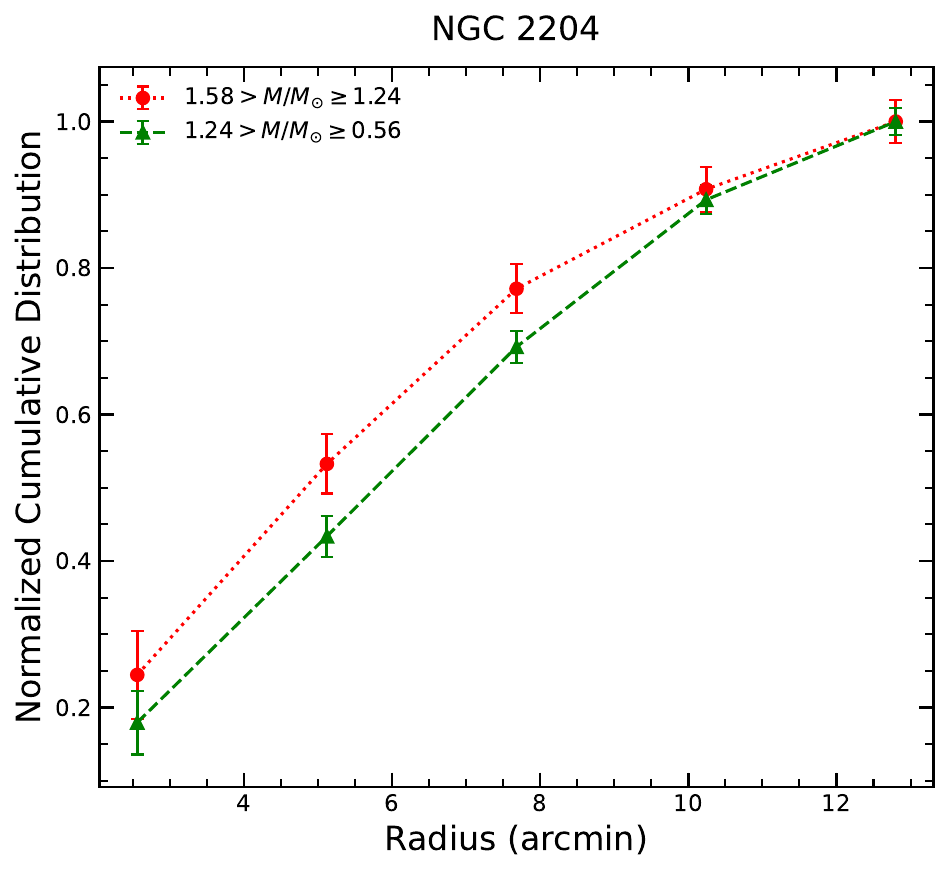}
    \includegraphics[width=4.0cm,height=4.0cm]{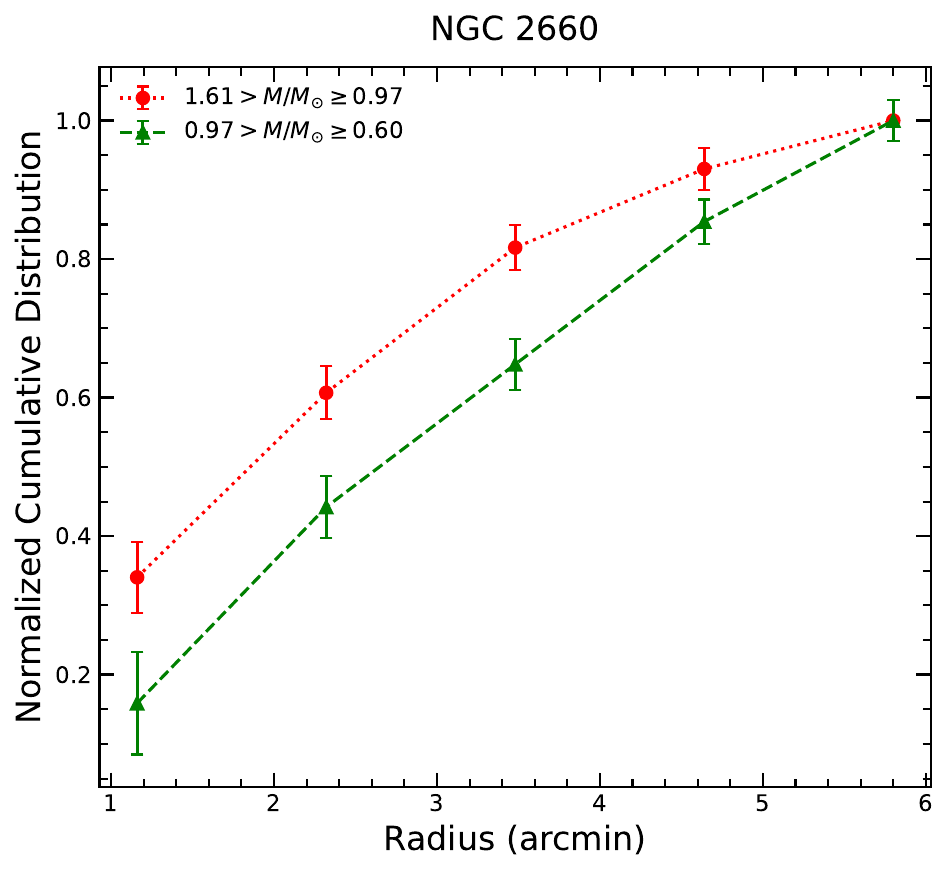}
    \includegraphics[width=4.0cm,height=4.0cm]{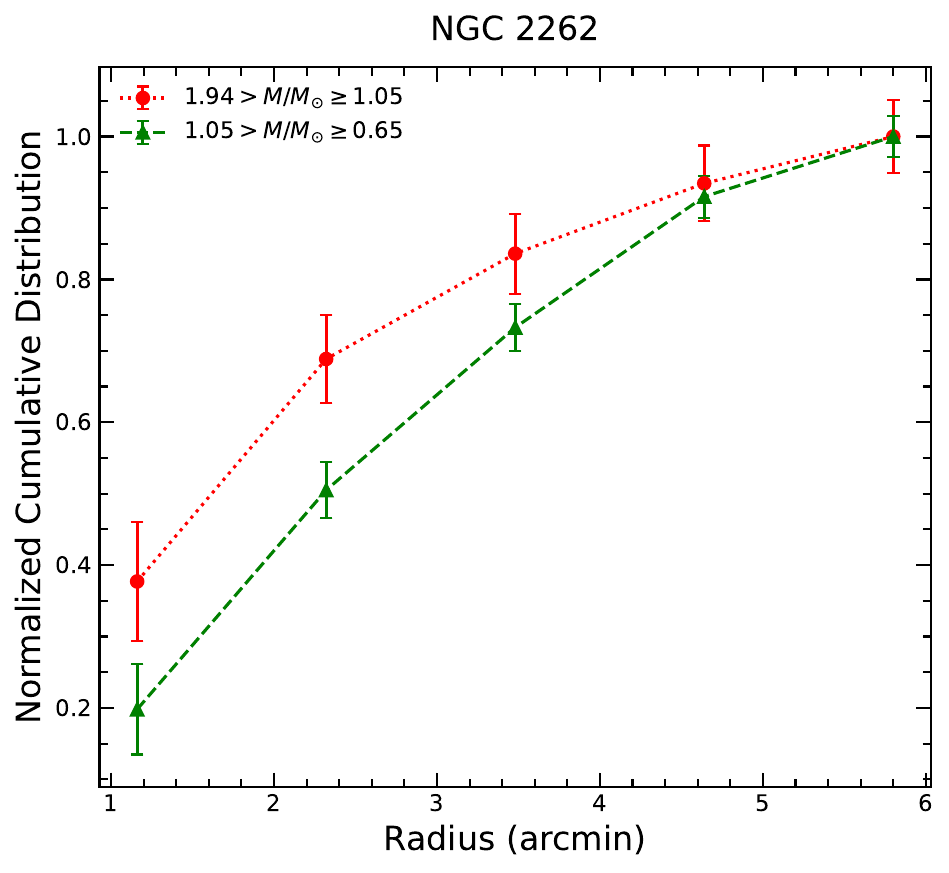}
    \includegraphics[width=4.0cm,height=4.0cm]{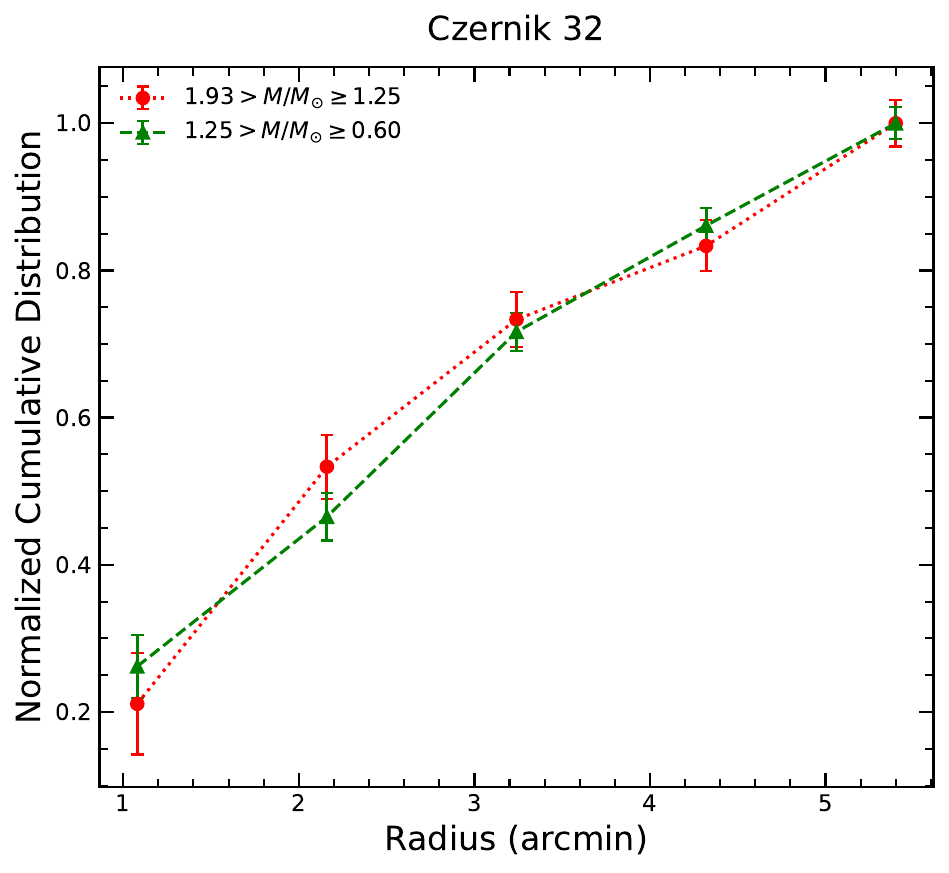}
    \includegraphics[width=4.0cm,height=4.0cm]{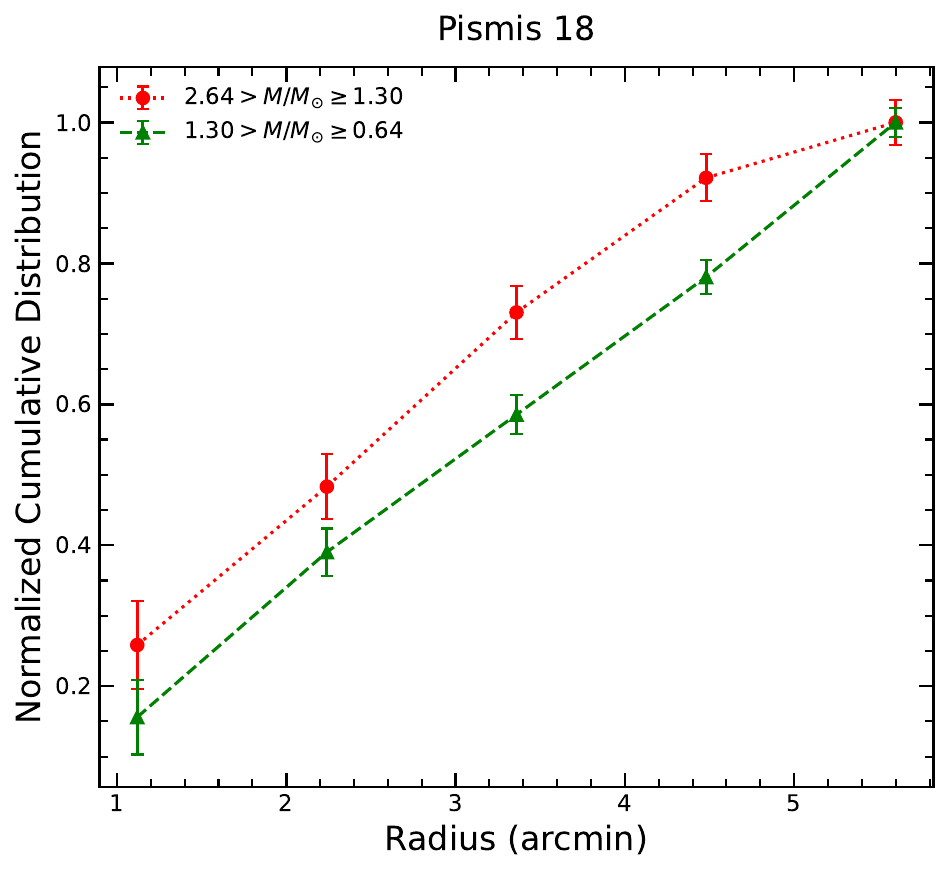}
    \includegraphics[width=4.0cm,height=4.0cm]{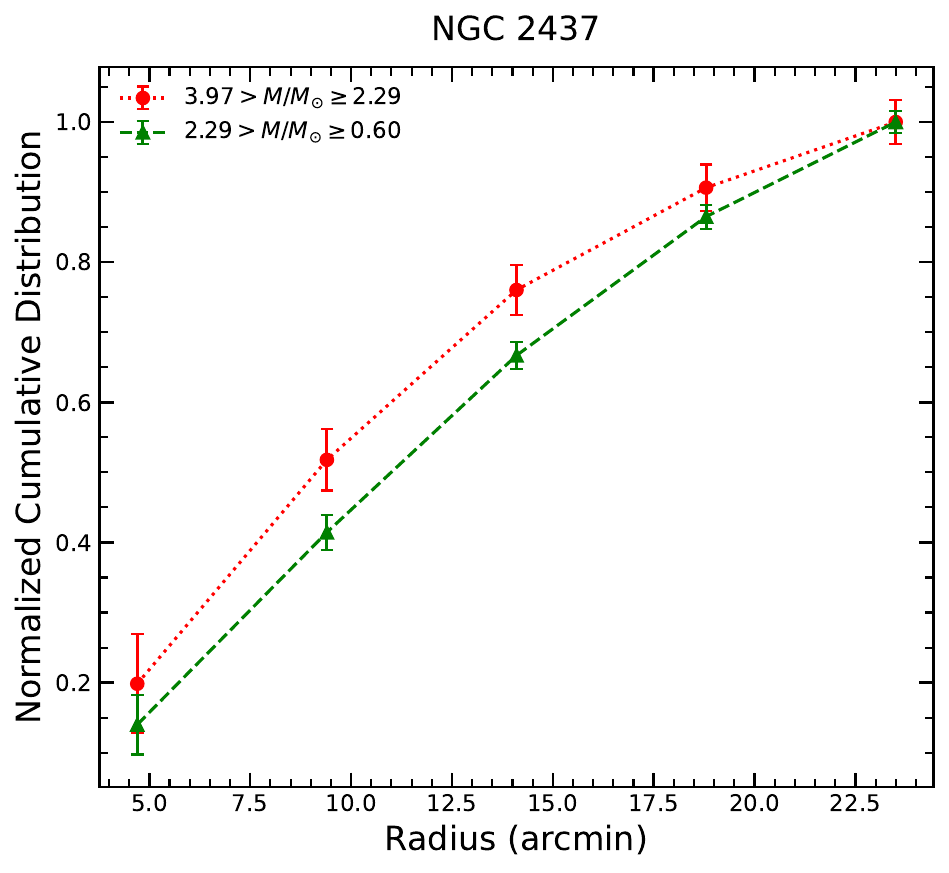}
    \caption{Cumulative radial distribution of stars in different mass ranges for six open clusters: NGC~2204, NGC~2660, NGC~2262, Czernik~32, Pismis~18, and NGC~2437. The red and green dashed lines represent the high-mass and low-mass stellar populations, respectively, with the specific mass ranges indicated in the legend of each panel. A steeper distribution of high-mass stars toward the centre suggests evidence of mass segregation in these clusters.}
    \label{fig: Mass Segregation}
\end{figure*}

In the study of mass segregation, it is suggested that high-mass stars migrate toward the cluster center while low-mass stars shift toward the halo region. This phenomenon has been observed in various clusters by several authors, including \citet{Kroupa2002MNRAS.336.1188K}, \citet{dib2007origin}, and \citet{bisht2020comprehensive}, \citet{bisht2022deep11}. This occurs due to the equipartition of energy: high-mass stars, which have lower velocities and experience stronger gravitational forces, concentrate near the cluster center. Conversely, low-mass stars with higher velocities can more easily escape the gravitational pull of the central region, resulting in their distribution toward the cluster's outskirts. In this analysis, we used only true cluster members to investigate the effect of mass segregation. 
To analyze mass segregation in different stellar regimes, the stars in each cluster were divided into two mass bins: high-mass and low-mass stars. The mass ranges for each bin were defined as follows: NGC~2204 ($1.58 > M/M_{\odot} \geq 1.24$ and $1.24 > M/M_{\odot} \geq 0.56$), NGC~2660 ($1.61 > M/M_{\odot} \geq 0.97$ and $0.97 > M/M_{\odot} \geq 0.60$), NGC~2262 ($1.94 > M/M_{\odot} \geq 1.05$ and $1.05 > M/M_{\odot} \geq 0.65$), Czernik~32 ($1.93 > M/M_{\odot} \geq 1.25$ and $1.25 > M/M_{\odot} \geq 0.60$), Pismis~18 ($2.64 > M/M_{\odot} \geq 1.30$ and $1.30 > M/M_{\odot} \geq 0.64$), and NGC~2437 ($3.97 > M/M_{\odot} \geq 2.29$ and $2.29 > M/M_{\odot} \geq 0.60$). These bins were chosen to capture the stellar distribution above and below the median mass range of each cluster's main-sequence population. Figure \ref{fig: Mass Segregation} shows the cumulative radial distribution of cluster members for all clusters.  We performed the Kolmogorov–Smirnov (K–S) test, a non-parametric statistical method, to evaluate the statistical significance of the mass-segregation effect. This test compares the cumulative distribution functions (CDFs) of two independent samples to determine whether they originate from the same distribution. Our analysis measures the maximum difference between the empirical CDFs of more massive and less massive stars. A significant result indicates a substantial difference in their spatial distributions, suggesting the presence of mass segregation within the cluster. This enables us to evaluate how stellar dynamics impact the arrangement of stars based on their masses. Our analysis yielded p-values ranging from 0.10 to 0.01, corresponding to confidence levels between 90\% and 99\%. These results provide statistically significant evidence that the spatial distributions of high- and low-mass stars differ across all six clusters. Such a trend strongly supports the presence of mass segregation, a dynamic process wherein more massive stars migrate toward the cluster center over time due to two-body relaxation effects.

\subsection{Relaxation Time}
The relaxation time provides a meaningful measure of the timescale over which a cluster loses any remnants of its initial conditions and reaches energy equipartition. It is characterised as the time when the stellar velocity distribution becomes Maxwellian. \citet{spitzer1971random} define the relaxation time T$_{R}$ using the following equation:
\begin{equation}
    T_{R} = \frac{8.9 \times 10^{5} \sqrt{N}\times R_{h}^{3/2}}{\sqrt{\bar{m}}\times log(0.4N)}
\end{equation}

N is the number of stars with a membership probability greater than $70\%$. $\bar{m}$ represents the average mass of the cluster stars in solar units, and R$_{h}$ denotes the half-mass radius in parsecs, as shown in Table \ref{tab: mass function slope}. R$_{h}$ specifically indicates the radius at which the cluster mass is half of the total cluster mass.

We have derived the value R$_{h}$ based on the transformation equation as given in \citet{Larsen2006},

\[ 
 R_{h} = 0.547 \times R_{c} \times \left( \frac{R_{t}}{R_{c}} \right)^{0.486}
\]

Where $R_{c}$ is the core radius and $R_{t}$ is the tidal radius, values mentioned in the table \ref{tab: kings model parameters}. The half-mass radius ($R_{h}$) ranges from 1.54 to 6.17 parsecs, as presented in Table \ref{tab: mass function slope}. Similarly, the estimated relaxation time ($T_{R}$) varies from 16 to 189 Myr, as shown in the same table. Since the age of each cluster exceeds its corresponding relaxation time, all clusters are dynamically relaxed.


\subsection{Orbits of the clusters}\label{orbits}

The orbital trajectories in the Galaxy were calculated for the six clusters under study using the \textbf{galpy} module developed by \citet{2015ApJS..216...29B}.
The detailed method for deriving orbits of open clusters using \textit{galpy} is described in \citet{2025JApA...46...52R}.
First, using Astropy, the heliocentric position and velocity components of all the OCs were transformed into their Galactocentric positions and velocities, which are listed in Table \ref{tab:orbits}. 

To perform an orbital analysis of clusters, we calculated the mean radial velocity (RV) using only the member stars of the cluster. For NGC 2204, we found 44 stars with RVs, yielding a mean value of $92.32 \pm 2.29$~km\,s$^{-1}$. For NGC 2660, 31 stars were identified with a mean RV of $22.66 \pm 5.43$~km\,s$^{-1}$. For NGC 2262, we found 13 stars with a mean RV of $63.94 \pm 3.76$~km\,s$^{-1}$. For Czernik 32, 15 stars were included, resulting in a mean RV of $72.29 \pm 4.69$~km\,s$^{-1}$. Pismis 18 had 22 stars with a mean RV of $-24.70 \pm 11.37$~km\,s$^{-1}$. Finally, for NGC 2437, 156 stars were considered, with a corresponding mean RV of $45.15 \pm 18.75$~km\,s$^{-1}$. 

For the present analysis, we also used \textbf{MWPotential2014} Galactic potential to derive the orbits. This axisymmetric Galactic potential consists of three main components: a power–law bulge with exponential cutoff, a Miyamoto–Nagai stellar disk, and a spherical Navarro–Frenk–White halo representing the dark matter. The model does not include non-axisymmetric features such as the Galactic bar or spiral arms; however, for OCs located in the thin disk at intermediate Galactocentric radii, this model provides a sufficiently accurate and widely used description of the Galactic potential. Since all clusters are located far from the central region of the galaxy, the influence of the Galactic bar will not be significant. The location of the Sun was taken at a distance of 8.178 kpc from the Galactic center and velocity 220 km s$^{-1}$ towards the Galactic rotation. 
The cluster orbits were back-integrated for a time interval equal to their age.

The calculated orbits are shown in Fig. \ref{fig:orbit}, where the birth and present positions of these OCs are denoted by a red dot and a triangle, respectively. 
The trajectories shown in green represent the path of cluster motion in the RZ projection. 
Here, R is the distance of the cluster from the Galactic center, and Z is the distance of the cluster from the Galactic disc.
The purple trajectories represent the path followed by the cluster around the galactic center in the projection of the x- and y-components of distance from the galactic center. 
The calculated orbital parameters, such as eccentricity, apogalactic and perigalactic distance, maximum vertical distance travelled by the clusters from the Galactic disc, and the time period, are listed in Table \ref{tab:orbits}.

We investigated the effect of uncertainties in the initial conditions, specifically proper motions, distance, and radial velocity on the derived orbital parameters using galpy. Error propagation was performed through Monte Carlo sampling with 500 random realizations. In each realization, the input parameters were drawn from respective probability distributions, incorporating the observational uncertainties, and resulting orbits were integrated backward over 1.9 Gyr, corresponding to the ages of the oldest cluster in our sample. From the distribution of orbital solutions, we estimated the maximum standard deviation of 0.50 kpc in R, 0.05 kpc in Z, 19.43 km/s in U, 15.33 km/s in V 10.6 km/s in W, 0.05 in eccentricity, 1.13 kpc in R$_{ap}$, 1.02 kpc in R$_{peri}$, 0.31 in Z$_{max}$, and 0.02 Myr in T. This indicates that the uncertainties are relatively small compared to the orbital integration scale.

\begin{figure*}
    \centering
    \includegraphics[width=0.23\linewidth]{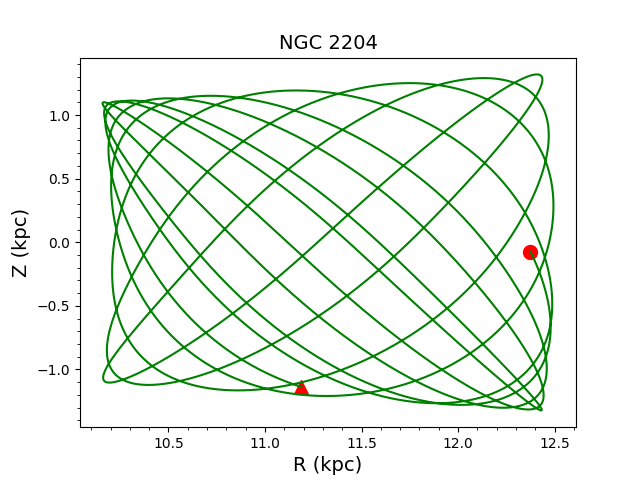}
    \includegraphics[width=0.23\linewidth]{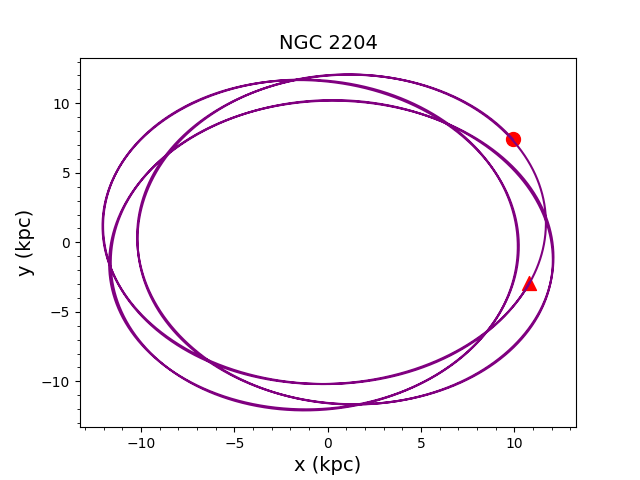}
    \includegraphics[width=0.23\linewidth]{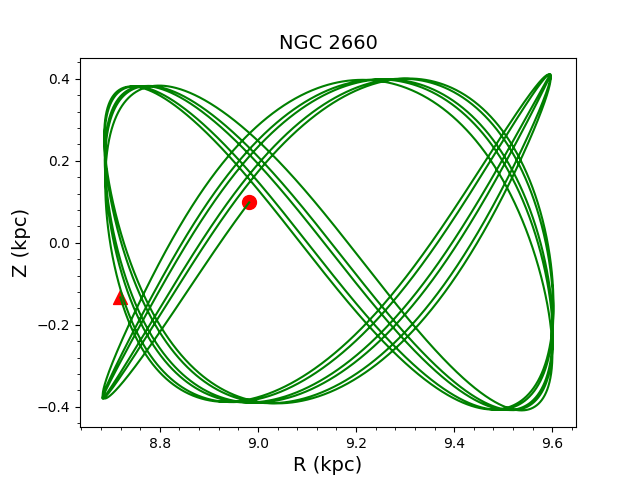}
    \includegraphics[width=0.23\linewidth]{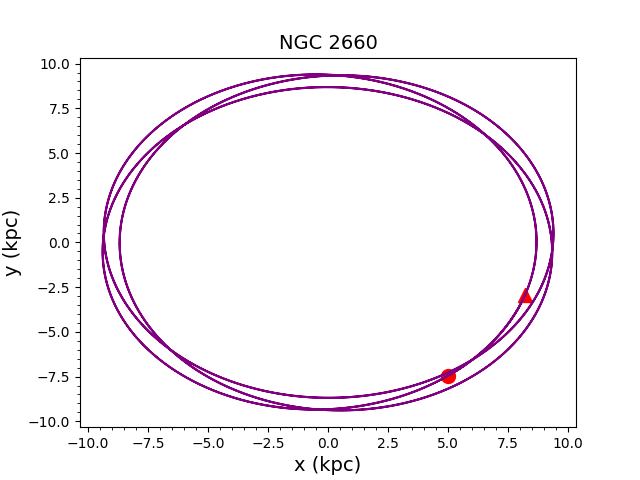}
    \includegraphics[width=0.23\linewidth]{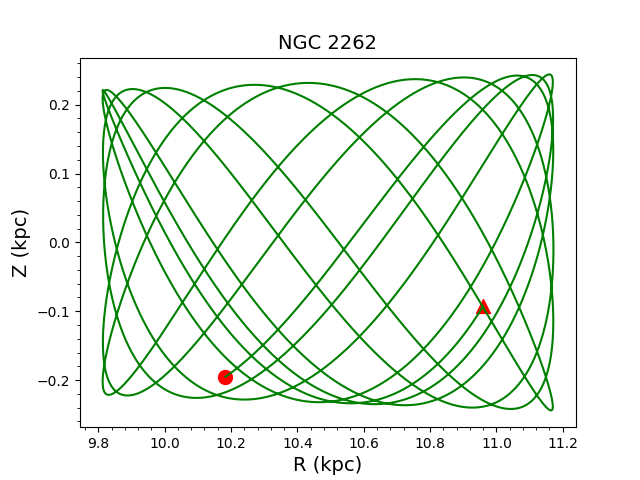}
    \includegraphics[width=0.23\linewidth]{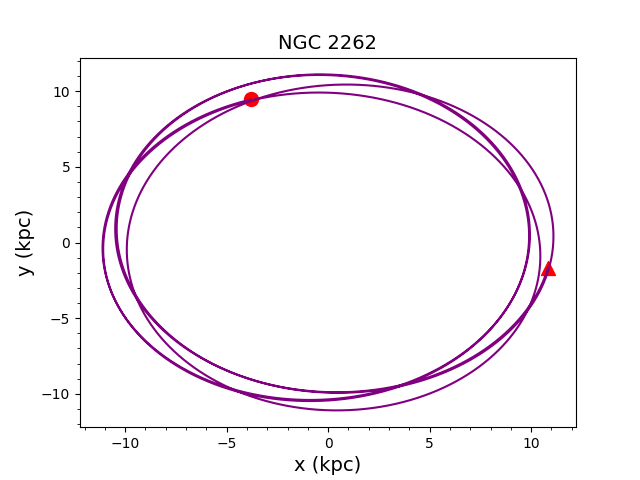}
    \includegraphics[width=0.23\linewidth]{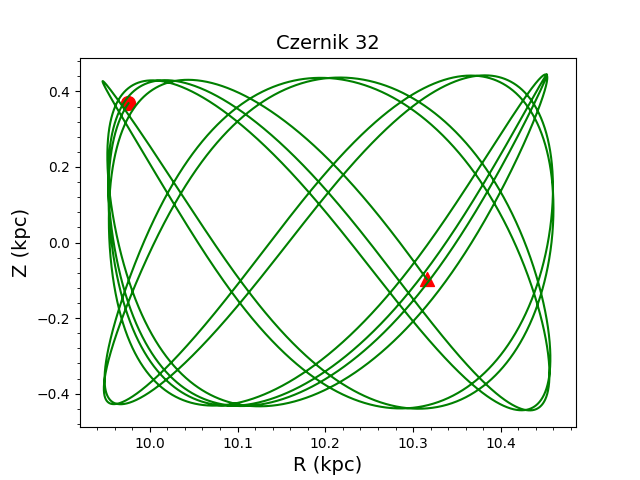}
    \includegraphics[width=0.23\linewidth]{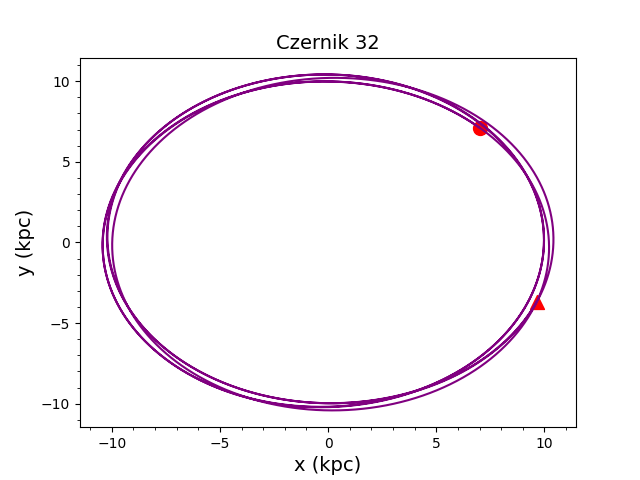}
    \includegraphics[width=0.23\linewidth]{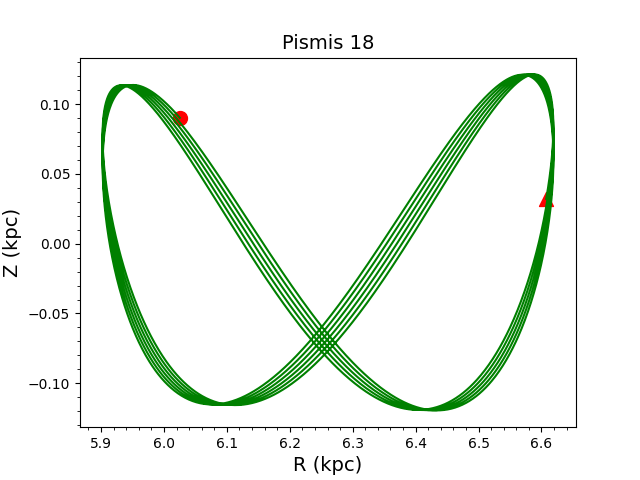}
    \includegraphics[width=0.23\linewidth]{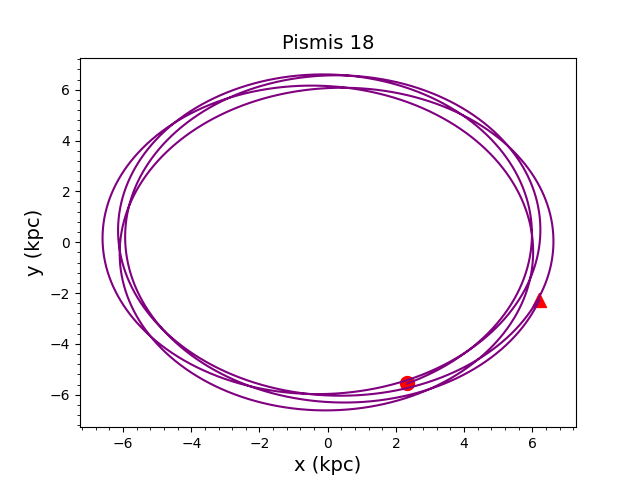}
    \includegraphics[width=0.23\linewidth]{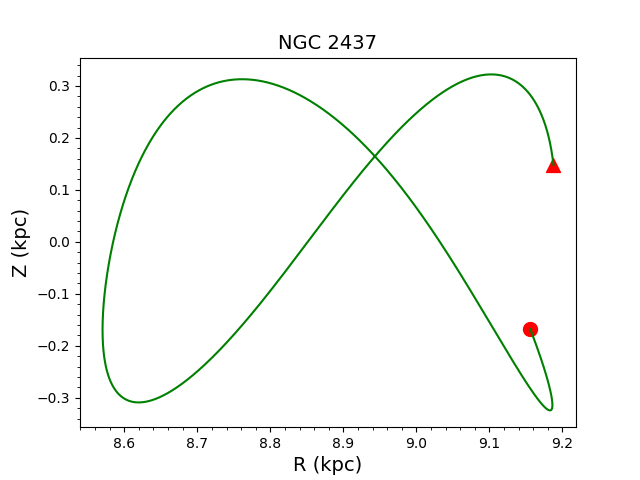}
    \includegraphics[width=0.23\linewidth]{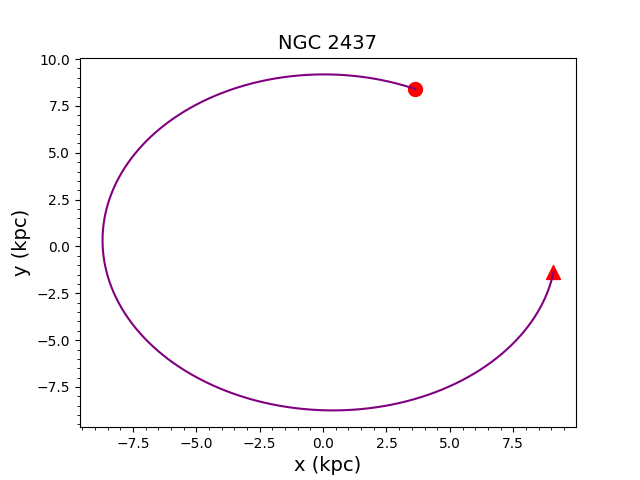}
    \caption{Orbital trajectories of the six open clusters under study—NGC~2204, NGC~2262, NGC~2660, Czernik~32, Pismis~18, and NGC~2437—plotted in the $XY$, $RZ$, and $YZ$ planes. These projections illustrate the Galactic orbits computed using the clusters’ kinematic and positional parameters. Here, $X$ and $Y$ denote the positions in the Galactic plane relative to the Galactic centre, $R$ is the radial distance from the Galactic centre, and $Z$ is the vertical distance from the Galactic mid-plane. The red dot indicates the birth position, while the red triangle marks the present-day positions of the clusters. These orbital maps help trace the dynamical evolution and vertical oscillations of the clusters in the Galaxy.} 
    \label{fig:orbit}
\end{figure*}

\begin{table*}
    \centering
    \caption{The orbital parameters of the clusters under study. Here, (R, Z) are the position coordinates, and (U, V, W) are the velocity components of the clusters. R$_{ap}$ and R$_{peri}$are the apogalactic and perigalactic positions in the orbits of the cluster. Z$_{max}$ is the maximum distance travelled by the cluster from the Galactic disk, and T is the time of the revolution. }
    \begin{tabular}{|lllllllllll|}
    \hline
      Name & R & Z & U & V & W & eccentricity & R$_{ap}$ & R$_{peri}$ & Z$_{max}$ & T \\
      \hline
      & (kpc) & (kpc) & (km s$^{-1}$) & (km s$^{-1}$)  & (km s$^{-1}$) &  & (kpc) & (kpc) & (kpc) & (Myr) \\
      \hline
      NGC 2204 & 11.19 & -1.14 & 80.58 & 195.42 & -13.17 & 0.10 & 12.50 & 10.21 & 1.32 &  0.25   \\
      NGC 2660 & 08.72 & -0.13 & 71.33 & 214.65 & 20.88 & 0.05 & 09.61 &  08.69 & 0.41 &  0.19 \\
      NGC 2262 & 10.96 & -0.09 &  43.73 & 199.07 & 10.12 & 0.06 & 11.17 & 09.81 &  0.24 & 0.23 \\
      Czernik 32 & 10.32 & -0.10 & 80.75 &  194.99 & -19.71 & 0.02 &  10.46 & 9.95 &  0.45 & 0.22\\
      Pismis 18 &  06.61 &  0.03 & 69.34 & 200.23 & -10.82 & 0.06 & 6.62 &  5.90 & 0.12 &  0.12 \\
      NGC 2437 & 9.19 &  0.15 &  32.72 &  206.95 & -15.93 & 0.03 & 9.19 & 8.57 &  0.32 &  0.19\\
      \hline
    \end{tabular}
    \label{tab:orbits}
\end{table*}

Fig. \ref{fig:orbit} shows that all the clusters under study are tracing a circular trajectory around the Galactic center in a boxy pattern, a well-known characteristic of the thin disk Galactic clusters. The orbital analysis shows that all the clusters are born in the thin disk, hence moving very close to the Galactic disc, tracing a small vertical height. The oldest cluster of the sample, NGC 2204, is reaching the maximum vertical distance of 1.32 kpc, and all the other younger clusters are not reaching further than 0.45 kpc from the Galactic disc. This indicates that all these clusters are highly perturbed by the tidal forces present in the Galactic disk. This may be the reason for losing their low mass stars in a short time duration, as indicated in the luminosity functions of these clusters shown in Fig. \ref{fig: Luminosity Function}. NGC 2437 is the youngest cluster in the sample and has not completed one revolution around the Galactic center. This is also the largest, with a radius of 23.5 arcmin, and the most populous cluster among all the clusters under study. The luminosity function suggests that low-mass stars have begun to evaporate, which may be attributed to their proximity to the Galactic disk.

\section{Identification and Classification of Variable Stars}\label{sec: Variable_stars_classification}

Variable stars are celestial objects whose brightness changes over time, providing valuable insights into the physical processes that drive stellar evolution. In this study, we analyzed TESS data to identify variable stars within the regions of the clusters NGC 2204, NGC 2262, Pismis 18, and NGC 2437.

We identified one variable star in NGC 2204 (TIC 59711686), one in NGC 2262 (TIC 369672843), and two variable stars in Pismis 18 (TIC 318170016 and TIC 318170024). In NGC 2437, eight variable stars were detected: TIC 94229743, TIC 94690694, TIC 94322635, TIC 94408904, TIC 94409484, TIC 94409620, TIC 94410959, and TIC 94519429. We analyzed the light curves obtained from the TESS mission to investigate the variability characteristics of these stars. The periods were determined using the Lomb-Scargle algorithm \citep{lomb1976least,scargle1982studies}, which is well-suited for analyzing unevenly spaced time-series data, as shown in the figure \ref{fig: Lomb_scargle_periodogram1}, \ref{fig: Lomb_scargle_periodogram2}. The estimated period and other parameters are shown in the Table \ref{tab: Variable_Stars_detials}.

The Hertzsprung-Russell (H-R) diagram primarily illustrates the empirical relationship between a star's spectral type and its luminosity. Alternative versions of the H-R diagram may plot temperature against luminosity or color index versus absolute magnitude. The position of a star on the diagram is influenced not only by its initial mass and age but also by factors such as stellar winds, magnetic fields, rotation, and chemical composition. For variable stars, their location on the H-R diagram provides valuable insight into their evolutionary stage. Moreover, the H-R diagram has proven highly effective in classifying variable stars, as different types tend to occupy well-defined regions within it.

In the present study, we analyzed the H–R diagram by plotting log($T_{\rm eff}$) versus log($L/L_{\odot}$) to determine the evolutionary stages of the identified variable stars. The effective temperature ($T_{\rm eff}$) and luminosity for seven stars were obtained from the \citet{2022yCat.1355....0G}. For an additional five member stars, the values of log($T_{\rm eff}$) and log($L/L_{\odot}$) were taken from the theoretical isochrone of \citet{marigo2017new}. The H–R diagram was constructed using these adopted values, as shown in Figure~\ref{fig:HR diagram}. Two distinct theoretical instability strips, corresponding to slowly pulsating B-type (SPB) stars and $\gamma$ Doradus stars, are also marked on the diagram. Within the instability strip corresponding to high-mass stars such as SPB stars, we identified one host star. Additionally, one variable star was found to lie within the $\gamma$ Doradus instability strip. The positions of these variable stars on the H–R diagram have been used to classify their variability types.

\subsection{Classification}
\subsubsection{$\gamma$ Dor Variable Stars}

$\gamma$~Doradus ($\gamma$~Dor) stars are non-radial gravity-mode (g-mode) pulsators with spectral types ranging from A7 to F5. They are found within the $\gamma$~Dor instability strip on the H–R diagram. This strip designates a transitional zone where stellar energy transport changes from convective cores with radiative envelopes to radiative cores with convective envelopes. As such, $\gamma$~Dor stars are useful for constraining theoretical models of stellar heat transfer. Their pulsation periods typically range from 0.3 to 3.0 days, with amplitudes of around 0.1 mag. These stars generally have masses between 1.5 and 1.8~$M_\odot$, making them slightly less massive than $\delta$~Scuti stars.

One of the variable stars in our sample, TIC~369672843 in the open cluster NGC~2262, shows characteristics consistent with a $\gamma$~Dor classification. It is located within the $\gamma$~Dor instability strip on the H-R diagram, with a pulsation period of 0.62375 days and an amplitude typical of $\gamma$~Dor variables. Its effective temperature, 6693.4~K, as reported by \citet{2022yCat.1355....0G}, is barely lower than the expected range of 6700--7400~K for $\gamma$~Dor stars. Based on its pulsation period, amplitude, position on the CMD, and effective temperature, we classify TIC~369672843 as a $\gamma$~Dor variable.

\subsubsection{SPB Variable Stars}

Slowly pulsating B-type (SPB) stars occupy the upper main-sequence region of the H–R diagram and show multi-periodic, non-radial g-mode pulsations driven by the $\kappa$ mechanism \citep{fedurco2020pulsational}. Their typical pulsation periods range from about half a day to several days \citep{stankov2005catalog}. In our study, we identified one bright main-sequence star in NGC 2437, TIC 94519429, with a pulsation period of 1.225416 days, which is well above the 0.42-day threshold, and an effective temperature of 9654.67 K, slightly below 10,000 K. Although the temperature and luminosity values were taken from theoretical isochrones, the star's position on the H–R diagram is consistent with the SPB instability region. Therefore, we classified it as an SPB variable. The corresponding instability strip is marked by the blue line in Figure~\ref{fig:HR diagram}.

\subsubsection{Eclisping Binary}

\begin{figure*}
    \centering
    \includegraphics[width=5.9cm,height=7.8cm]{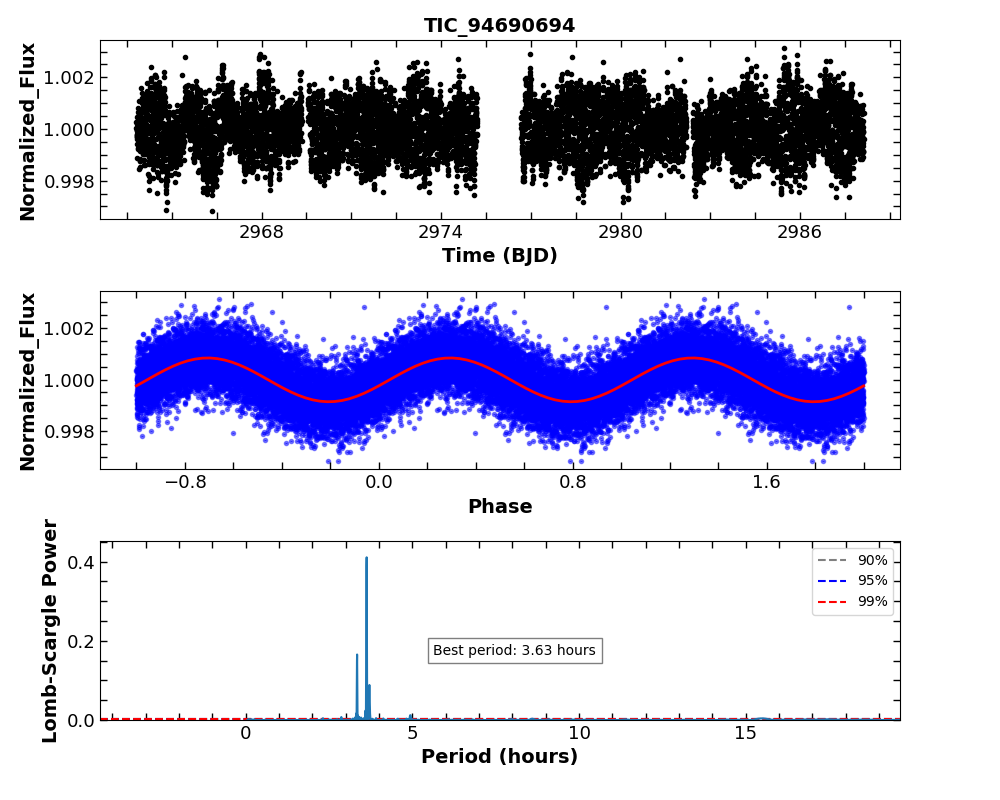}
    \includegraphics[width=5.9cm,height=7.8cm]{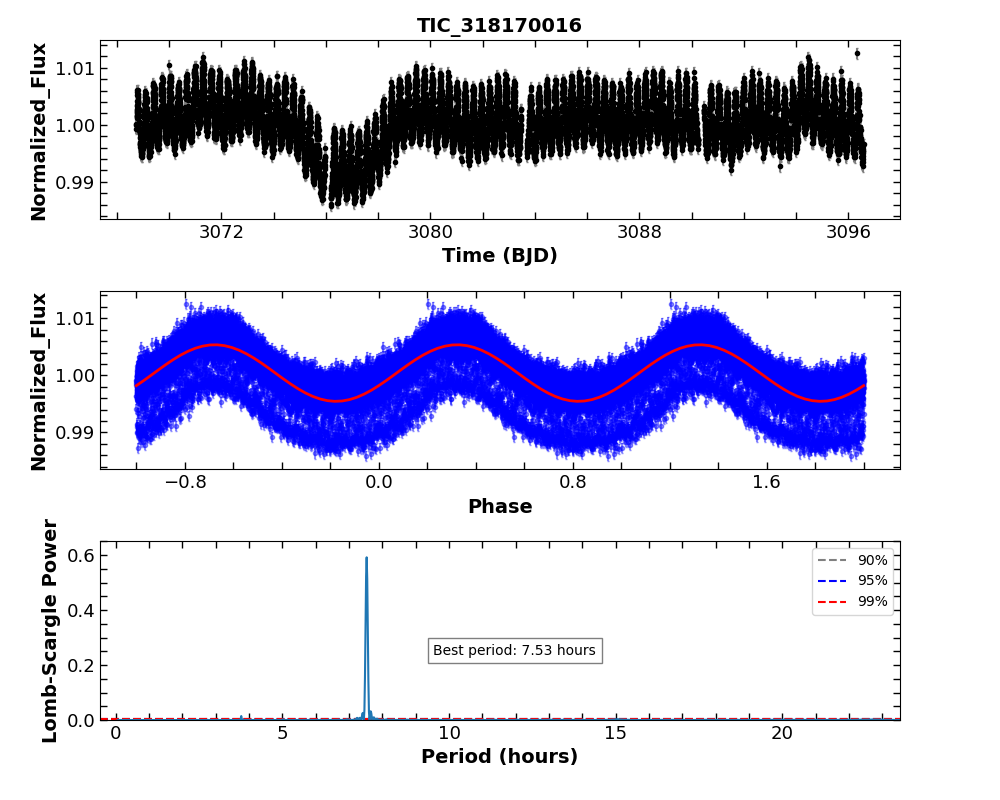}
    \includegraphics[width=5.9cm,height=7.8cm]{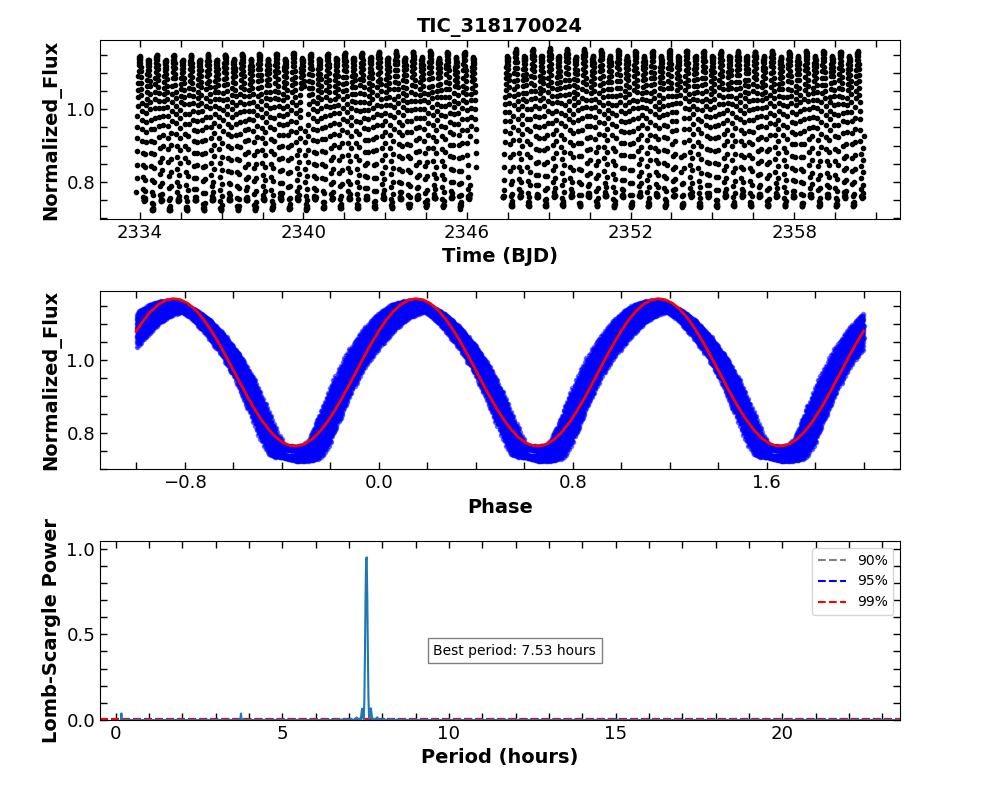}
    \caption{Light curves and periodogram analyses of the identified variable stars in the clusters NGC 2204, NGC 2262, Pismis 18, and NGC 2437. For each star, three plots are shown vertically: the observed TESS light curve (top), the phase-folded light curve with the fitted model overlaid in blue (middle), and the corresponding Lomb-Scargle periodogram (bottom). The panels correspond to the following stars (from top to bottom, left to right): TIC 59711686 (NGC 2204), TIC 369672843 (NGC 2262), TIC 318170024 and TIC 318170016 (Pismis 18), and TIC 94609694, TIC 94409620, TIC 94322635, TIC 94409484, TIC 94519429, TIC 94322074, and TIC 94229743 (NGC 2437). The detected periods, indicated by peaks in the periodograms, aid in classifying the variability types and provide constraints for stellar parameter estimation through light curve modeling.}
    \label{fig: Lomb_scargle_periodogram2}
\end{figure*}

Eclipsing binaries play a crucial role in accurately determining stellar masses, luminosities, radii, and distances \citep{stassun2021parallax}. In this study, two eclipsing binaries were identified through visual inspection of their light curves. These systems can be recognised photometrically by characteristic features, particularly the depth and shape of the primary and secondary minima, which can be interpreted using Roche lobe geometry. Algol-type (EA) binaries typically involve one component that has filled or overflowed its Roche lobe. Well-defined beginnings and ends of eclipses distinguish their light curves.
In contrast, $\beta$ Lyrae-type (EB) systems, which are semi-detached with tidally distorted components, exhibit continuously varying light curves without distinct eclipse boundaries \citet{maurya2023investigating}. W Ursae Majoris-type (EW) binaries, also known as W UMa stars, are contact systems with ellipsoidal components and nearly equal depths in their primary and secondary minima. Their light curves lack clearly defined eclipse phases and generally have periods shorter than one day.\\
We found 1 EA reported in \citet{avvakumova2013eclipsing, watson2006international} and 1 EW reported in \citet{watson2006international} binaries in the region of NGC 2437 and Pismis 18, respectively. The light curve of eclipsing binaries is shown in the figure \ref{fig: Lomb_scargle_periodogram1}, \ref{fig: Lomb_scargle_periodogram2}. A detailed analysis of this star is presented in Section \ref{sec: Phoebe_modeling}, along with one other eclipsing binary that is part of the field star population.

\subsubsection{Yellow Stragglers}

\begin{figure}
    \centering
    \includegraphics[width=8.0cm,height=8.0cm]{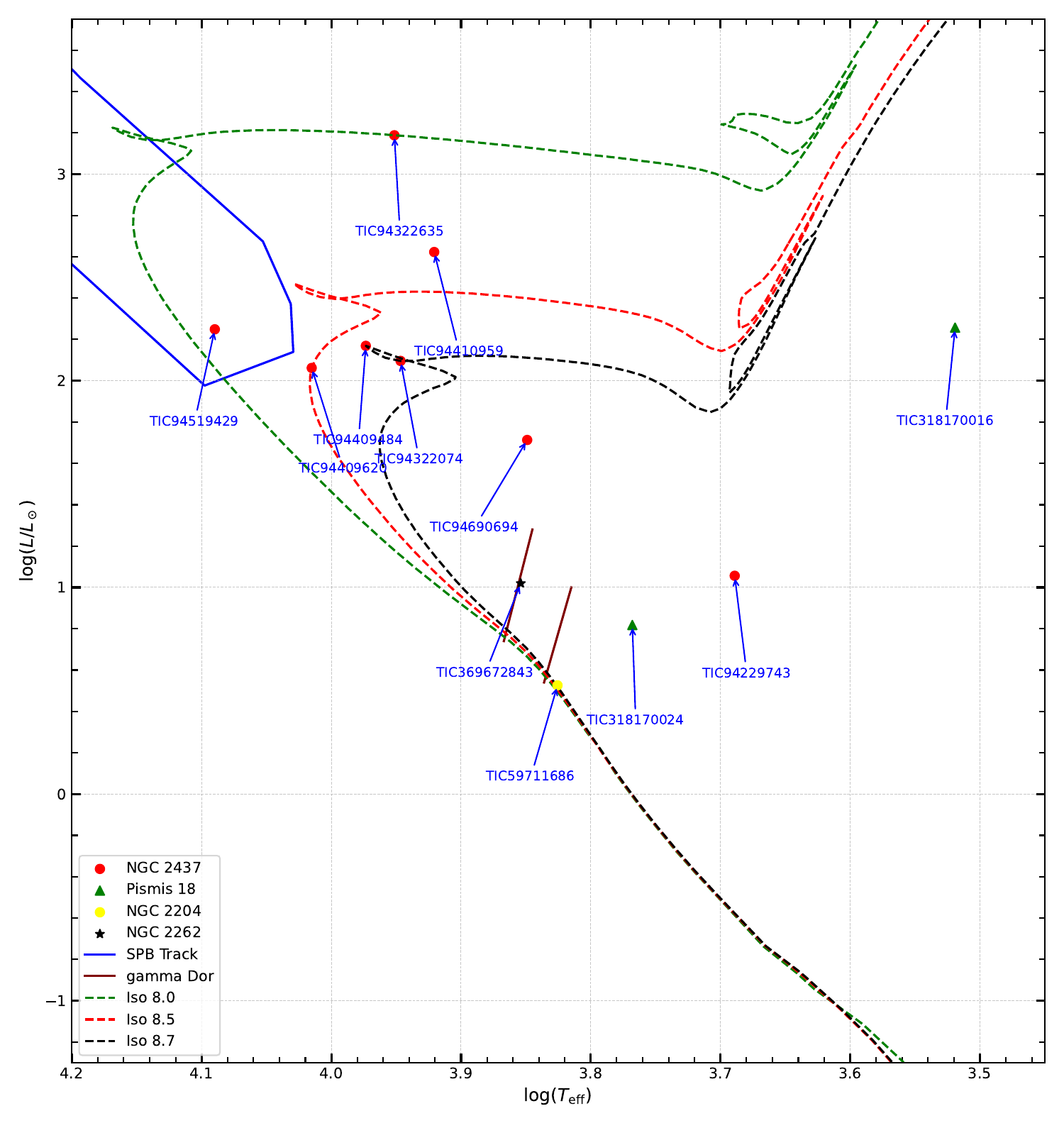}
    \caption{HR diagram (log($T_{\mathrm{eff}}$) vs. log($L/L_\odot$)) illustrating the positions of candidate and confirmed variable stars in the open clusters NGC~2437, Pismis~18, NGC~2204, and NGC~2262. Colored symbols denote different clusters. Theoretical instability strips for SPB (solid blue), $\gamma$~Doradus (brown solid) stars are overplotted, and isochrones of log age 8.0,  8.5, and 8.7 (dashed line).}
    \label{fig:HR diagram}
\end{figure}

In the H–R diagram shown in Figure~\ref{fig:HR diagram}, the star TIC~94322635 seems to satisfy the criteria for a Yellow Straggler Star (YSS) as summarized by \citet{rain2021new}. It is found above the main sequence turnoff and to the red side of the blue stragglers, which is where YSS usually appear in cluster CMDs. Our identified star is a confirmed member of NGC 2437, with an effective temperature of 7,870 K and a pulsation period of 47.04 hours. The presence of a YSS in NGC 2437 is significant because this type of star is very rare in open clusters (OCs) and may have formed through complex processes, such as interactions between two stars or mass transfer. Further spectroscopic analysis would help confirm its nature and refine its evolutionary status within the cluster.\\

Variable stars such as BSS, YSS, and binaries provide important clues to the dynamical evolution of star clusters. Their presence is closely linked to internal cluster processes: BSS and YSS are generally thought to form through mass transfer in binary systems or stellar mergers, both of which are driven by close stellar interactions. In particular, the detection of binaries among BSS underscores the role of binary evolution in shaping the observed stellar populations. Moreover, such interactions also influence the spatial distribution of stars, enhance mass segregation, and accelerate the dynamical evolution of clusters. Consequently, investigating the properties and membership of these variable stars yields valuable insights into stellar interactions, the origin of BSS and YSS, and the overall evolutionary pathways of OCs.

\subsubsection{Miscellaneous}
 
A few stars in our sample could not be clearly classified because their positions on the H–R diagram do not match any known regions for variable stars. Seven such stars, TIC 94409620, TIC 94322074, TIC 94409484, TIC 94410959, TIC 59711686, TIC 318170016, and TIC 94690694, fall into this category. Particularly, TIC 94409620, TIC 94322074, and TIC 94409484 lie below the SPB region, which, according to standard models, is normally not associated with pulsating variables. These stars were classified as fast-rotating pulsating B-type (FaRPB) stars in spectroscopic studies by \citet{mowlavi2016stellar}, with reported periods shorter than or equal to 0.5 days. However, in our analysis, these stars exhibit periods longer than 0.5 days. Therefore, without any solid evidence, we categorize them as miscellaneous variables in this study.\\

Future observations should include multi-band photometry, long-term monitoring, and targeted spectroscopy to obtain secure classifications of the detected variables. In particular, multi-band photometry will help trace color variations, while long-term monitoring will further refine their periods. Additionally, spectroscopy will provide radial velocities and stellar parameters. The resulting light curves suggest several possible classifications for the unclassified stars: they may be $\delta$ Scuti or $\gamma$ Doradus pulsators if located near the main sequence, semi-regular variables if they are red giants, or eclipsing binaries (e.g., W UMa, Algol systems) if partial eclipses are apparent. The diversity of cluster ages, moreover, implies a wide range of variable-star populations. For example, intermediate-age clusters, such as NGC 2437, are likely to host short-period pulsators and binaries, whereas older clusters, like NGC 2204 and NGC 2660, may contain red giant variables. Notably, despite the richness of NGC 2437, only a few variable stars have been identified in the currently available TESS data. To address this gap, continuous CCD monitoring, combined with upcoming Gaia variability releases, will be crucial to uncovering additional variables and achieving reliable classifications.

\begin{figure*}
    \centering
    \includegraphics[width=5.9cm,height=7.8cm]{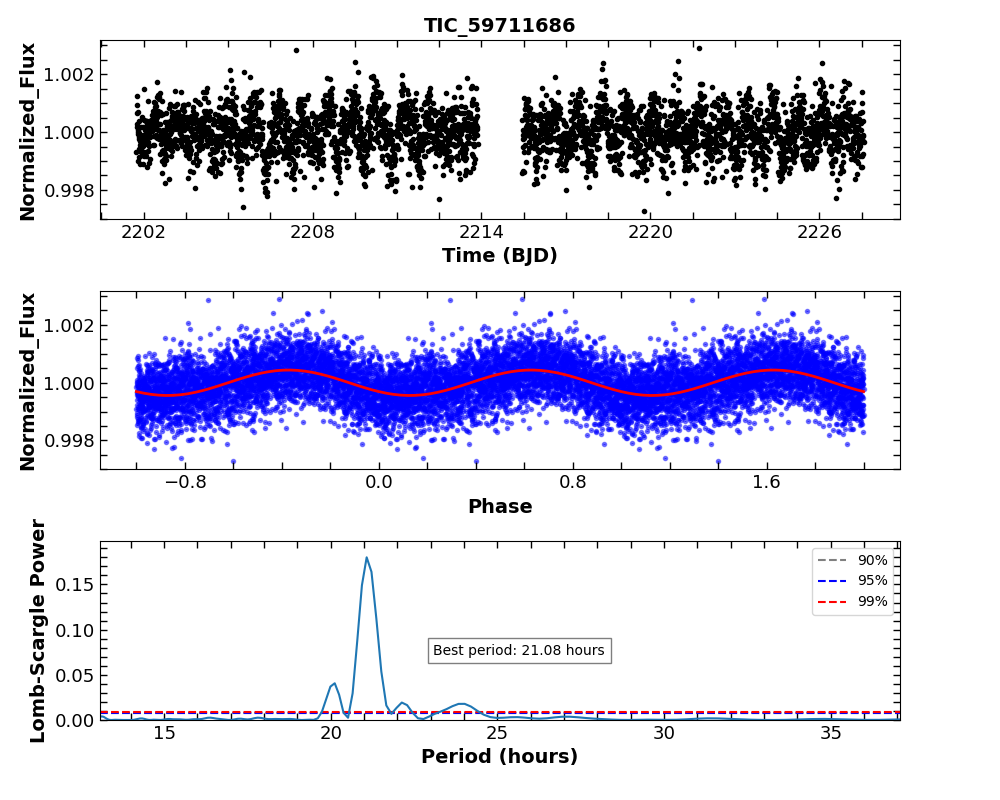}
    \includegraphics[width=5.9cm,height=7.8cm]{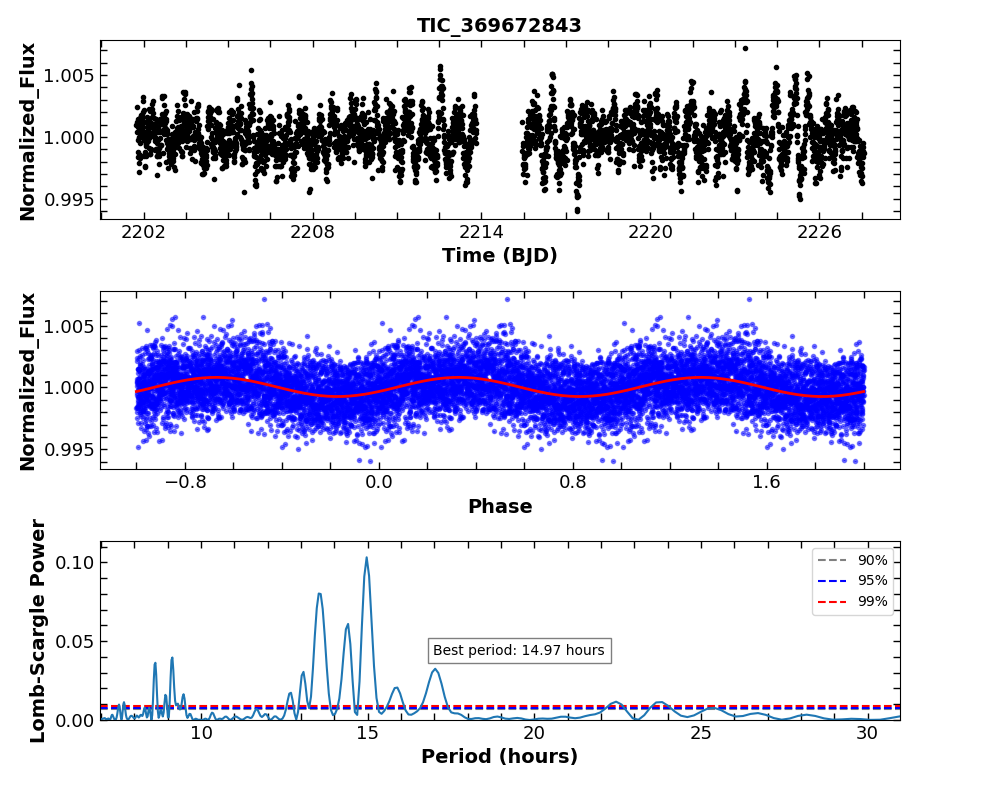}
    \includegraphics[width=5.9cm,height=7.8cm]{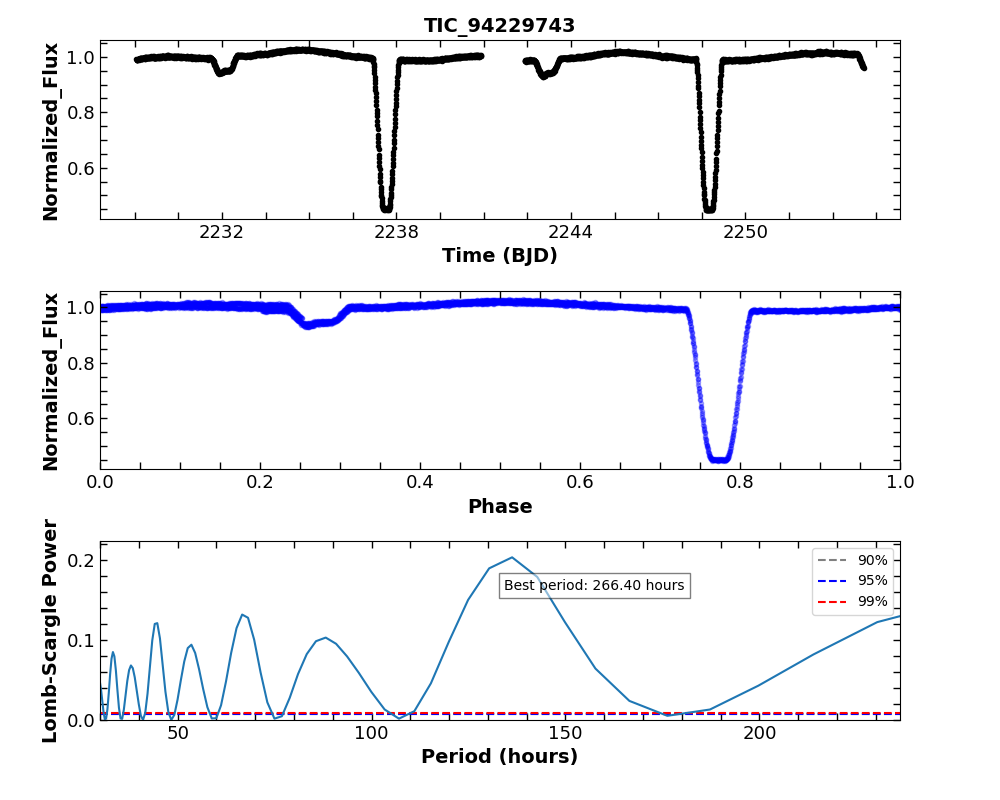}
    \includegraphics[width=5.9cm,height=7.8cm]{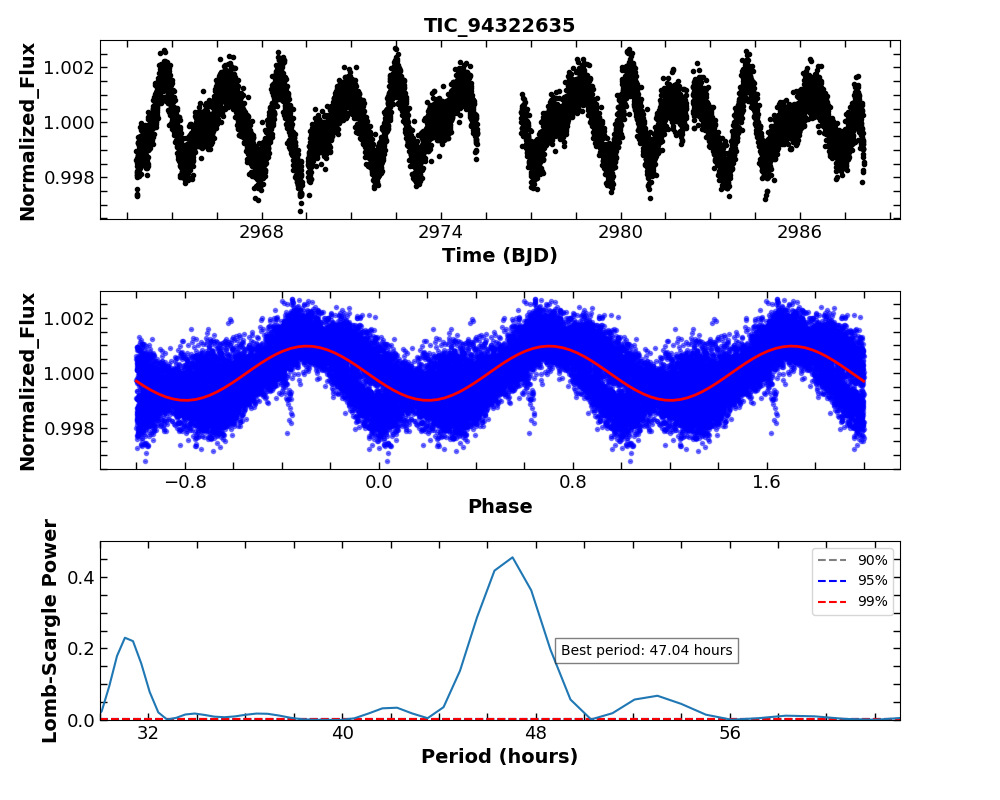}
    \includegraphics[width=5.9cm,height=7.8cm]{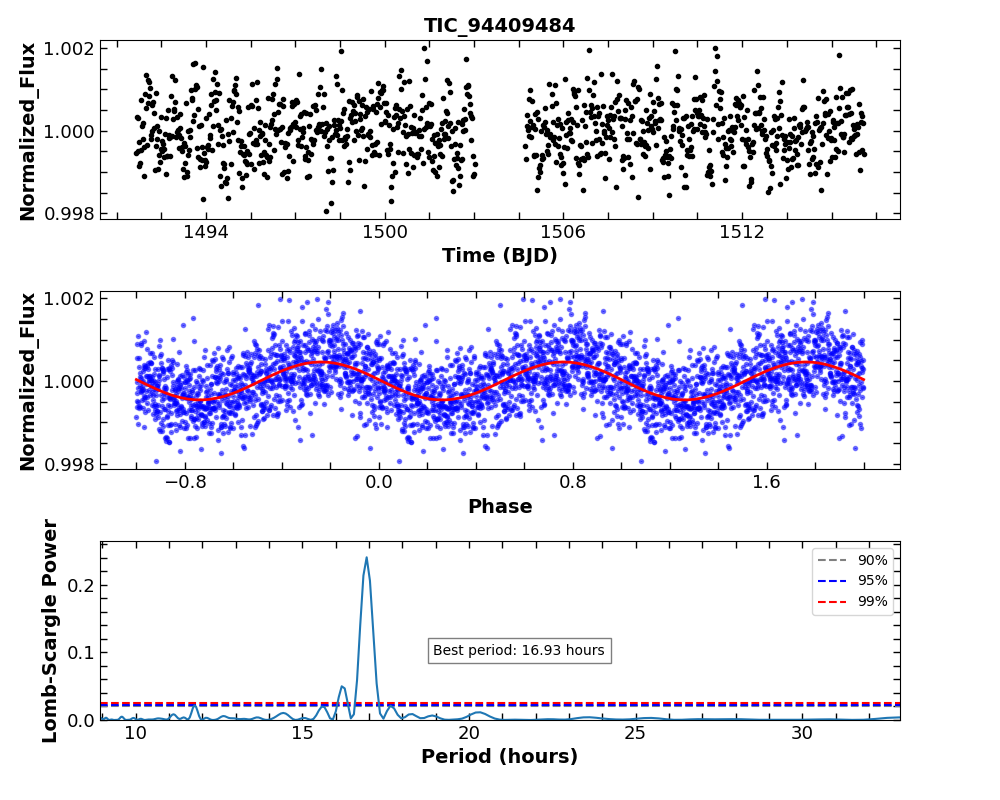}
    \includegraphics[width=5.9cm,height=7.8cm]{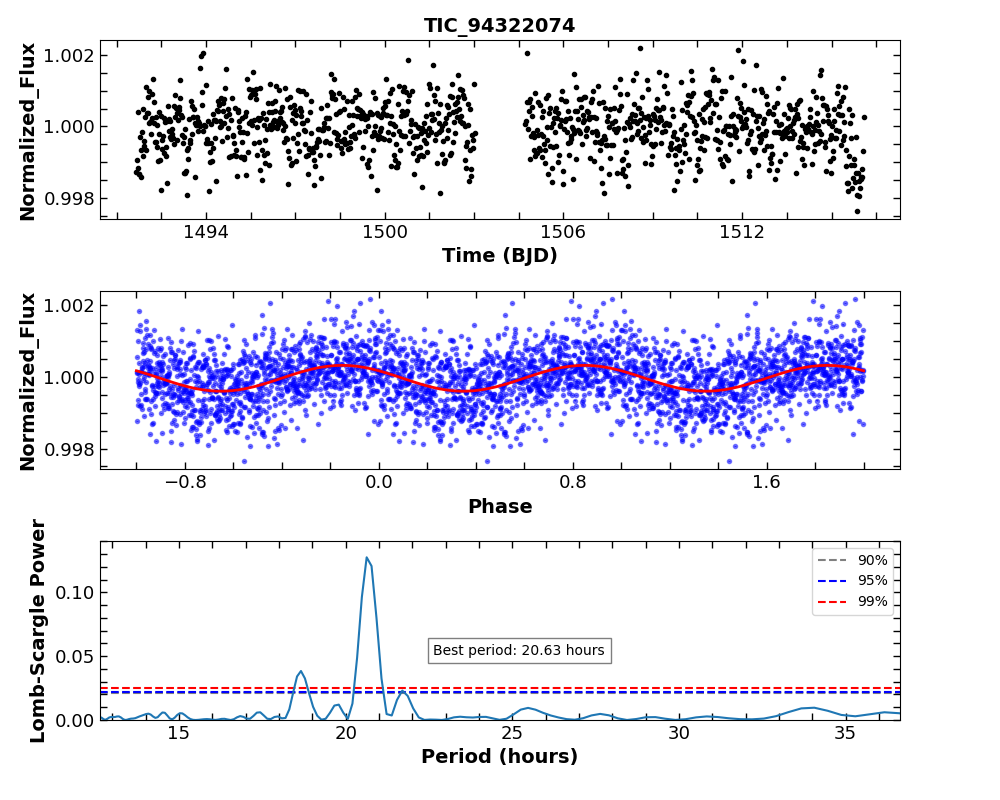}
    \includegraphics[width=5.9cm,height=7.8cm]{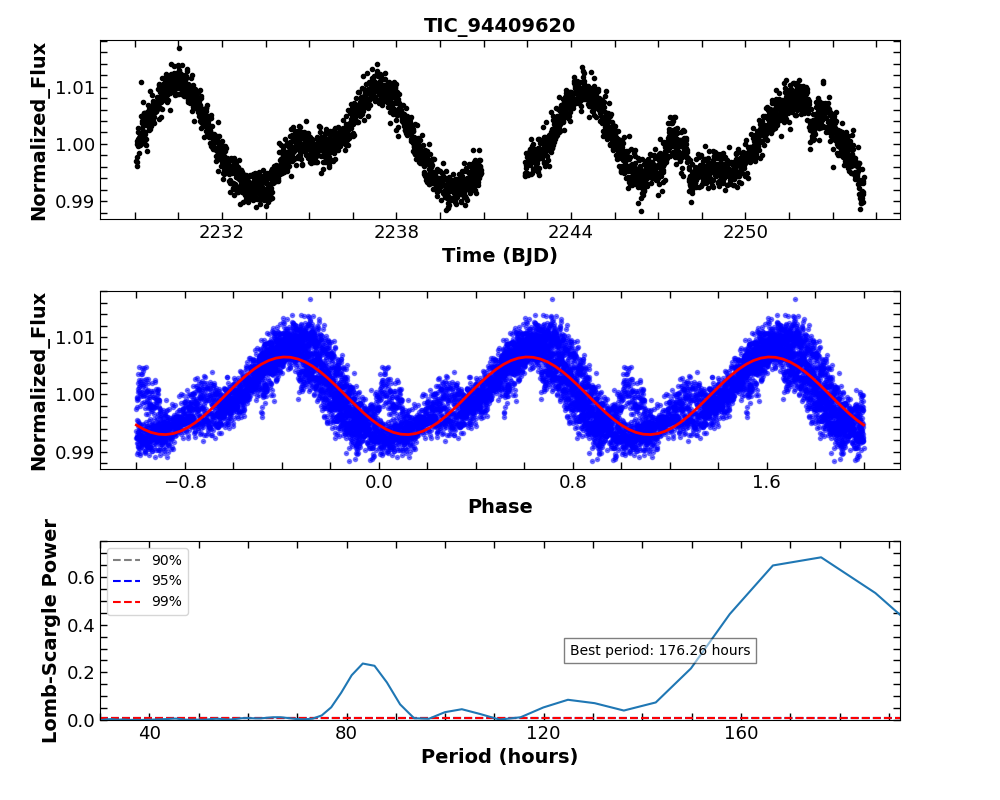}
    \includegraphics[width=5.9cm,height=7.8cm]{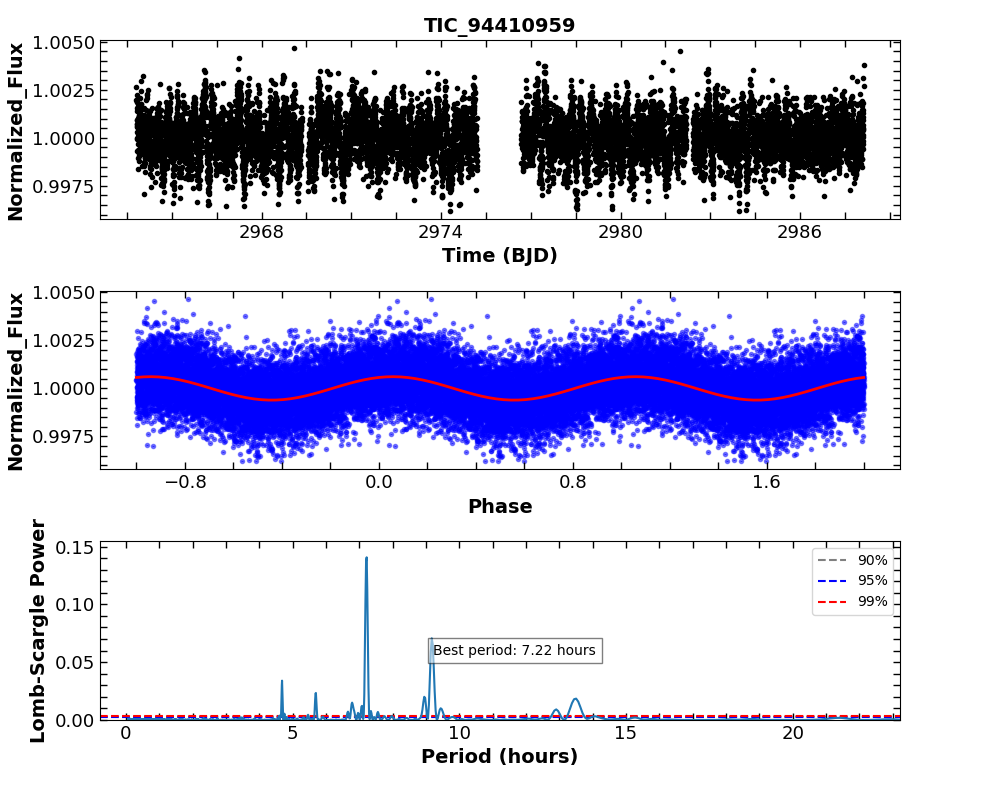}
    \includegraphics[width=5.9cm,height=7.8cm]{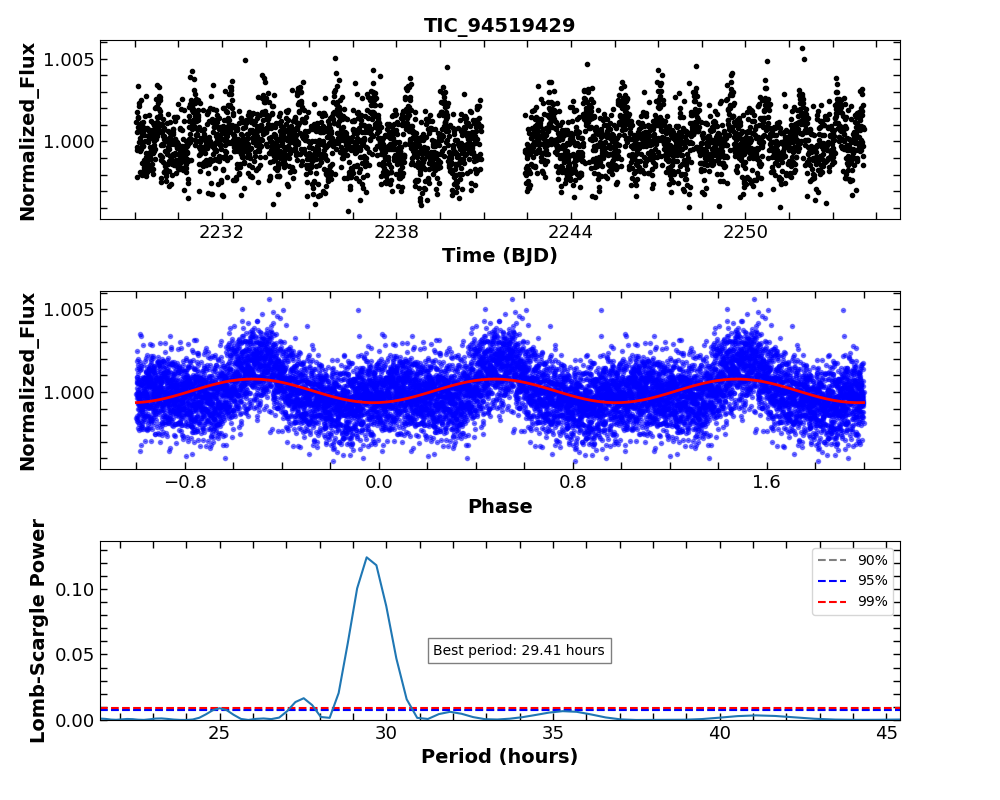}
    \caption{Continued.}
    \label{fig: Lomb_scargle_periodogram1}
\end{figure*}

\begin{table*}
\centering
\vspace{-0.3cm}

\caption{Basic parameters of the identified variable stars in the observed clusters. The columns list the TIC ID, right ascension (RA) and declination (DEC) in J2000 coordinates, period (in hours), effective temperature ($T_{\rm eff}$ in K), membership status, and type of variability. Membership is denoted as 'M' for cluster members and 'F' for field stars. The variability types include $\gamma$ Dor, SPB, and EW-type eclipsing binaries. The stars are grouped according to their respective clusters: NGC 2204, NGC 2262, Pismis 18, and NGC 2437.}

\scriptsize  
\setlength{\tabcolsep}{3pt}  
\renewcommand{\arraystretch}{1.1}  
\begin{tabular}{|c c c c c c c c c|}
\hline
ID & RA (J2000)  & DEC (J2000) & Period (h) & T$_{eff}$ K & log (T$_{eff}$) K   & log(L/L$_\odot$)   & Membership  & Type  \\
\hline
&  &   &  &   \bfseries NGC 2204 & &  &  & \\
\hline
 TIC 59711686 &  93.8615525358511 & -18.8423139753127  &  21.08 & 6693.41 $\pm$ 125  &   3.825  &  0.5258  &  F  & Misc \\
\hline

&  &  &  &  \bf{NGC 2262} & &  &  & \\
\hline
TIC 369672843 &  99.9949396680587  &  1.1005280367638  &  14.97  & 7144.00 $\pm$ 140 & 3.8539  & 1.0196 &  F  & $\gamma$ dorodus \\
\hline
&  &   &   & \bf{Pismis 18}  & &  &  & \\
\hline
TIC 318170024 & 204.26503754627 & -62.1932090248373 &  7.53 & 5861.0 $\pm$ 330 &  3.7679  & 0.8179   & F & EW    \\ 

TIC 318170016 & 204.208928184582 & -62.1923172179637 &  7.53 & 3305.0 $\pm$ 110 & 3.5192   & 2.2570     & F  & Misc  \\
\hline

&  & &  & \bf{NGC 2437}  & &  &  & \\
TIC 94690694	& 115.823510242526 & -14.8329139196625	& 12.399 & 7063.26 $\pm$ 160 & 3.8490     & 1.7134  &  F & Misc      \\
TIC 94409620	& 115.510616181402 & -14.7741055562133	& 176.26 & 9233.0 $\pm$ 350 & 4.0151 & 2.062 &  M & Misc     \\
TIC 94322635 & 115.345435822958 &  -14.8197453274406 & 47.04 & 7870.25 $\pm$ 240 & 3.9514  & 3.188  & M &  YSS     \\
TIC 94409484 & 115.470463684627 & -14.7423028495124 & 16.93  & 8541.67 $\pm$ 170  & 3.9734  &   2.169   &  M &  Misc      \\
TIC 94519429 & 115.647315663641 &  -15.0468249431276 &  29.41 & 9754.67 $\pm$ 370 & 4.0898   & 2.249 & M &  SPB     \\
TIC 94322074 & 115.378758817249 & -14.9394523881991 & 20.63 & 8745.67 $\pm$ 310 & 3.9466  & 2.096      & M & Misc    \\
TIC 94410959 & 115.560324659748 & 15.1116879741543 & 7.22 & 8330 $\pm$ 180 & 3.9207 & 2.623 & F & Misc\\
TIC 94229743 & 115.278194656648 & -14.9725638053055 & 266.4  & 4887.25 $\pm$ 50  &  3.8490   &  1.7134 & F &  EA \\
\hline
  \end{tabular}
\label{tab: Variable_Stars_detials}
\end{table*}

\subsection{PHOEBE Modeling}
\label{sec: Phoebe_modeling}
\begin{figure}
\vspace{-0.0cm}
\hspace{-0.0cm}
    \includegraphics[width=9.5cm,height=5.2cm]{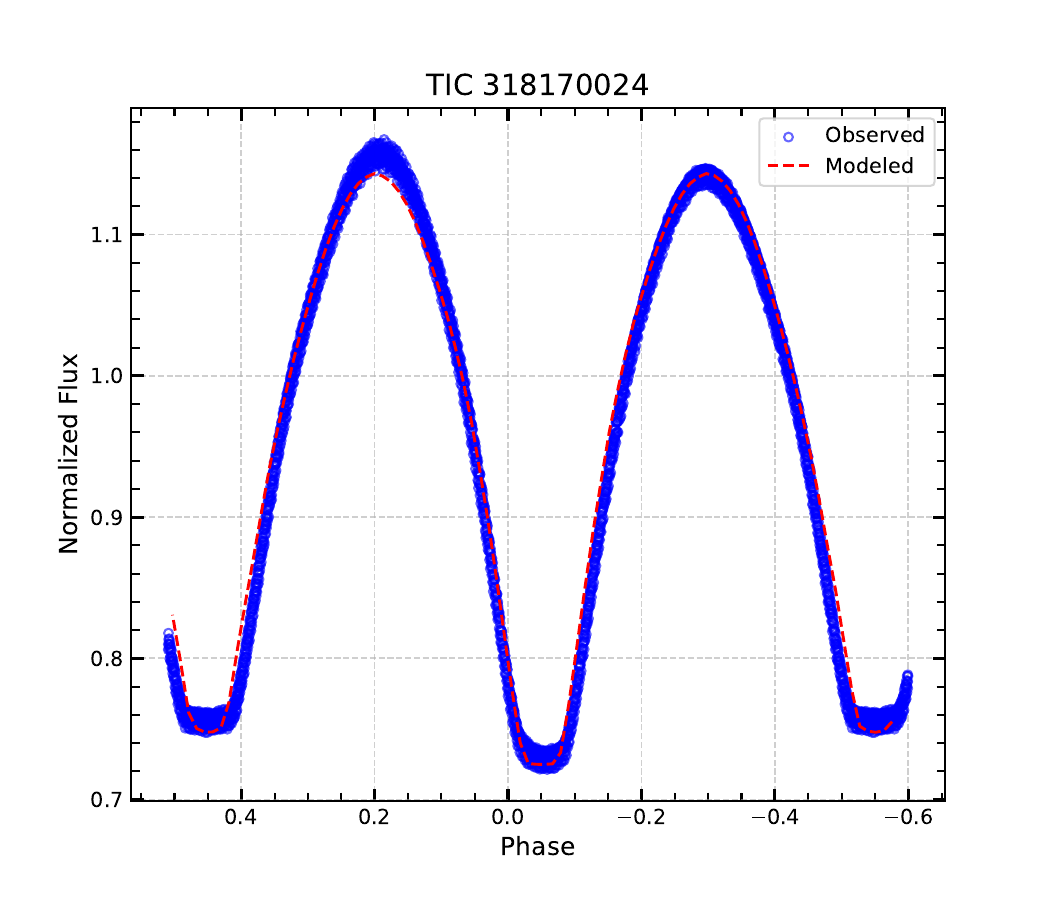}
\hspace{-0.0cm} 
    \includegraphics[width=9.5cm,height=5.2cm]{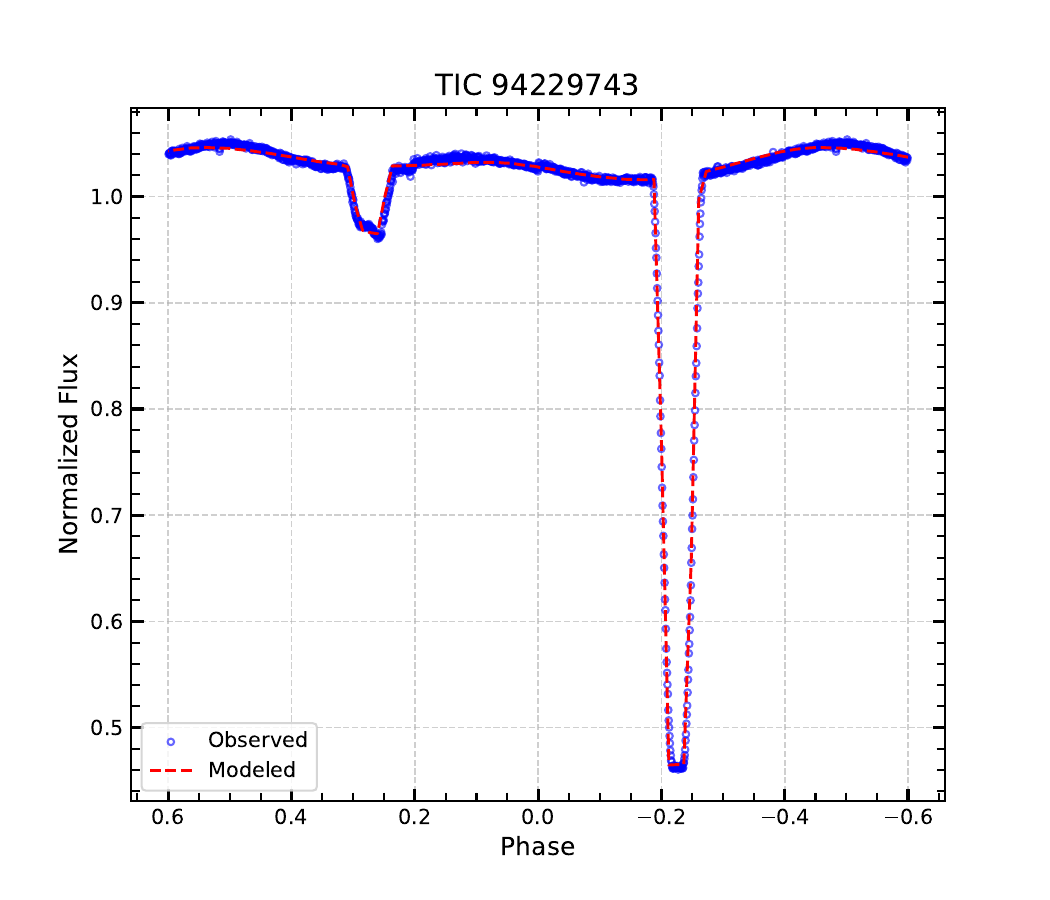}
    \caption{Phase-folded light curves of TIC~318170024 (Pismis~18; top panel) and TIC~94229743 (NGC~2437; bottom panel). Blue points represent the observed TESS photometric data, while the red dashed lines indicate the best-fit models generated using the \texttt{PHOEBE~1.0} software. These models were used to derive the orbital and physical parameters of the binary systems.}
    \label{fig:modeled_lightcurve}
\end{figure}

Eclipsing binaries show periodic variations in brightness due to their orbital geometry and alignment relative to the observer. Through photometric and spectroscopic techniques, key physical parameters such as orbital period, mass, distance, and stellar radii can be accurately determined. Continuous monitoring of these systems provides valuable insight into their formation and evolutionary processes. In our study, two eclipsing binaries were identified in the open clusters NGC 2437 and Pismis 18. The light curves of these systems were analyzed using the PHOEBE 1.0 (Physics of Eclipsing Binaries; \citep{prvsa2005computational}) modeling software. PHOEBE 1.0, which we used to model eclipsing binaries, is based on the Wilson–Devinney code originally written in FORTRAN. It provides both a graphical user interface (GUI) and a scripting tool, facilitating flexible and detailed analysis.\\ 
Two binary stars, TIC 94229743 (NGC 2437) and TIC 318170024 (Pismis 18), are selected as a detached binary and an overcontact binary, respectively, as the fitting models in PHOEBE. The effective temperature estimates available in \citet{2022yCat.1355....0G} are used as the T$_{eff}$ of the primary component for both sources.  Stars with effective temperatures (T$_{eff}$ $\leq$ 7200 K) are assumed to possess convective envelopes. For these stars, the surface albedo (A), representing the fraction of light reflected, is set to 0.5, whereas the gravity brightening coefficient (g), which accounts for the variation in surface brightness from pole to equator due to rotation, is taken as 0.32. In contrast, stars with radiative envelopes (T$_{eff}$ $\geq$ 7200 K) are assigned values of A = 1.0 and g = 1.0. The software automatically updates the limb darkening coefficients after each iteration, based on the tables provided by \citet{van1993new}. Determining the mass ratio $ (q = m_{2}/m_{1})$ is the first step in modeling the photometric light curve. In the absence of radial velocity data, we employed the q-search method to estimate the mass ratio. Many authors have used this approach \citet{panchal2021photometric,li2023five,panchal2023optical}. In this approach, known parameters such as the time of minimum light, orbital period, and effective temperature (T$_{eff}$) of the primary component are held constant. Meanwhile, other parameters, including the orbital inclination ($i$), the effective temperature of the secondary component, the surface potential ($\Omega$), and the luminosities of both components ($l_{1}$ and $l_{2}$) are treated as free parameters during the fitting process. q was varied in small increments, and for each value, a best-fit model was obtained through multiple iterations. The corresponding cost function was evaluated, and the q value that minimized the cost function was adopted as the optimal mass ratio. This q-search technique was applied to both binary systems, yielding estimated mass ratios of 1.37 for TIC 94229743  and 2.16 for TIC 318170024. Based on these values, we ran the PHOEBE scripter iteratively to refine the final binary star models. The fitted light curves are displayed alongside the observed data in Figure \ref{fig:modeled_lightcurve} and the corresponding parameters are shown in Table~\ref{tab: Phoebe_parameters}.
 
In the dataset, the light curve of system TIC 94229743 exhibits a slight brightness asymmetry. To account for this, one dark spot was placed on each stellar component—one on the primary and one on the secondary. The position of a spot on a stellar surface is defined by its colatitude and longitude. The spot colatitude refers to the angular distance from the star’s North pole (ranging from 0 to 180$^{\circ}$). In contrast, the spot longitude is measured counterclockwise from the direction of the companion star (ranging from 0 to 360$^{\circ}$). The spot radius (in degrees) defines the spot's size, and the temperature ratio is the ratio of the spot temperature to the local photospheric temperature. For the spot on the primary component, the adopted parameters are: colatitude = 90$^{\circ}$ (fixed), longitude = 78$^{\circ}$ $\pm$ 4$^{\circ}$, spot radius = 22$^{\circ}$ $\pm$ 2$^{\circ}$, and temperature ratio (T$_{spot}$ / T$_{star}$) = 1.035. For the secondary component, the spot parameters are: colatitude = 90$^{\circ}$ (fixed), longitude = 164$^{\circ}$ $\pm$ 4$^{\circ}$, spot radius = 20$^{\circ}$ $\pm$ 4$^{\circ}$, and temperature ratio (T$_{spot}$ / T$_{star}$) = 0.920.\\
The evaluated parameters from the PHOEBE light curve modeling should be considered preliminary; improved accuracy will require long-term, multiband photometric observations of these systems. It will also allow for the detection of phenomena such as period changes and stellar surface activity.

\begin{table}
\centering
\vspace{-0.0cm}
\caption{Absolute parameters derived from the light curve modeling of TIC 94229743 (NGC 2437) and TIC 318170024 (Pismis 18) using the PHOEBE 1.0 software. The parameters include: orbital period (in days), mass ratio ($q = M_2/M_1$), semi-major axis ($a$ in $R_\odot$), orbital inclination ($i$ in degrees), effective temperatures of the primary ($T_{1\mathrm{eff}}$) and secondary ($T_{2\mathrm{eff}}$) components in Kelvin, luminosities of the components ($l_1$, $l_2$ in $L_\odot$), surface potentials ($\Omega_1$, $\Omega_2$), and radii ($R_1$, $R_2$ in $R_\odot$) and masses ($M_1$, $M_2$ in $M_\odot$) of the components.}
\begin{tabular}{ |c |c |c|}
\hline
Parameters & \bf{NGC 2437} & \bf{Pismis 18} \\
           & TIC 94229743  & TIC 318170024 \\
\hline
\hline
q &  1.37  &  2.16   \\
i $\deg$ & 86.6   &  89.8     \\
T$_{1eff}$ K (fixed) & 4887    &  5861  \\
T$_{2eff}$ K & 6670   &  6100 \\  
l$_{1}$ (L$_\odot$)  & 3.958    &  5.47   \\
l$_{2}$ (L$_\odot$)  & 3.404    &  4.87   \\
$\Omega_{1}$ & 16.30  &  5.05   \\
$\Omega_{2}$ & 7.55   &  $\Omega_{1}$  \\  
R$_{1}$ (R$_\odot$) & 1.20    & 1.446      \\
R$_{2}$ (R$_\odot$) & 3.12    & 1.953      \\
M$_{1}$ (M$_\odot$) & 0.308    & 0.641      \\
M$_{2}$ (M$_\odot$) & 0.348    & 1.385      \\

\hline
  \end{tabular}
\label{tab: Phoebe_parameters}
\end{table}

\section{Comparative Discussion of the Six Clusters} \label{sec: discussion}
A comparative assessment of the six clusters (NGC 2204, NGC 2660, NGC 2262, Czernik 32, Pismis 18, and NGC 2437) based on the results in Tables \ref{tab: kings model parameters}, \ref{tab: mass function slope} and \ref{tab:orbits} highlights both shared evolutionary behaviour and cluster-specific differences. We find only a weak tendency for mass function (MF) slopes to flatten in older systems when considering age. Clusters such as NGC 2204 ($\sim$1.9 Gyr) and NGC 2660 ($\sim$1.5 Gyr) have flatter MF slopes. They also show more centrally concentrated stellar distributions. In contrast, younger clusters like NGC 2437 ($\sim$200 Myr) and Pismis 18 ($\sim$0.7 Gyr) retain larger fractions of low-mass stars in their halos. This finding broadly supports the idea that dynamical evolution gradually depletes low-mass members and enhances mass segregation. However, the trend remains modest and is not statistically significant across the limited age range of our sample. The mass segregation indices appear nearly constant. This suggests that segregation may be primordial or has reached saturation in these intermediate-age clusters. The comparison of total masses (Table 4) with structural parameters (Table 2) reveals that mass and size are not tightly correlated. The most massive cluster, NGC 2437 ($\sim$2450 $M_{\odot}$), is also the most extended system, with a large half-mass radius. Clusters of lower mass, such as NGC 2262 ($\sim$ 268$M_{\odot}$) and Czernik 32 ($\sim$ 350$M_{\odot}$), are markedly more compact. This suggests that cluster size is not determined solely by mass. Instead, initial formation conditions and subsequent dynamical processing also influence size. Clusters with similar masses can show substantial differences in compactness. This points to the role of environmental effects, such as tidal shocks or disk crossings. The orbital parameters (Table \ref{tab:orbits}) provide further context. All six clusters are confined to nearly circular, thin-disk orbits (eccentricities 0.02–0.10; |z| $\leq$ 132 pc). This configuration helps ensure survival over several hundred Myr. Within this broad similarity, some contrasts are evident. Clusters such as NGC 2204, located at larger Galactocentric distances on circular orbits, retain higher masses and more extended structures. In contrast, NGC 2660, which is closer to the Galactic plane, is more compact. NGC 2262 and Czernik, having similar age and location in the Galaxy, follow similar orbits and have similar sizes. This likely reflects the stronger tidal stresses and repeated disk crossings experienced in its environment. Taken together, these comparisons indicate that mass segregation and MF flattening are universal dynamical signatures. Cluster longevity and present-day structure are determined by an interplay of intrinsic properties (mass, compactness, relaxation time) and extrinsic conditions (orbital parameters, Galactic tides). No single parameter governs cluster evolution across the sample. The results illustrate the complexity of open cluster survival in the Galactic disk. They also point to the importance of expanding such analyses to larger and more diverse cluster sets.

\section{Conclusions}
\label{sec: Conclusion}
In this work, we carried out a comprehensive investigation of six Galactic open clusters—NGC~2204, NGC~2660, NGC~2262, Czernik~32, Pismis~18, and NGC~2437—by combining high-precision astrometric data from \textit{Gaia} DR3 with time-series photometry from \textit{TESS}. Our study focused on determining accurate cluster memberships, structural and fundamental parameters, dynamical states, stellar content (including stragglers and variables), and orbital properties. Below, we summarize the key results of our analysis:

\begin{enumerate}
    \item We have applied Bayesian and Gaussian Mixture Model (GMM) approaches to determine high-probability cluster members in all studied clusters. The GMM method proved more reliable in minimizing field star contamination, particularly for faint stars with magnitudes between 18 and 20, ensuring a cleaner and more precise selection of cluster membership.
    
    \item After fitting King's formula, the structural parameters were calculated for all the clusters, with core radii ranging from 1 to 10 arcminutes and overall cluster radii spanning 5 to 24 arcminutes. Additionally, the density contrast parameter was estimated to quantify the degree of central concentration. The results show that all the clusters in this study exhibit a relatively high degree of compactness.  
    
    \item Fundamental parameters, including distance, age, and Galactocentric coordinates, were accurately determined using Gaia DR3 data, demonstrating consistency with previous studies and validating the methodologies adopted. The cluster ages, estimated using theoretical isochrones from \citet{marigo2017new}, range from 0.2 to 2 Gyr, highlighting a diverse set of evolutionary stages. Distances, derived using the Bailer-Jones method, range from 1.76 to 4.20 kpc. All clusters in this study are classified as intermediate-aged and are located within -1.1 to 1.2 kpc from the Galactic plane.
    
    \item We identified a total of 13 Blue Straggler Stars (BSS) and 3 Yellow Straggler Stars (YSS) across multiple OCs, with 7 BSS and 2 YSS in NGC 2204, 2 BSS in NGC 2660, 2 BSS in Czernik 32, 1 BSS in Pismis 18, and 1 BSS along with 1 YSS in NGC 2437. Our findings include previously reported and newly identified BSS, expanding the current catalog of these evolved stellar populations. The observed central concentration of BSS within their respective clusters suggests that their formation is primarily driven by mass transfer in binary systems or stellar collisions. These results offer valuable insights into the dynamic evolution of OCs and stellar interaction processes.
    
    \item The LFs and MFs were analyzed using the most probable cluster members for the six clusters studied. The MF slopes, estimated separately for the core and outer regions, range from 0.96 to 1.19, indicating a shallower distribution than the canonical Salpeter slope (1.35). The observed increase in the MF slope toward the outer regions suggests mass segregation, likely driven by dynamical relaxation processes. This flattening of the MF further implies a higher retention of low-mass stars, providing insights into the dynamical evolution of these clusters.

    \item Our orbit analysis shows that all six clusters follow low-eccentricity orbits ($  e\approx 0.23$--$0.10$), with perigalactic and apogalactic distances ranging from 5.9 to 12.5~kpc and maximum vertical excursions $|z| \leq 132$~pc. These nearly circular, planar orbits indicate that the clusters have remained confined to the Galactic thin disk over long timescales, thereby sustaining their observed dynamical relaxation and resisting disruptive forces. Their confinement in the thin disk is resulting in the quick evaporation of faint stars from the clusters.

    \item We identified twelve variable stars across four clusters, including one SPB (TIC 94519429), one $\gamma$ Doradus (TIC 369672843), two eclipsing binaries (TIC 94229743, TIC 318170024), a yellow straggler candidate (TIC 94322635), and several unclassified variables. In the HR diagram, the place and variability properties confirmed their classifications. PHOEBE modeling of the eclipsing binaries demonstrated detached and overcontact configurations, with mass ratios of 1.37 and 2.16, orbital inclinations of 86.6$^\circ$ and 89.8$^\circ$, and component temperatures in the range of 6100--6760 K. These binaries offer robust constraints on stellar parameters and binary evolution in cluster environments.

\end{enumerate}

\section*{Acknowledgements}
 We thank the referee for the constructive comments and suggestions that have helped to improve this manuscript. We are also grateful to the editor for valuable comments regarding the presentation of figures. Ing-Guey Jiang acknowledges the support from the National Science and Technology Council (NSTC), Taiwan, with a grant  NSTC 113-2112-M-007-030. This work has utilized data from the European Space Agency (ESA) mission \emph{Gaia} (https://www.cosmos.esa.int/gaia), processed by the \emph{Gaia} Data Processing and Analysis Consortium (DPAC, https://www.cosmos.esa.int/web/gaia/dpac/consortium). The DPAC is funded by national institutions, particularly those participating in the \emph{Gaia} Multilateral Agreement.
The Digitized Sky Surveys were produced at the Space Telescope Science Institute under U.S. Government grant NAG W-2166. This study also utilizes data obtained by the Transiting Exoplanet Survey Satellite (\emph{TESS}) mission, funded by NASA's Explorer Program.

 Python modules: NUMPY (1.19.0;\citep{harris2020array}), SCIPY (v1.3.1 \citep{Eri01}), and MATPLOTLIB (v3.1.1 \citep{hunter2007matplotlib})

\section*{Data Availability}

This work is based on two publicly available data sets: \textit{Gaia} DR3 and \textit{TESS}, for the clusters NGC 2204, NGC 2660, NGC 2262, Czernik 32, Pismis 18, and NGC 2437. The \textit{Gaia} DR3 data can be accessed at: https://gea.esac.esa.int/archive/. The \textit{TESS} data used in this study are publicly available through the Mikulski Archive for Space Telescopes (MAST): https://mast.stsci.edu/.



\bibliographystyle{mnras}
\bibliography{ref} 




\appendix

\section{Membership Plot}\label{appendix: appendixA1}

We conducted a comprehensive photometric and kinematic study of six intermediate-age open clusters (OCs): NGC 2204, NGC 2660, NGC 2262, Czernik 32, Pismis 18, and NGC 2437, utilizing data from Gaia DR3. Membership probabilities were evaluated within the regions of these clusters, identifying the most probable members, defined as those with membership probabilities exceeding 70$\%$.

\begin{figure*}
    \centering
    \vspace{-0.3cm}
    \includegraphics[width=16cm,height=4.7cm]{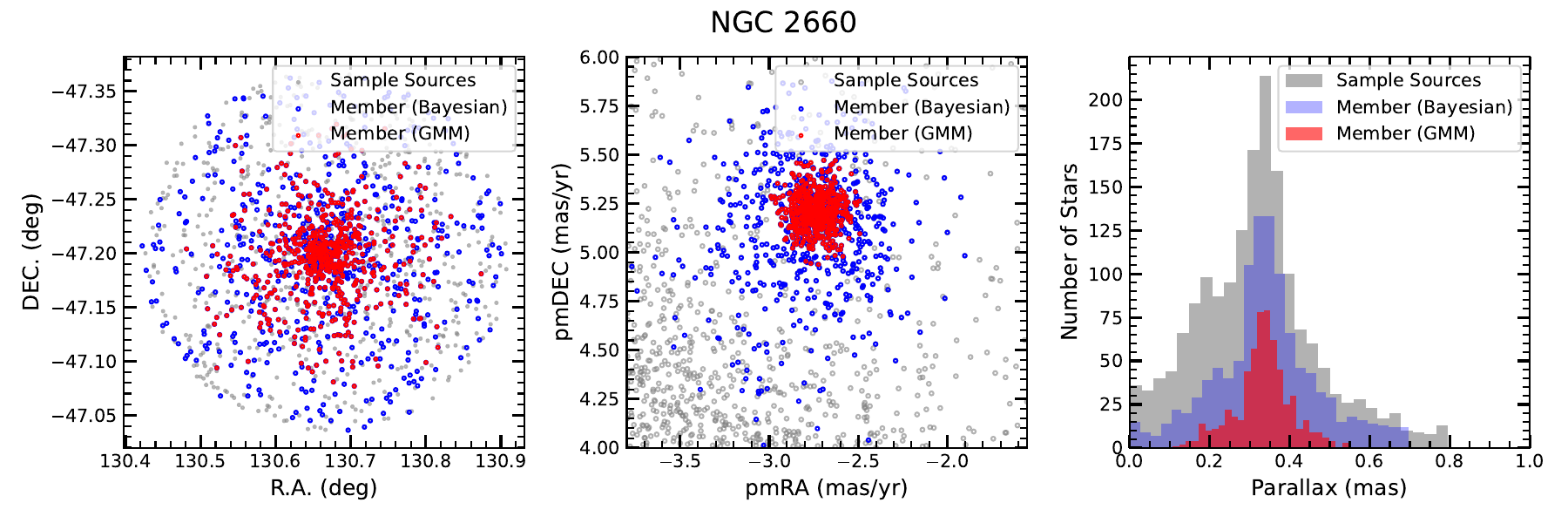}
    \hspace{0.3cm}\includegraphics[width=15.8cm,height=4.7cm]{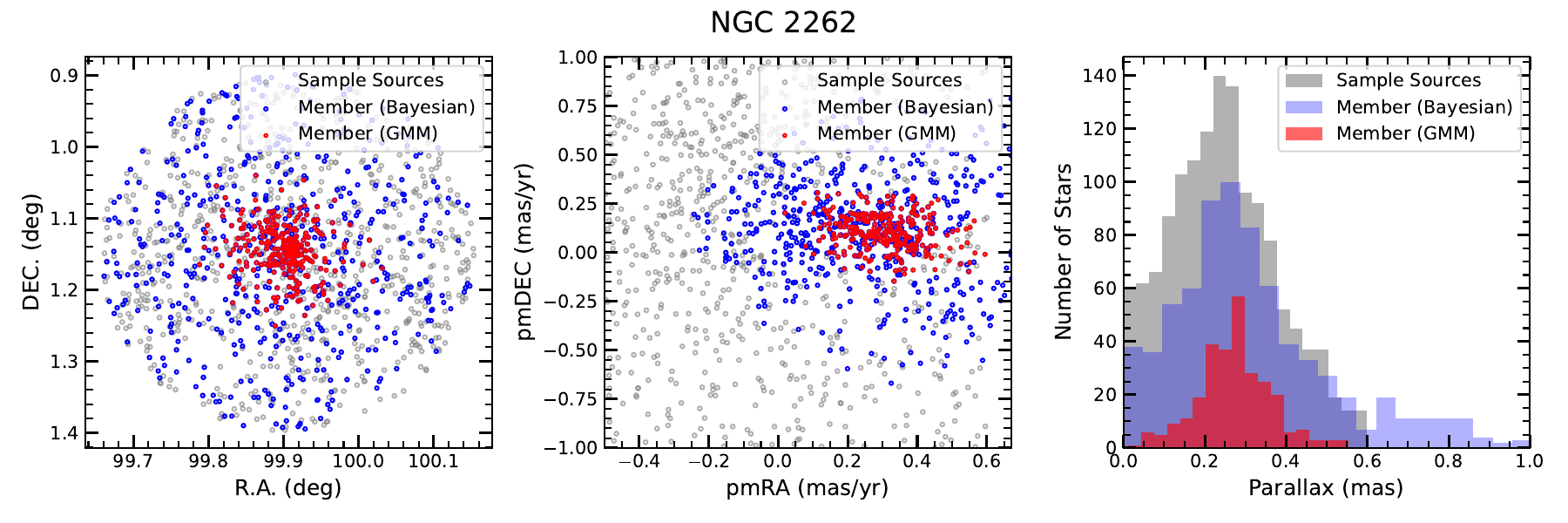}
    \includegraphics[width=16cm,height=4.7cm]{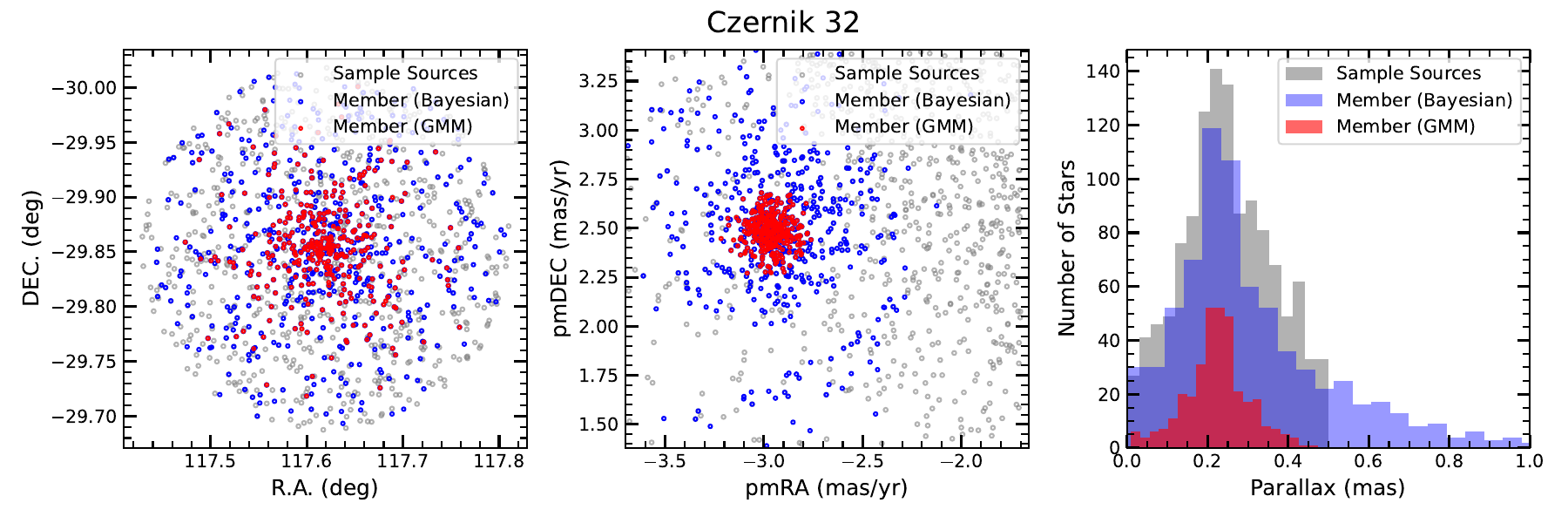}
    \includegraphics[width=16cm,height=4.7cm]{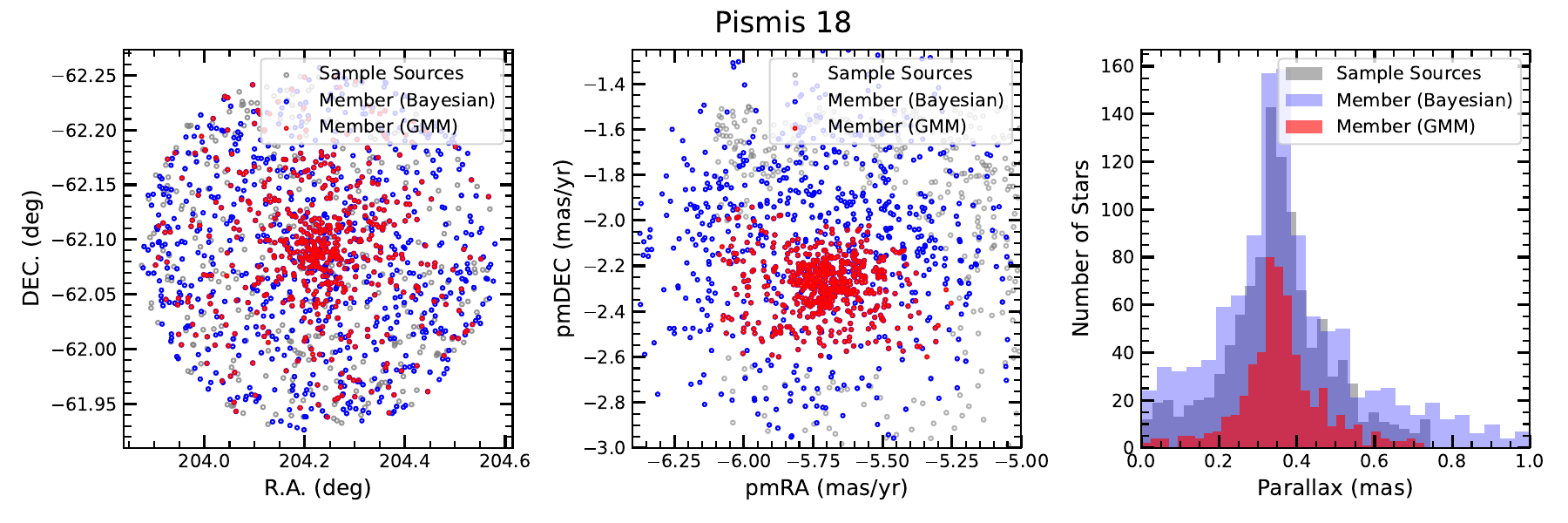}
    \includegraphics[width=16cm,height=4.7cm]{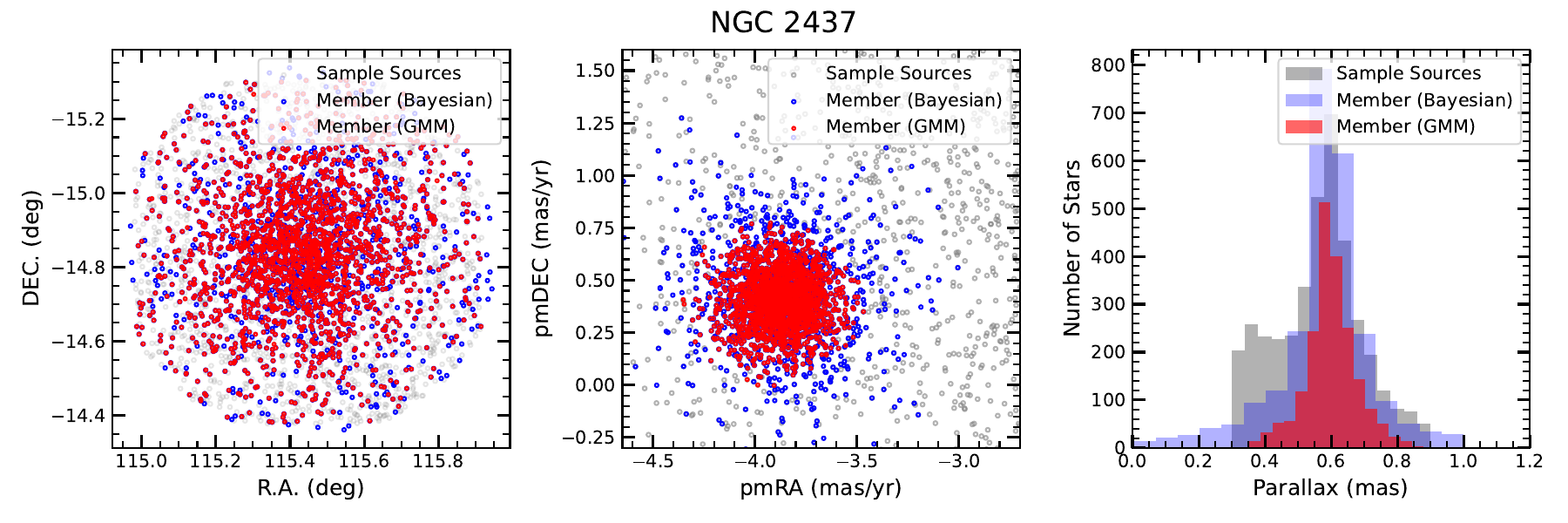}
    \caption{(a) The spatial distribution, (b) proper motion distribution, and (c) parallax distribution using ML-MOC algorithm and Bayesian approach on Gaia DR3 data.}
    \label{fig: gmm_member_plot}
\end{figure*}

\section{Comparsion Plot}\label{appendix: appendixA2}

\begin{figure*}
    \centering
     \vspace{-0.0cm}
    \hspace{-0.6cm}
    \includegraphics[width=4.6cm,height=5.2cm]{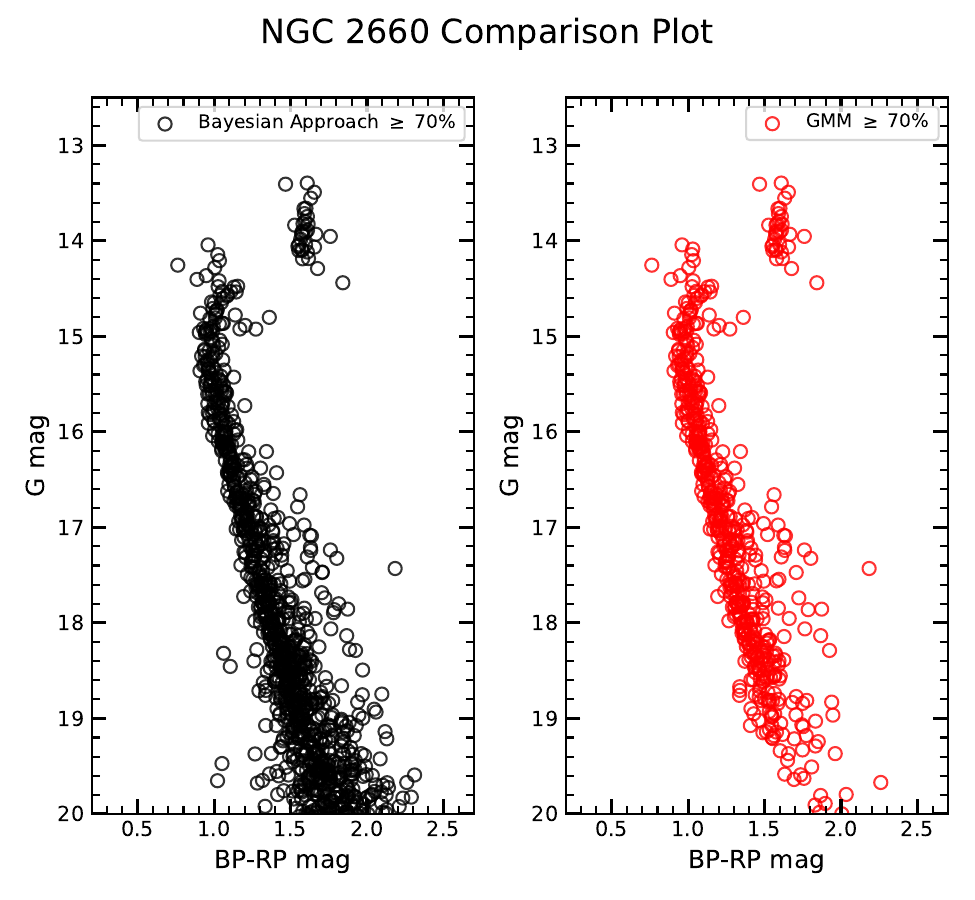}
    \includegraphics[width=4.6cm,height=5.2cm]{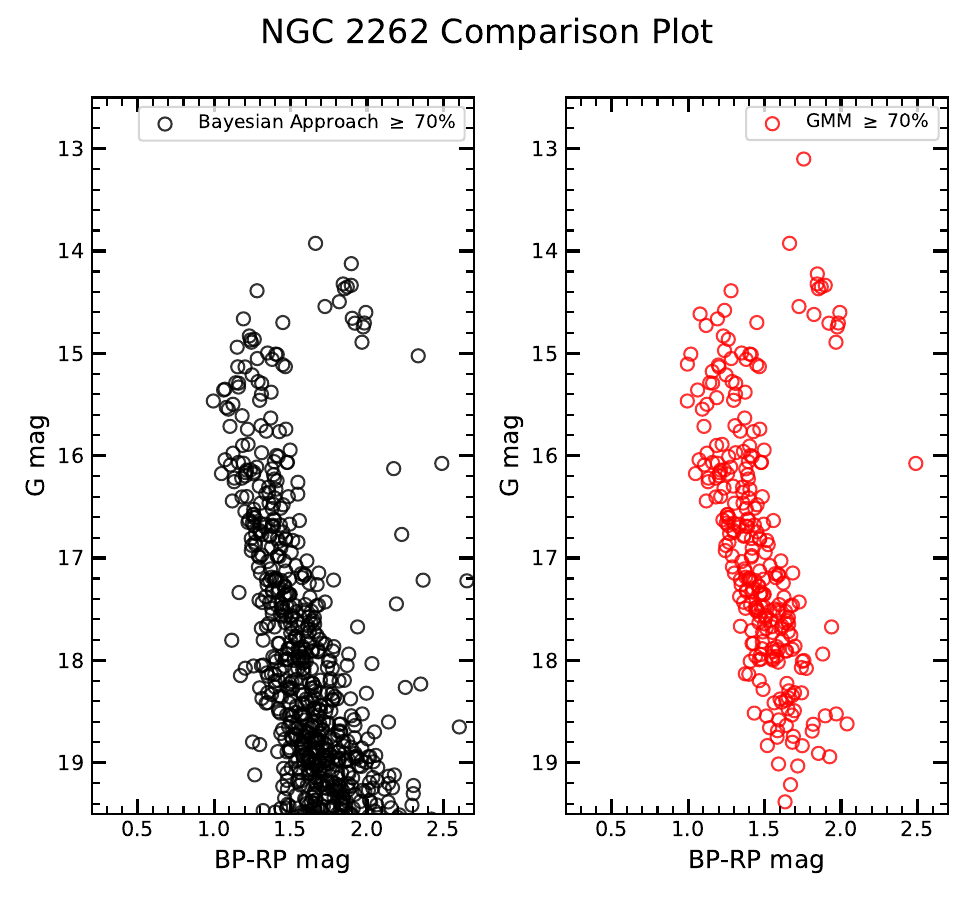}
    \includegraphics[width=4.6cm,height=5.2cm]{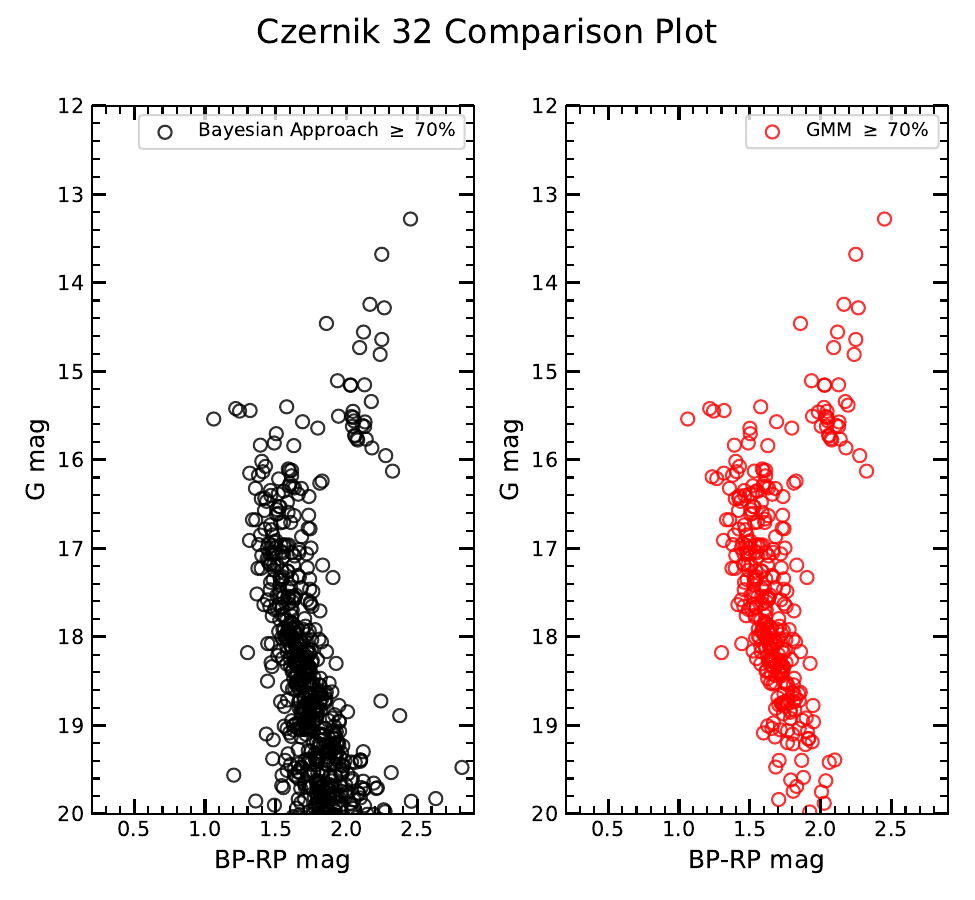}

    \hspace{-0.3cm}
    \includegraphics[width=4.4cm, height= 3.75cm]{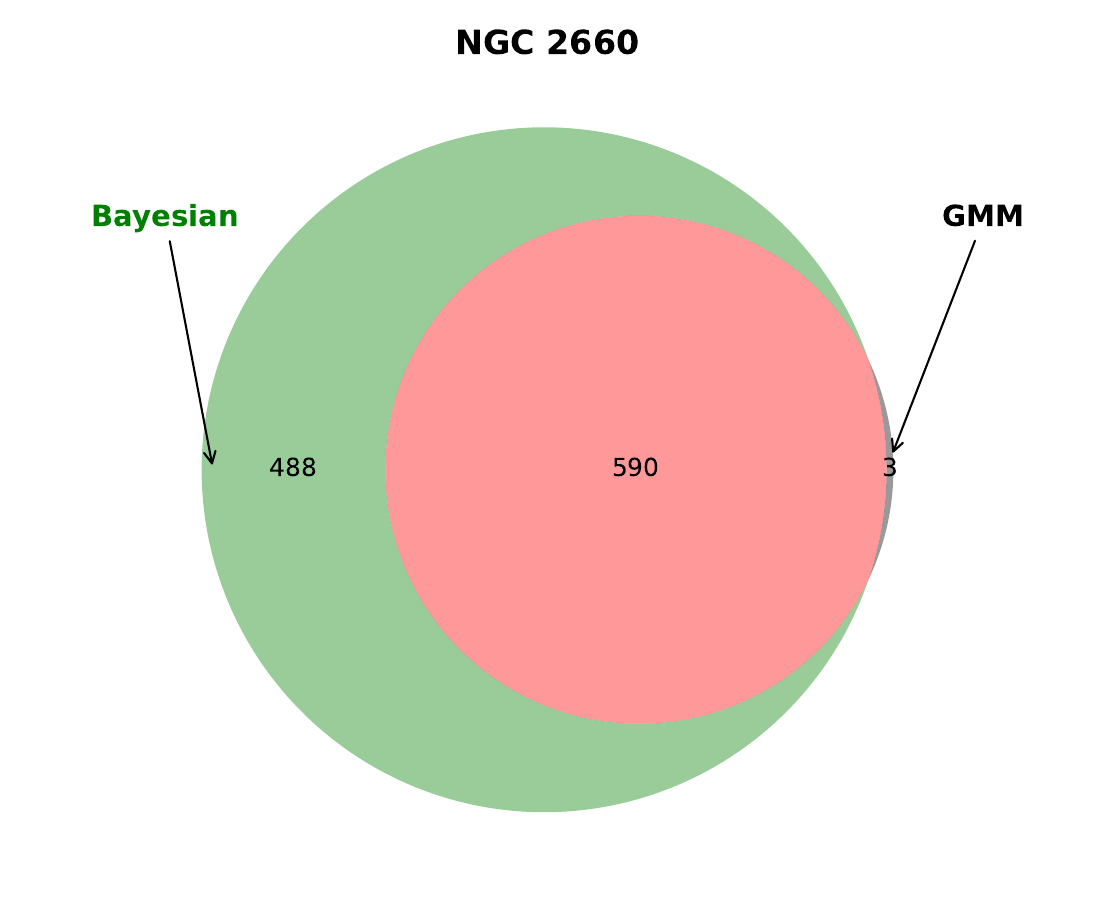}
    \includegraphics[width=4.4cm, height= 3.75cm]{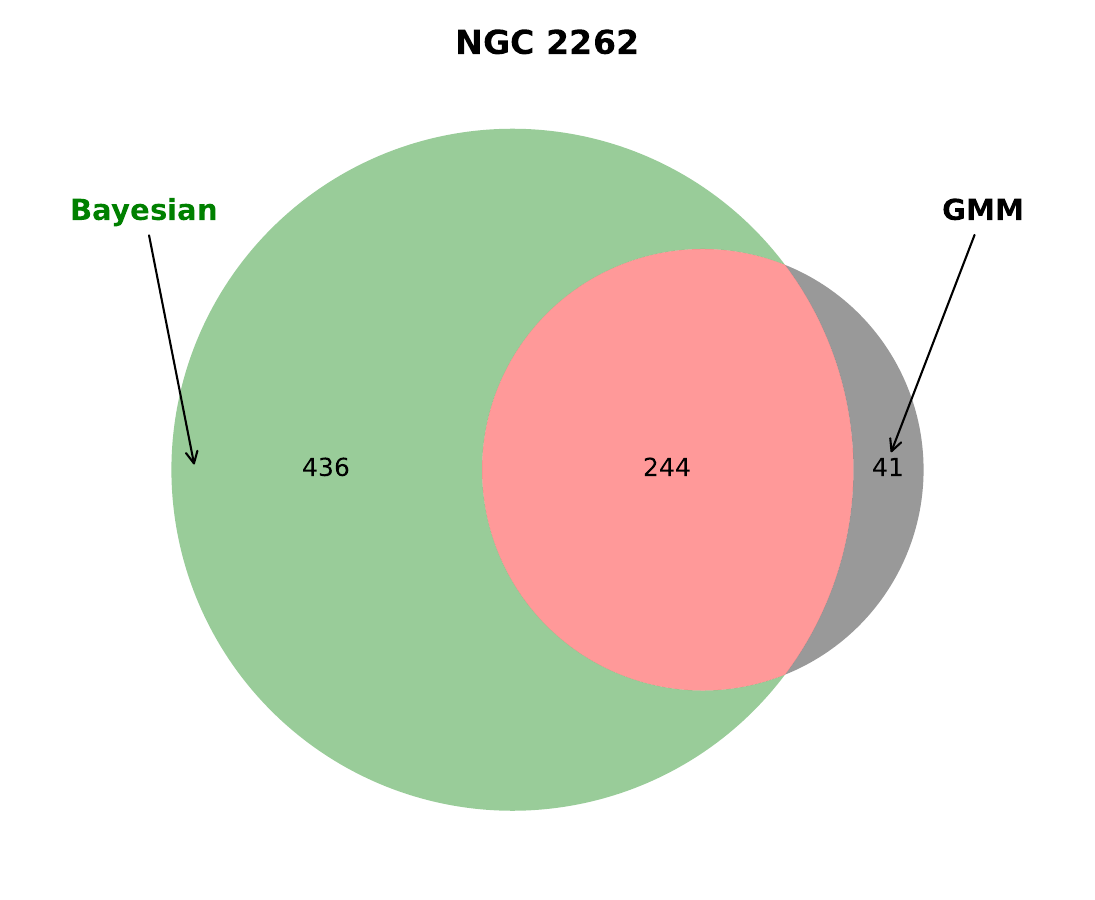}
    \includegraphics[width=4.4cm, height= 3.75cm]{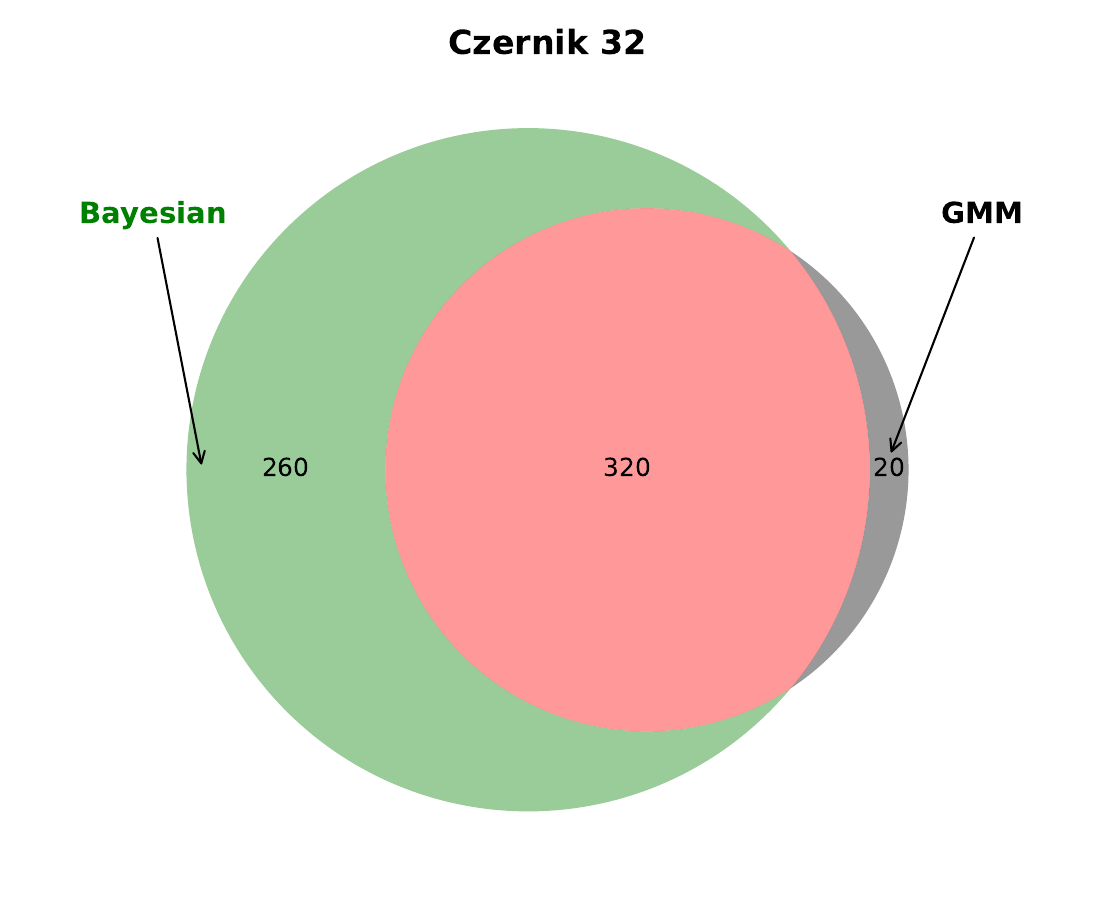}

\end{figure*}

\begin{figure*}
    \centering
    \vspace{-0.1cm}
    \hspace{-0.6cm}
    \includegraphics[width=4.6cm,height=5.2cm]{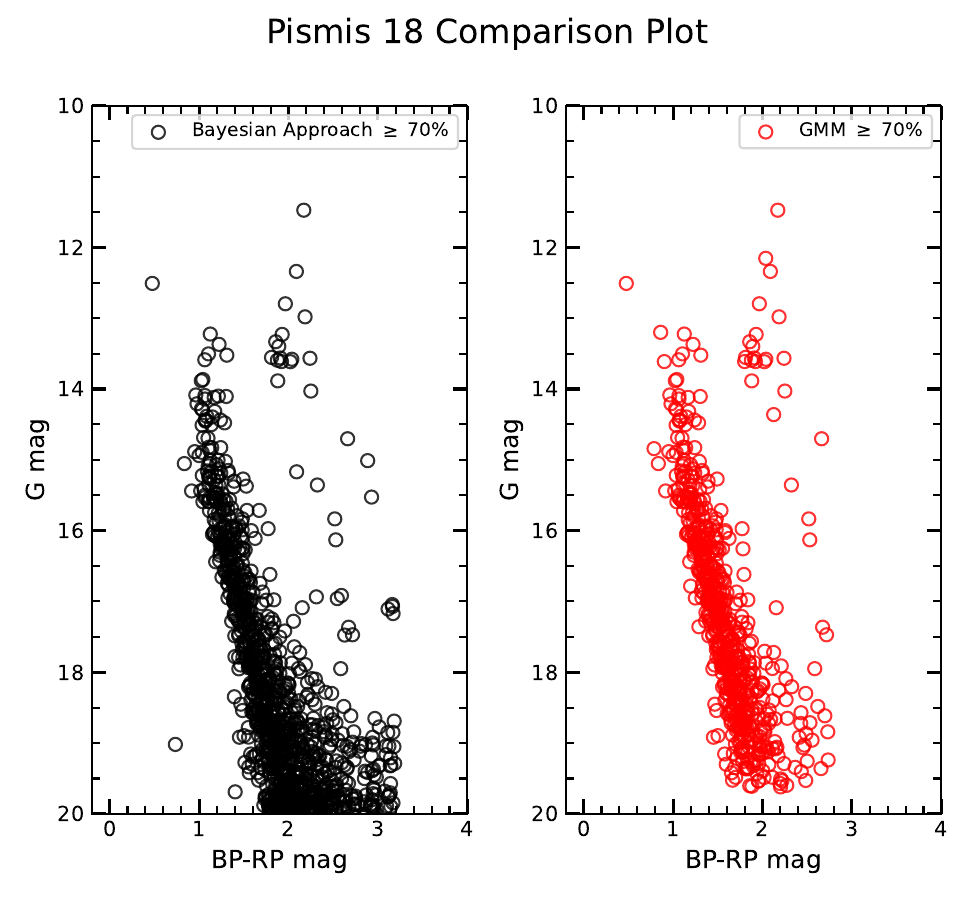}
    \includegraphics[width=4.6cm,height=5.2cm]{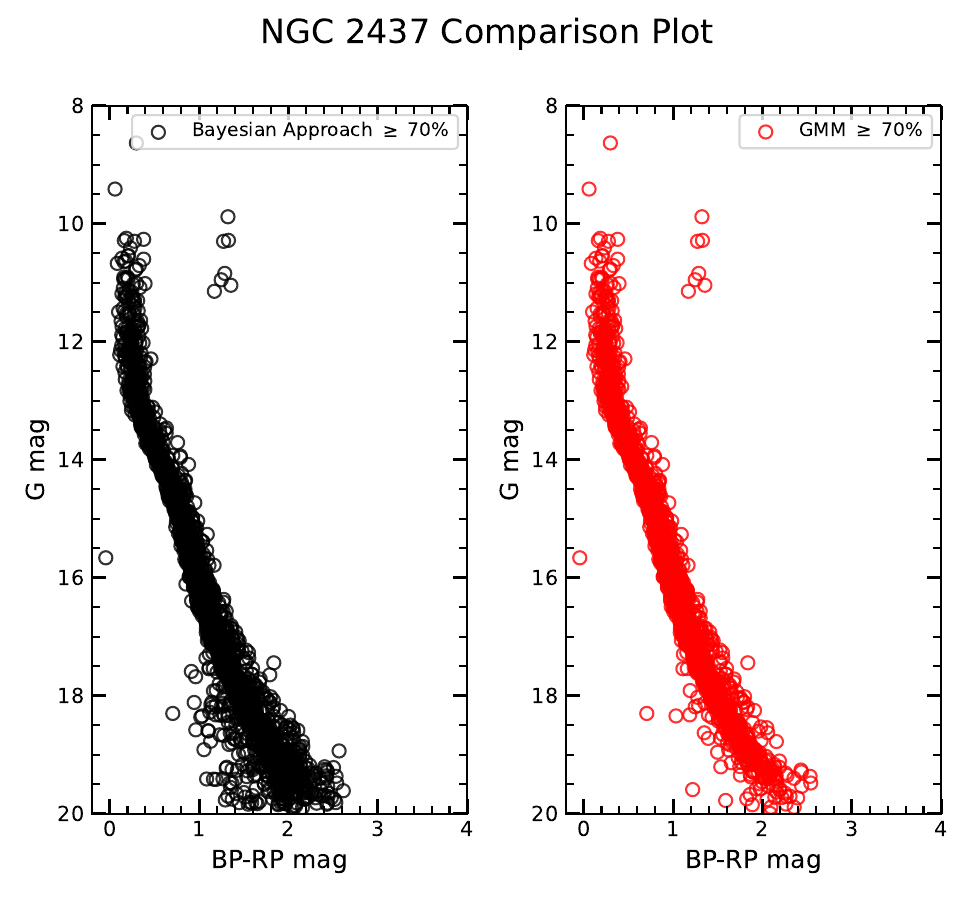}
    \includegraphics[width=4.4cm, height= 3.75cm]{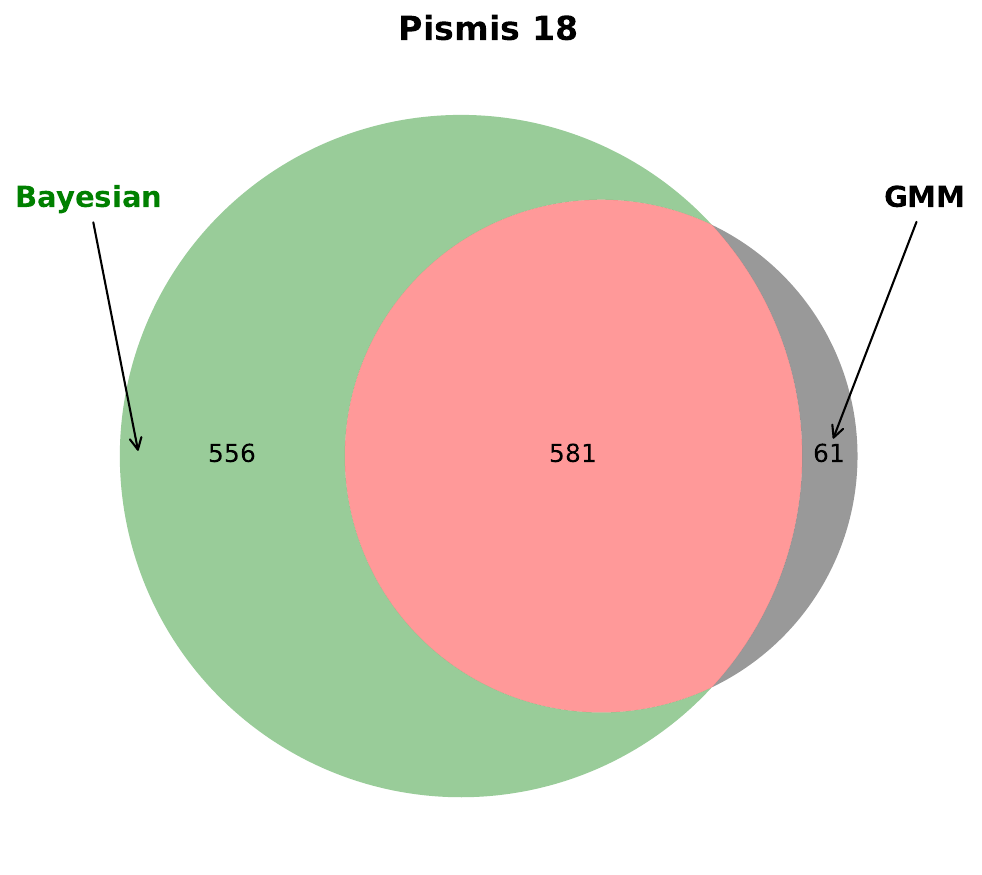}
    \includegraphics[width=4.4cm, height= 3.75cm]{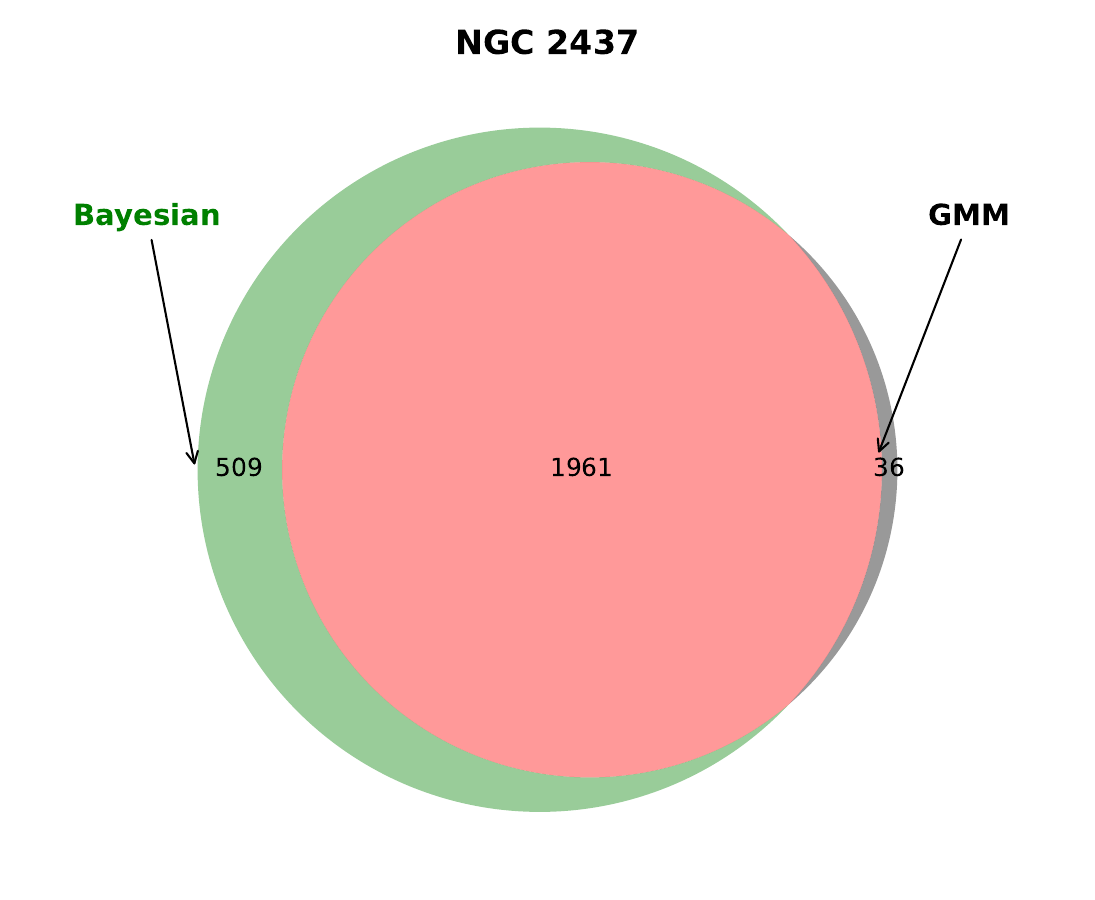}
    \caption{Comparison plot and Venn diagram for the Bayesian approach and Gaussian Mixture Model using cluster members with a probability of $\geq$ 70$\%$.}
    \label{fig: comparsion_plot}
\end{figure*}


\bsp	
\label{lastpage}
\end{document}